%% file: CFD_MPC_paper.tex
\documentclass[review]{elsarticle}
\makeatletter
\def\ps@pprintTitle{%
 \let\@oddhead\@empty
 \let\@evenhead\@empty
 \def\@oddfoot{\centerline{\thepage}}%
 \let\@evenfoot\@oddfoot}
\makeatother

\pdfoutput=1
\usepackage{graphicx}
\usepackage{amsfonts, amsmath, amssymb}
\usepackage{mathabx}
\usepackage{booktabs,siunitx}
\usepackage[svgnames,table]{xcolor}
\usepackage[tableposition=above]{caption}
\usepackage{pifont}
\usepackage{bm}
\usepackage{cancel}
\usepackage{caption}
\usepackage{color}
\usepackage{enumerate}
\usepackage{float}
\usepackage{hyperref}
\usepackage[english]{babel}
\usepackage[margin=1in]{geometry}
\usepackage{mathtools}
\usepackage{multirow}
\usepackage{setspace}
\usepackage{subfigure}
\usepackage{lineno}
\usepackage[ruled,linesnumbered]{algorithm2e}
\RestyleAlgo{ruled}
\SetKwComment{Comment}{/* }{ */}

\SetCommentSty{mycommfont}

\renewcommand \d [2]{\frac{{\rm d} #1}{{\rm d} #2}}
\renewcommand \d [1]{\rm{d} #1}
\newcommand \D [2]{\frac{\partial #1}{\partial #2}}

\renewcommand{\vec}[1]{\bm{\mathrm{#1}}}
\newcommand{\V}[1]{\bm{\mathrm{#1}}}

\renewcommand \tt [1]{\texttt{#1}}

\def \div{\nabla \cdot \mbox{}}
\def \grad{\nabla}

\def \x{\vec{x}}

\def \n{\vec{n}}

\def \u{\vec{u}}

\def \N{\mathbb{N}}

\def \U{\vec{U}}
\def \L{\vec{L}}

\def \cJu{\bm{\mathcal{J}_{\boldsymbol{u}}}}
\def \cJv{\bm{\mathcal{J}_{\boldsymbol{v}}}}

\def \xr{{\boldsymbol{x}}_{\boldsymbol{r}}}
\def \Ar{{\bf{A}}_{\boldsymbol{r}}}
\def \Br{{\bf{B}}_{\boldsymbol{r}}}
\def \Cr{{\bf{C}}_{\boldsymbol{r}}}
\def \Ac{{\bf{A}}_{\boldsymbol{c}}}
\def \Bc{{\bf{B}}_{\boldsymbol{c}}}
\def \Cc{{\bf{C}}_{\boldsymbol{c}}}
\def \Xc{{\bf{X}}_{\boldsymbol{c}}}
\def \Zc{{\bf{Z}}_{\boldsymbol{c}}}
\def \Ad{{\bf{A}}_{\boldsymbol{d}}}
\def \Bd{{\bf{B}}_{\boldsymbol{d}}}
\def \Fd{{\bf{F}}_{\boldsymbol{d}}}
\def \Cd{{\bf{C}}_{\boldsymbol{d}}}
\def \Xd{{\bf{X}}_{\boldsymbol{d}}}
\def \Zd{{\bf{Z}}_{\boldsymbol{d}}}
\def \Zdmax{{\bf{Z}}_{\boldsymbol{d}}^{\rm max}}
\def \Zdmin{{\bf{Z}}_{\boldsymbol{d}}^{\rm min}}
\def \Deltau{{\boldsymbol{\underline{\Delta u_d}}}}
\def \Deltav{{\boldsymbol{\underline{\Delta v_d}}}}
\def \Sb{S_\text{b}}

\def \C{\vec{C}}

\def \g{\vec{g}}

\def \N{\mathbb{N}}
\def \Nx{N_x}
\def \Ny{N_y}
\def \Nz{N_z}

\def \Omegag{\Omega_{\text{g}}}

\def \cP{{\mathcal{P}}}
\def \vcP{\vec{\mathcal{P}}}

\def \R{\mathbb{R}}
\def \U{\vec{U}}

\def \W{\vec{W}}

\def \X{\vec{X}}
\def \Xcom{\X_{\text{com}}}

\def \cH{\mathcal{H}}
\def \cT{\mathcal{T}}

\def \cO{\mathcal{O}}

\def \f{\vec{f}}

\def \half{\frac{1}{2}}
\def \3half{\frac{3}{2}}
\def \5half{\frac{5}{2}}

\def \n{\vec{n}}

\def \nref{n_{\text{ref}}}
\def \ncells{n_{\text{cells}}}
\def \ncycles{n_{\text{cycles}}}

\def \rhow{\rho_{\text{w}}}

\def \u{\vec{u}}

\def \x{\vec{x}}

\def \div{\nabla \cdot \mbox{}}
\def \grad{\nabla}

\def \dt{\Delta t}
\def \dx{\Delta x}
\def \dy{\Delta y}
\def \dz{\Delta z}

\newcommand{\upperRomannumeral}[1]{\uppercase\expandafter{\romannumeral#1}}

\newcommand{\REVIEW}[1]{{#1}}

\begin{document}
\let\today\relax

\begin{frontmatter}
	
\title{A model predictive control (MPC)-integrated multiphase immersed boundary (IB) framework for simulating wave energy converters (WECs)}
\author[SDSU]{Kaustubh Khedkar}
\author[SDSU]{Amneet Pal Singh Bhalla\corref{mycorrespondingauthor}}
\ead{asbhalla@sdsu.edu}

\address[SDSU]{Department of Mechanical Engineering, San Diego State University, San Diego, CA}
\cortext[mycorrespondingauthor]{Corresponding author}

\begin{abstract}
\input{Abstract}
\end{abstract}


\end{frontmatter}

\section{Introduction}
\input{Introduction}

\section{Dynamical model and model predictive control of WEC devices}\label{sec_wec_model_mpc_form}

Model-based optimal control is only possible when a dynamical model describing the system/plant is available and is computationally efficient.  For WEC devices, the linear potential theory, also known as the Airy wave theory, provides a dynamical model that is computationally efficient. Thus, in this section we first outline the linear dynamical model of the converter, which is well-suited for its optimal control. Discussion includes pros and cons of the linear model and improvements in terms of incorporating non-linear wave excitation forces. Afterwards,  model predictive control of WEC is presented using the first-order hold method. We also discuss some of the key concepts of a WEC's MPC, including defining the cost function, device constraints, regularizing/penalizing the cost function, and predicting future wave excitation forces.

\input{WEC_modeling}
\section{Wave dynamics}\label{sec_wave_eqs}
\input{Wave_dynamics}
\section{Numerical model based on the incompressible Navier-Stokes equations}\label{sec_wsi_eqs}
\input{IBAMR}

\section{Validation of BEM and MPC solvers and motivation behind this work}
\label{sec_validation_and_motivation}
\input{Validation_and_motivation}

\section{Spatial and temporal resolution tests}
\label{sec_spatiotemporal_tests}
\input{Spatial_temporal_resolution_tests}

\section{Results and discussion}
\label{sec_results_and_discussion}
\input{Results_and_discussion}

\section{Conclusions}
\label{sec_conclusions}
\input{Conclusions}


\section*{Acknowledgements}
K.K  and A.P.S.B~acknowledge support from NSF award OAC 1931368. NSF XSEDE and SDSU Fermi compute resources 
are particularly acknowledged. The authors also acknowledge helpful discussions with Barry Smith related to the PETSc-MATLAB interface and enabling the glue code to run on clusters without requiring any intervention from the cluster admin. 


\section{Bibliography}
\begin{flushleft}
 \bibliography{CFD_MPC_paper}
\end{flushleft}

\end{document}

%% file: Abstract.tex
In this work, we present a novel MPC-integrated multiphase IB framework that can compute the optimal energy-maximizing control force ``on-the-fly" by dynamically interacting with a high-fidelity numerical wave tank (NWT).  The computational model closely mimics the working setup of the device at its site of operation.  Due to the requirement of solving a constrained optimization problem at each time step of the IB simulation, the MPC algorithm utilizes a low-dimensional dynamical model of the device that is based on the linear potential theory (LPT). The multiphase IB solver, on the other hand, is based on the high-dimensional fictitious domain Brinkman penalization (FD/BP) method, which fully-resolves the hydrodynamic non-linearities associated with the wave-structure interaction (WSI).
A time-series forecasting auto-regressive model is implemented that predicts wave heights (from the past NWT data) to estimate the future wave excitation/Froude-Krylov forces for the MPC algorithm. Moreover, we also experiment with non-linear Froude-Krylov (NLFK) forces for the first time in an MPC formulation. The NLFK forces are computed efficiently using a static Cartesian grid, in which the WEC geometry is implicitly represented by a signed distance function.  Under varying sea conditions, the predictions of the MPC-integrated multiphase IB solver are compared to the widely popular LPT-based solvers. In agitated sea conditions and/or under aggressive control, the LPT-based WSI solvers produce too optimistic (and misleading) power output values. Overall, six WSI/MPC solver combinations are compared for a heaving vertical cylinder to determine the reasons for discrepancies between high- and low-fidelity predictions.  We also determine the pathway of energy transfer from the waves to the power take-off (PTO) system and verify the relationships using IB simulations. Additionally, three different sea states are simulated within the IB simulation to test the adaptive capability of MPC for WECs. MPC is demonstrated to adapt to changing sea conditions and find the optimal solution for each sea state.

The interaction between the distributed-memory parallel multiphase IB solver (written in C++) and the serial MPC solver (written in MATLAB) is fully described to facilitate reproducibility. A bespoke communication layer between the two solvers is developed, which can be easily modified by the WEC community to experiment with other optimal controllers and computational fluid dynamics (CFD) solvers. All codes for this work are made open-source for pedagogical and research purposes.


%% file: Introduction.tex
Global warming is on the rise and is likely to breach the 1.5$^\circ$C limit in the coming decades. It is imperative to switch to clean renewable energy, including hydro, solar, and wind, in order to mitigate the effects of climate change and meet the growing energy demands. A combination of renewable energy technologies and existing energy sources is necessary to accelerate the transition from carbon-based sources. This can be achieved, in part, through ocean energy, which remains a largely untapped energy resource. It has been demonstrated that wave energy can be harvested, but commercial devices have yet to be developed. This is mainly because wave energy converters (WEC) operate in harsh marine environments, which causes salt water corrosion, marine growth, sub-system failure, and high maintenance costs. The highly irregular nature of sea waves further complicates device design and the controller's task of optimizing performance. The testing of expensive WEC devices and power take-off units (PTO) in physical wave tanks is another challenge.

Numerical modeling of WECs is an efficient way to compare different designs and control strategies. A widely popular modeling approach in WEC research is the boundary element method (BEM) or its time-domain variant, the Cummins equation~\cite{Cummins1962} based on the linear potential theory (LPT)~\cite{book_Offshore,book_Waves_in_oceanic_and_coastal_waters} due to its simplicity, low computational cost, and flexibility in simulating the wave structure interaction (WSI) of a variety of WEC devices and control strategies. The linear models, which were created originally to model large sea vessels, ships, and similar seakeeping applications, assume small body motion with respect to the wave amplitudes and lengths. Additionally, inviscid, irrotational, and incompressible flows are assumed. The BEM solvers perform exceptionally well for relatively calm sea states with small wave amplitudes. Nevertheless, the assumptions upon which linear methods are based are severely challenged in conditions of agitated seas or aggressive control. We demonstrate that, under these operating conditions, linear methods overestimate the converter's dynamics and power consumption. Additionally, BEM solvers use low-dimensional dynamical models that do not provide insights into fluid dynamics resulting from fluid-structure interaction (FSI), such as vortex shedding, wave breaking, and wave overtopping.


In recent years, models based on the non-linear potential flow theory (NLPT) have been proposed~\cite{Davidson2020, Penalba2017}. By simulating the actual free water surface and including large body displacements, these models provide more accurate power estimates of the device than LPT-based models. The NLPT-based models are computationally expensive and are not easily applicable to the model-based control of WECs. An acceptable compromise, which is also sufficiently accurate, is the partially non-linear BEM model, which accurately resolves the hydrodynamical interactions between waves and devices~\cite{Merigaud2012, Penalba2015_review, Giorgi2016, Giorgi2017}. This can be accomplished by modifying the wave excitation force in the linear time-domain Cummins equation. In particular, the wave excitation or Froude-Krylov (FK) force is computed by integrating the incident wave pressure force over an instantaneous wetted surface area instead of assuming it is stationary at its mean equilibrium position. In this work, the Cummins equation-based WSI solver employing the non-linear Froude-Krylov (NLFK) method is referred to as the BEM-NLFK solver, and its linear counterpart as the BEM-LFK solver.


Although the NLPT-based models are more accurate than those based on LPT, they still do not account for the viscous phenomenon or other major hydrodynamical non-linearities, such as wave-breaking and vortex shedding.  Computational fluid dynamics (CFD)  provides the most accurate description of WSI of WECs~\cite{Penalba2017, Agamloh2008, Ghasemi2017, Anbarsooz2014, Dafnakis2020, Khedkar2021}. Some groups have recently begun performing control-integrated CFD simulations of WEC devices. These studies, however, are mostly limited to classical control laws, such as reactive control (also called proportional-derivative control) or latching control (also called phase control or bang-bang control); see for example~\cite{Penalba2018high,Agamloh2008,Giorgi2016_latching,Windt2021}. Agamloh et al.~\cite{Agamloh2008} performed CFD simulations of a cylindrical buoy, in which the PTO was modeled as an ideal linear damper to generate a control force proportional to the device velocity, that is, the derivative control law. In~\cite{Agamloh2008}, the optimal damping coefficient was estimated offline and kept constant throughout the simulation. To accurately capture the motion of the body, their CFD technique remeshed the domain at every time step. Giorgi et al.~\cite{Giorgi2016_latching}  used the latching control strategy for a 2D heaving cylinder subject to regular waves and compared BEM-LFK and CFD solvers. This is the first paper to implement a latching control for a WEC device within a CFD framework. The authors computed the optimal latching period offline using a combination of analytical techniques and free decay tests of the WEC device in the CFD-based numerical wave tank (NWT). According to Giorgi et al.~\cite{Giorgi2016_latching}, the BEM-LFK solver overestimates heave amplitude (and therefore power production) compared to the CFD solver. Recently, Windt et al.~\cite{Windt2021} compared the performance of a heaving WEC using BEM-LFK and CFD solvers. The predictions for three controllers were compared: (1) classical resistive (derivative) control; (2) classical reactive (proportional-derivative) control; and (3) moment-matching optimal control~\cite{Faedo2018}. As for the resistive and reactive controllers, their optimal coefficients/gains were computed offline and kept constant throughout the simulation, while the moment-matching controller used a pre-computed/offline optimal control force sequence. Similarly to Giorgi et al., Windt et al. also found that the BEM-LFK solver over-predicts power absorption of the WEC device (for all three controllers).


Unlike previous control-integrated CFD studies that used pre-computed controller gains or optimal control force sequences, this work uses the model predictive control (MPC) algorithm to compute the optimal energy-maximizing control force online. Due to its ability to handle many types of device and PTO topologies, model predictive control of WECs has been dubbed the ``Tesla" of controllers~\cite{WEC-white-paper}. In our modeling approach, the MPC interacts with the CFD-based NWT that sends the wave elevation and device dynamics data to the controller, which then solves a constrained optimization problem to find the optimal control force sequence. In the NWT, both regular and irregular sea conditions are modeled. A time-series forecasting auto-regressive model is implemented to predict wave heights (from past NWT data) to estimate the future wave excitation forces required by the MPC. Due to the requirement of solving a constrained optimization problem at each time step of the CFD simulation, the MPC algorithm is formulated using the computationally efficient LPT. Moreover, we include NLFK forces for the first time in an MPC formulation. The NLFK forces are computed efficiently using a static Cartesian grid, in which the WEC geometry is implicitly represented by a signed distance function.  The predictions of the MPC-integrated CFD solver are compared to the MPC-integrated BEM solvers under varying sea conditions. For a heaving 3D vertical cylinder device, six WSI/MPC solver combinations are compared. The current study is the first of its kind and comprehensively examines the reasons for prediction discrepancies between different solvers. We also determine the pathway of energy transfer from the waves to the power take-off (PTO) system and verify the relationships using IB simulations.  Additionally, three different sea states are simulated within a CFD simulation to test the adaptive capability of MPC of WECs. MPC is demonstrated to adapt to changing sea conditions and find the optimal solution for each sea state.

Our CFD solver is based on the multiphase fictitious domain Brinkman penalization (FD/BP) technique. FD/BP is a fully-Eulerian version of the immersed boundary (IB) technique~\cite{Angot1999} which solves a single set of equations in the entire domain, including the air, water, and solid WEC regions. In comparison with body conforming grid techniques that have previously been used to simulate WEC dynamics, the FD/BP method is computationally efficient, since it eliminates the need to remesh the domain to account for body motion. To accurately resolve the wave and WEC dynamics in the specific regions of interest, we also make use of locally refined Cartesian grids. As a result, the computation costs of 3D simulations are low. For reproducibility of the technique, the interaction between the distributed-memory parallel CFD solver (written in C++) and the serial MPC solver (written in MATLAB) is fully described here. Using the open-source PETSc library~\cite{petsc-web-page}, a custom communication layer is developed between the solvers.  Furthermore, the communication layer can be easily customized to experiment with other optimal controllers and CFD solvers by the WEC community. We have made all code freely available at \url{https://github.com/IBAMR/cfd-mpc-wecs}.

The paper is structured as follows. \REVIEW{In Table~\ref{tab_abbreviation}, we list the abbreviations that are frequently used throughout the paper.} Sec.~\ref{sec_wec_model_mpc_form} discusses the LPT-based dynamical models, MPC formulation with device constraints and regularization/penalization of the objective function, and LFK/NLFK force estimation. In Sec.~\ref{sec_wave_eqs}, we describe the numerical wave tank setup to simulate regular and irregular sea conditions. Sec.~\ref{sec_wsi_eqs} describes continuous equations of motion and their spatiotemporal discretizations. The section also deduces the pathway for energy transfer from the waves to the PTO system. Sec.~\ref{sec_software} discusses the interactions between MPC and CFD codes. Sec.~\ref{sec_validation_and_motivation} simulates a benchmarking example from the literature to validate our implementations of the BEM and MPC solvers. The same section includes a motivation example illustrating the stark differences between the power predictions of the BEM and CFD solvers. We conduct a spatial and temporal grid resolution study in Sec.~\ref{sec_spatiotemporal_tests} in order to determine the optimal mesh spacing and time-step size for the IB solver. In the results and discussion Sec.~\ref{sec_results_and_discussion}, a systematic comparison is conducted. Lastly, Sec.~\ref{sec_conclusions}  summarizes the findings and draws the main conclusions of this study. 

\begin{table}[]
\centering
\caption{\REVIEW{Frequently used abbreviations in the paper.}}
\rowcolors{2}{}{gray!10}
\begin{tabular}{ccc}
 \toprule
\REVIEW{Abbreviation} & \REVIEW{Entity}\\
 \midrule
 \REVIEW{AR} & \REVIEW{Auto-regression}	\\
 \REVIEW{BEM} & \REVIEW{Boundary element method}	\\
 \REVIEW{CFD} & \REVIEW{Computational fluid dynamics}	\\
 \REVIEW{FSI} & \REVIEW{Fluid-structure interaction}	\\
 \REVIEW{LFK} & \REVIEW{Linear Froude-Krylov}	\\
 \REVIEW{LPT} & \REVIEW{Linear potential theory}	\\
\bottomrule
\end{tabular}
\quad
\rowcolors{2}{}{gray!10}
\begin{tabular}{cc}
 \toprule
\REVIEW{Abbreviation} & \REVIEW{Entity} \\
 \midrule
 \REVIEW{MPC} & \REVIEW{Model predictive control}	\\
 \REVIEW{NLFK} & \REVIEW{Non-linear Froude-Krylov}	\\
 \REVIEW{NWT} & \REVIEW{Numerical wave tank}	\\
 \REVIEW{PTO} & \REVIEW{Power take-off}	\\
 \REVIEW{WEC} & \REVIEW{Wave energy converter}	\\
 \REVIEW{WSI} & \REVIEW{Wave-structure interaction}	\\
\bottomrule
\end{tabular}
\label{tab_abbreviation}
\end{table}


%% file: WEC_modeling.tex
\subsection{Linear potential theory-based WEC dynamical model} \label{sec_wec_model}

\begin{figure}[]
 \centering 
  \includegraphics[scale = 0.35]{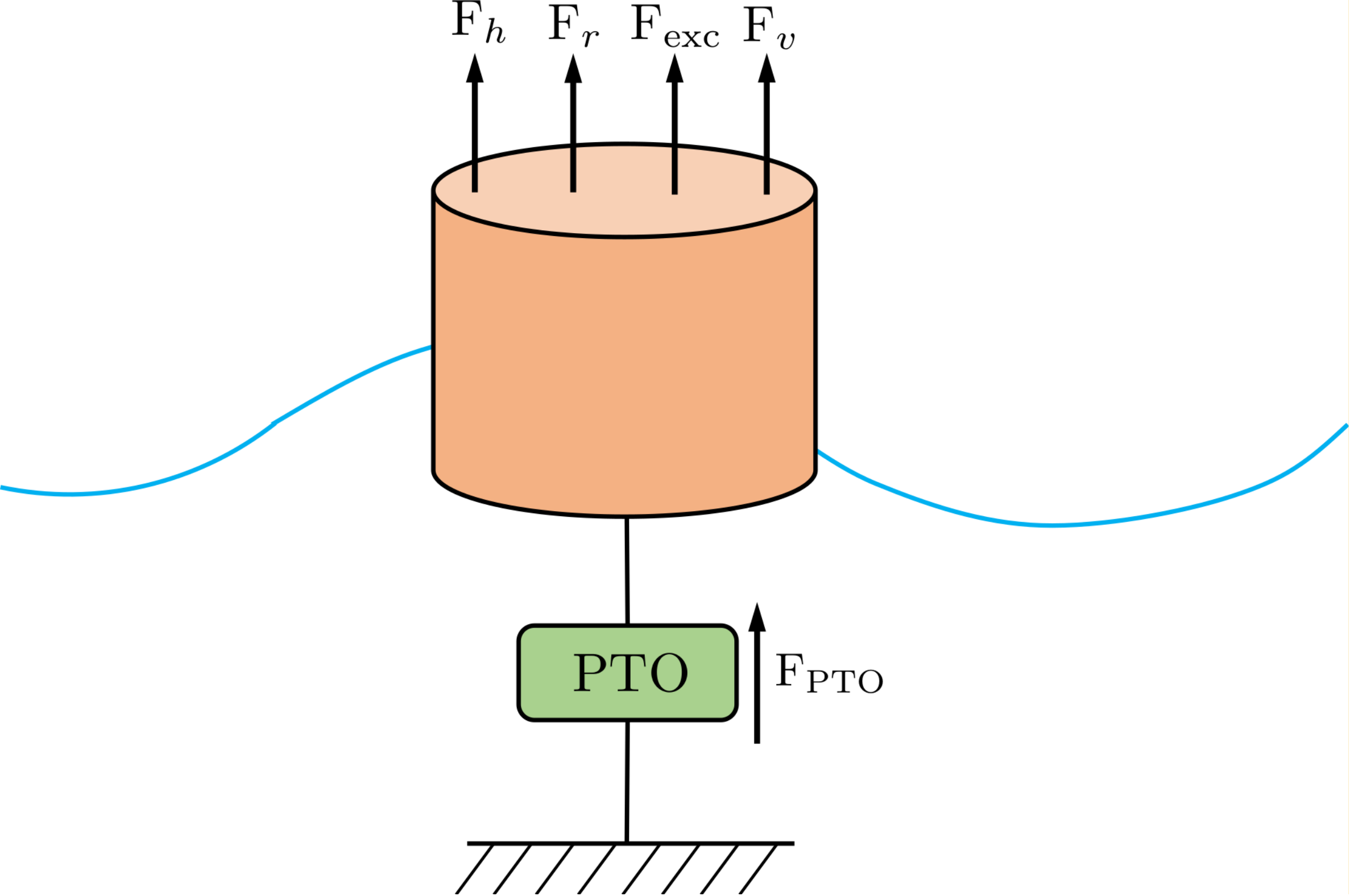}
 \caption{Schematic representation of a 1 DOF heaving cylindrical wave energy converter device. 
 }
 \label{fig_WEC_schematic}
\end{figure}

The WEC device considered in this study is a one degree of freedom (DOF) cylindrical point absorber\footnote{Point absorber is a WEC device whose characteristic dimensions are much smaller than the sea/ocean wavelength.} that heaves on the air-water interface. A schematic representation of the device is shown in Fig.~\ref{fig_WEC_schematic}. Axisymmetric point absorbers are among the most common WEC-types that mainly absorb wave energy due to their heaving motion. Therefore, for such devices, motion in the other DOFs can be neglected (or is constrained). If the amplitude of the motion of the device is significantly smaller than the wave height, then according to the LPT, the total force acting on the body is a linear sum of the hydrostatic restoring force $\text{F}_h$, radiation force $\text{F}_r$, wave excitation (including wave diffraction) force $\text{F}_\text{exc}$, and the viscous drag force $\text{F}_v$. The wave-induced motion of the device is retarded by the controller to extract the electrical energy. The WEC controller is typically embedded within a power take-off unit, which exerts the actuator/control force $\text{F}_\text{PTO}$ on the device. 

Using the Newton's second law of motion, the dynamics of the device in the heave direction ($z$) can be written as 
\begin{equation}
m\ddot{z}(t) = \text{F}_h(t) + \text{F}_r(t) + \text{F}_\text{exc}(t) + \text{F}_v(t) + \text{F}_\text{PTO}(t),
\label{eqn_Newton_2nd_law}
\end{equation}
in which $m$ is the mass of the cylinder, $z(t), \dot{z}(t)$, and $\ddot{z}(t)$ are the displacement (from the mean equilibrium position), velocity, and acceleration of the device in the heave direction, respectively. 

The hydrostatic restoring force due to buoyancy is given by
\begin{equation}
\text{F}_h(t) = -k_\text{stiff} \cdot z(t),
 \label{eqn_hydro_force}
\end{equation}
in which $k_\text{stiff}$ is the hydrostatic stiffness coefficient. For a cylindrical shaped body, the hydrostatic stiffness coefficient is given by $k_\text{stiff} = \rho_w  g  \pi  R_\text{cyl}^2$, in which  $\rho_w$ is the density of water, $g$ is the acceleration due to gravity, and $R_\text{cyl}$ is the radius of the cylinder. The length of the vertical cylinder is $L_\text{cyl}$. For a vertical heaving cylinder,  $k_\text{stiff}$ does not change with time because the water plane area of the body does not change.  A possible means of modeling nonlinear buoyancy forces for floating bodies whose water plane areas differ is discussed in Sec.~\ref{sec_NLFK_calculation}.  In addition, Giorgi et al.~\cite{Giorgi2016, Giorgi2017} describe an analytical approach to model nonlinear buoyancy forces.

The radiation force $\text{F}_r(t)$ in Eq.~\eqref{eqn_Newton_2nd_law} is written as
\begin{equation}
\text{F}_r(t) = -m_\infty \ddot{z}(t) - \int_{0}^{t} \text{K}_r(t-\tau) \dot{z}(\tau) \; \text{d}\tau.
\label{eqn_radiation_force}
\end{equation}
Here, $m_\infty$ is the added mass~\footnote{The added mass represents the additional inertia of the system due to the motion of the surrounding fluid.} at infinite frequency. The radiation force in Eq.~\eqref{eqn_radiation_force} also includes a convolution integral of the radiation impulse response function (RIRF) $\text{K}_r(t)$ with the velocity of the body. Physically, RIRF explains how kinetic energy is dissipated by the water waves produced by the oscillation of the body, which began its motion at time $t = 0$ and continues to do so until current time $t$.

Excitation forces due to incident/incoming waves can be computed on either a mean or instantaneous wetted surface of the device. In the former case, excitation forces can be expressed as a convolution integral between the wave impulse response function (WIRF) $\text{K}_e(t)$ and the undisturbed wave surface elevation $\eta_\text{wave}(t; x_{B})$ at the device location $x_{B}$:
\begin{equation}
\text{F}_\text{exc}(t) =  \text{K}_e *\eta_\text{wave} =  \int_{-\infty}^{\infty} \text{K}_e(\tau) \eta_\text{wave}(t-\tau; x_{B}) \; \text{d}\tau.
\label{eq_exc_force}
\end{equation}
From Eq.~\eqref{eq_exc_force} it can be seen that $\text{F}_\text{exc}(t)$ is non-causal because future surface elevations affect the current motion of the body. The non-causality of WIRF has practical implications when it comes to the implementation of MPC of WECs, since the wave elevations at the device location must be forecasted. Discussion of wave prediction is deferred to Sec.~\ref{sec_LFK_calculation}. In Sec.~\ref{sec_NLFK_calculation}, we discuss the evaluation of wave excitation forces using the instantaneous wetted surface. 

Lastly, the viscous drag force acting on the body can be written using the non-linear Morison equation~\cite{Sarpkaya1986} as
\begin{equation}
\text{F}_v(t) = -\frac{1}{2} \rho_w C_d \pi R_\text{cyl}^2 \lvert{\dot{z}(t)}\rvert \dot{z}(t), 
  \label{eqn_nonlinear_Fv}
\end{equation}
in which $C_d$ is the coefficient of drag. Estimating an accurate value of $C_d$ for Eq.~\eqref{eqn_nonlinear_Fv} is a non-trivial task. This work estimates $C_d$ by equating the work done $\left( \int_0^T \text{F}_v \dot{z} \; \rm{d} \tau \right)$ by viscous forces on a freely decaying cylinder that heaves on an air-water interface in a NWT with the work done by viscous forces defined according to Eq.~(\ref{eqn_nonlinear_Fv}). We chose one period of the damped oscillation for the integral.

Putting all terms together, the governing equation for the 1 DOF heaving WEC reads as
\begin{equation}
\ddot{z}(t) + \frac{1}{m + m_\infty} \int_{0}^{t} \text{K}_r(t-\tau) \dot{z}(\tau)\text{d}\tau + \frac{1}{m + m_\infty} k_\text{stiff} \cdot z(t) = u(t) + v(t) + \frac{\text{F}_v(t)}{m + m_\infty},
\label{eqn_Cummins_eq}
\end{equation}
in which 
\begin{equation}
u(t) = \frac{\text{F}_\text{PTO}(t)}{m + m_\infty},~~~~v(t) = \frac{\text{F}_\text{exc}(t)}{m + m_\infty}. \nonumber
\end{equation}
To obtain $ \text{K}_\text{e} (t)$ and $m_\infty$, we use the boundary element method software ANSYS AQWA~\cite{ANSYS_AQWA}.  The radiation convolution integral given by Eq.~(\ref{eqn_radiation_force}) is approximated in a state-space form~\cite{Falnes1995} with velocity of the device $\dot{z}(t)$ as input and the approximated convolution integral as output. The state-space representation offers both computational efficiency~\cite{Taghipour2008} and representational convenience for matrix-based MPC control.  Following Yu and Falnes~\cite{Falnes1995}, the state-space representation of the radiation convolution integral reads as  
\begin{align}
&\boldsymbol{\dot{x}_r}(t) = \Ar \xr(t) + \Br \dot{z}(t)  \nonumber \\
&\int_{0}^{t} \text{K}_r(t-\tau) \dot{z}(\tau)\text{d}\tau \approx \Cr \xr(t),
  \label{eqn_linear_Fr}
\end{align}
in which $\xr \in \R^{n_r \times 1}$, $\Ar \in \R^{n_r \times n_r}$, $\Br \in \R^{n_r \times 1}$, $\Cr \in \R^{1 \times n_r}$, and $n_r = 3$ is the approximation order of the radiation force used in this work. The viscous drag force acting on the cylinder is linearized around the current velocity of the cylinder $\dot{z_0}(t)$ and is approximated as
\begin{equation}
\text{F}_v(t) \approx  -\beta \left |\dot{z_0}\right | \dot{z_0} + 2\beta \left |\dot{z_0}\right | \dot{z},
\label{eqn_linear_Fv}
\end{equation}
in which $\beta = -\frac{1}{2}\rho_w C_d \pi  R_\text{cyl}^2$. Using Eqs.~(\ref{eqn_Cummins_eq})-(\ref{eqn_linear_Fv}), a continuous-time, linear state-space form governing the dynamics of the WEC device is obtained as
\begin{align}
   \label{eqn_cont_SS_sysa}
  & \dot{\bf{X}}_{\boldsymbol{c}}{(t)} = \Ac \Xc(t) + \Bc \left(u_c(t) + v_c(t) -\beta \left |\dot{z_0}\right | \dot{z_0} \right),  \\
  & \Zc(t) = \Cc \Xc(t),  \label{eqn_cont_SS_sysb}
\end{align}
in which the subscript $c$ denotes the continuous-time quantities and
\begin{align}
&\Ac = \begin{bmatrix}
  0      &      1       &        \bf{0}        \\  
-\frac{k_\text{stiff}}{(m+m_\infty)}  &  \frac{2\beta\left |\dot{z_0}(t)\right |}{(m+m_\infty)} & -\frac{\Cr}{(m+m_\infty)}   \\ 
0   &    \Br    &       \Ar       \\
\end{bmatrix} \in \R^{(n_r+2) \times (n_r+2)}, 
~~~~~\Bc = \begin{bmatrix}
0  \\
1  \\
\bf{0}
\end{bmatrix} \in \R^{(n_r+2) \times 1}, \nonumber \\
&\Cc = \begin{bmatrix}
1    &      0      &     \bf{0}  \\
0    &      1      &     \bf{0}  \\
\end{bmatrix} \in \R^{2 \times (n_r+2)},  
~~~\Xc(t) = \begin{bmatrix}
z(t)     \\
\dot{z}(t)    \\
\xr(t)
\end{bmatrix}\in \R^{(n_r+2) \times 1},
~~~\Zc(t) = \begin{bmatrix}
z(t)     \\
\dot{z}(t)
\end{bmatrix}\in \R^{2 \times 1}.   \nonumber
\end{align}
Let us note that except for the linearized drag coefficient, all entries of matrices  $\Ac$ and $\Bc$ are time invariant. Therefore, the dynamical system described by Eqs.~\eqref{eqn_cont_SS_sysa} and~\eqref{eqn_cont_SS_sysb} is quasi linear time invariant (QLTI). The dynamical system is reduced to an LTI one if the drag coefficient is linearized around a fixed point, e.g., around the mean equilibrium position of the device.     

\subsection{Model Predictive Control of WECs} \label{sec_mpc_form}
Having discussed the control-oriented dynamical model of the WEC device, we now focus our attention on model predictive control for WECs. Its basic principles are straightforward. For each control sequence, the controller uses the dynamical model of the plant to predict the plant's future trajectory over a prediction horizon (time period) of $\cT_h$. Out of a large set of possible outcomes, MPC selects the control sequence which extremizes (maximizes or minimizes) a predefined objective function. The extremization of the objective function is typically achieved by solving an optimization problem numerically.  The first part/signal of the optimal control sequence is used to control the plant, while the rest is discarded. This process is repeated again and again by receding/moving the prediction horizon forward. With WECs, the control objective is to maximize the device's energy output. Thus, to implement MPC for WECs, we require:
\begin{enumerate}
 \item A discrete-time dynamical model of the device to predict the future dynamics over a finite time horizon $\cT_h$. In this work we use the first order hold (FOH) method of Cretel et al.~\cite{Cretel2011} to obtain the discrete-time model~\cite{book_DigitalControl} from the continuous-time Eqs.~\ref{eqn_cont_SS_sysa} and~\ref{eqn_cont_SS_sysb}. More specifically, if $\Delta t$  denotes the discrete time step size and $k \in \N$ denotes the (discrete) time index, then the current state $\Xd(k)$ is advanced to the next time level $\Xd(k+1)$ as 
\begin{align}
\label{eqn_disc_SS_sysa}
 &\Xd(k+1) = \Ad \Xd(k) +\Bd \Delta u_d(k+1) + \Fd \Delta v_d(k+1), \\
  &\Zd(k) = \Cd \Xd(k),
  \label{eqn_disc_SS_sysb}
\end{align}
in which the subscript $d$ denotes the discrete-time quantities and 
\begin{align}
&~~~~\Ad = \begin{bmatrix}
  \V{\phi}(\Delta t)  &  \V{\Upsilon}  &  \V{\Upsilon}   \\  
     0     &    1     &      0   \\ 
     0     &    0     &       1       \\
\end{bmatrix} \in \R^{(n_r+4) \times (n_r+4)}, 
~~~\Bd = \begin{bmatrix}
\V{\Lambda}  \\
1  \\
0
\end{bmatrix} \in \R^{(n_r+4) \times 1},
~~~\Fd = \begin{bmatrix}
\V{\Lambda}  \\
0  \\
1
\end{bmatrix} \in \R^{(n_r+4) \times 1}, \nonumber  \\
&\Cd = \begin{bmatrix}
     1     &    0     &   0  &   ...  &  0   &   0   &  0    \\  
     0     &    1     &   0  &   ...  &  0   &   0   &  0    \\ 
     0     &    0     &   0  &   ...  &  0  &    1    &  0    \\
\end{bmatrix} \in \R^{3 \times (n_r+4)},  
~~\Xd(k) = \begin{bmatrix}
\Xc(k \Delta t)     \\
u_d(k)          \\
v_d(k)
\end{bmatrix}\in \R^{(n_r+4) \times 1}, 
~\Zd(k) = \begin{bmatrix}
\Zc(k \Delta t)     \\
u_d(k)  
\end{bmatrix}\in \R^{3 \times 1}.  
\end{align}  
Here, $\Xc $ and $\Zc$ denote the possibility of initializing data from a continuous-time solver at the beginning of the time step $k$. For example, in many cases presented in this work, we use the continuous-time multiphase IB solver that sends the device state $\Xc(k \Delta t)$ to the MPC algorithm.  In the matrices defined above, the following definitions are used: 
\begin{align}
\label{eq_uc_interpolation}
&\V{\phi}(\Delta t) = e^{\Delta t \Ac} \in \R^{(n_r+2) \times (n_r + 2)},
~~~\V{\Upsilon} = \Ac^{-1} \left( \V{\phi}(\Delta t)-{\bf{I}} \right)\Bc  
~\text{and~}\V{\Lambda} = \frac{1}{\Delta t}\Ac^{-1} \left( \V{\Upsilon} - \Delta t \Bc \right) \in \R^{(n_r+2)\times 1}, \nonumber \\
&u_c(t) = u_d(k) + \left(\frac{t-k\Delta t}{\Delta t}\right)\Delta u_d(k+1),
~~v_c(t) = v_d(k) + \left(\frac{t-k\Delta t}{\Delta t}\right)\Delta v_d(k+1), \nonumber \\
&\Delta u_d(k+1) = u_d(k+1) - u_d(k),  
~~\Delta v_d(k+1) = v_d(k+1) - v_d(k).     
\end{align}

 \item A receding strategy in which only the first part/signal of the optimal control sequence is used for actuating the device, and the prediction horizon is moved forward in time to compute the next optimal control sequence (by taking into account the latest device state and wave measurements). We use a prediction horizon of one wave period in this work, unless stated otherwise.

Assuming that a $N_p$-step prediction horizon is employed, i.e., $\cT_h = N_p \cdot \Delta t_p$, the output vector, $\underline{\Zd}(k)$, is obtained from the discrete-time model by time marching Eqs.~\ref{eqn_disc_SS_sysa} and~\ref{eqn_disc_SS_sysb} through the prediction horizon as~\cite{Cretel2011,book_DigitalControl} 
 \begin{equation}
\underline{\Zd}(k) = \vcP \Xd{(k)} + \cJu \; \Deltau(k) + \cJv \; \Deltav(k).
\label{eqn_output_pred_vector}
\end{equation}
In the equation above
\begin{align}
\underline{\Zd}(k) = \begin{bmatrix}
\Zd(k+1|k)     \\
\Zd(k+2|k)      \\
.     \\
.     \\
\Zd(k+N_p|k)
\end{bmatrix} \in \R^{(3N_p \times 1)}, 
~~\cJu &= \begin{bmatrix}
 \Cd \Bd  & 0 & ... & 0 \\ 
 \Cd \Ad \Bd & \Cd \Bd & ... & 0 \\ 
 .&  .&  .& .\\ 
 .&  .&  .& . \\
 \Cd \Ad^{(N_p-1)} \Bd & \Cd \Ad^{(N_p-2)} \Bd & ... & \Cd \Bd 
\end{bmatrix} \in \R^{3N_p \times N_p}, \nonumber \\
\vcP = \begin{bmatrix}
\Cd \Ad       \\
\Cd \Ad^2      \\
.     \\
.     \\
\Cd \Ad^{N_p}
\end{bmatrix} \in \R^{3N_p \times (n_r+4)}, 
~~~\cJv &= \begin{bmatrix}
 \Cd \Fd  & 0 & ... & 0 \\ 
 \Cd \Ad \Fd & \Cd \Fd & ... & 0 \\ 
 .&  .&  .& .\\ 
 .&  .&  .& . \\
 \Cd \Ad^{(N_p-1)} \Fd & \Cd \Ad^{(N_p-2)} \Fd & ... & \Cd \Fd 
\end{bmatrix} \in \R^{3N_p \times N_p}
\end{align}

Sec.~\ref{sec_Fexc_prediction} describes the methods for obtaining the future wave excitation force values stored in the vector $\Deltav(k)$. Note that the (WSI) solver time step size $\Delta t$ is generally different from the MPC time step size $\Delta t_p$. In many of the examples presented in this work, we employ a continuous-time CFD solver with a much smaller time step of $\Delta t$  than $\Delta t_p$ in order to accommodate the convective Courant-Friedrichs-Levy (CFL) number restriction.

 \item An objective function to determine the optimal control sequence over the prediction horizon. Here, the goal is to maximize the amount of energy absorbed by the WEC device, which can be expressed by the relation
 \begin{equation}
J_0 = -(m+m_\infty)~\int_{t}^{t+
\cT_\text{h}}u(\tau)\cdot \dot{z}(\tau) \text{d}\tau.
\label{eqn_energy_abs}
\end{equation}
The negative sign in the objective function indicates the flow of energy from the device to the power grid. Using the trapezoidal rule to evaluate the definite integral of Eq.~\ref{eqn_energy_abs}, we obtain
\begin{equation}
J_0 = -(m+m_\infty)\Delta t_p\left( \frac{1}{2}u_d(k)\dot{z}(k) + \sum_{i=k+1}^{k+N_p-1}u_d(i|k)\dot{z}(i|k) + \frac{1}{2}u_d(k+N_p|k)\dot{z}(k+N_p|k)\right)
\label{eq_trapz_energy_abs}
\end{equation}
For purposes of extremization of $J_0$, we can remove the constant pre-factor and the known term at time level $k$ ($u_d(k)\dot{z}(k)$) from the discrete summation and redefine the objective function to be
\begin{equation}
J_1(k) = \sum_{i=k+1}^{k+N_p-1}u_d(i|k)\dot{z}(i|k) + \frac{1}{2}u_d(k+N_p|k)\dot{z}(k+N_p|k)
\label{eq_J1_k}
\end{equation}
Since the (constant) negative pre-factor $-(m+m_\infty)\Delta t_p$ has been dropped from $J_0$ to obtain $J_1$, the initial maximization problem is now a minimization problem. Moreover, the objective function can be expressed in terms of the output vector as follows:
\begin{equation}
J_1(k) = \frac{1}{2}\underline{{\bf{Z}}_{\boldsymbol{d}}^T}(k)~\V{Q}~\underline{\Zd}(k),
\label{eq_J1}
\end{equation}
in which
\begin{equation}
\V{Q} = \begin{bmatrix}
\V{M} &    &      &  \\
    & \ddots  &      &  \\
    &    & \V{M}  &  \\
    &    &      & \frac{1}{2}\V{M} \\
\end{bmatrix} \in \R^{3N_p \times 3N_p}~~\text{and}
~~~\V{M} = \begin{bmatrix}
 0 &  0  &  0   \\
 0 &  0  &  	1   \\
 0  &  1  &  0 
\end{bmatrix} \in \R^{3 \times 3} \nonumber
\end{equation}
By substituting Eq.~\ref{eqn_output_pred_vector} into Eq.~\ref{eq_J1} and expanding the terms, we get
\begin{equation}
J_1 = \frac{1}{2}\Deltau^T \cJu^T \V{Q} \cJu \Deltau + \Deltau^T \cJu^T \V{Q} (\vcP \Xd + \cJv \Deltav) + \frac{1}{2}(\vcP \Xd + \cJv \Deltav)^T \V{Q} (\vcP \Xd + \cJv \Deltav)
\label{eq_J1_expn1}
\end{equation}
The minimization of $J_1$ with respect to the unknown control sequence $\Deltau$ yields the optimal control $\Deltau^\star$ for the entire prediction horizon. Observe that the last term of Eq.~\eqref{eq_J1_expn1} does not contribute to the evaluation of $\Deltau^\star$ and can be safely dropped. Therefore, the objective or in this case the cost function to minimize reads as
\begin{equation}
J_1 = \frac{1}{2}\Deltau^T \cJu^T \V{Q} \cJu \Deltau + \Deltau^T \cJu^T \V{Q} (\vcP \Xd + \cJv \Deltav).
\label{eq_J1_expn2}
\end{equation}
The cost function $J_1$ is quadratic in $\Deltau$ and is assumed to be positive semi-definite. We use the quadratic programming (QP) methods available in MATLAB~\cite{MATLAB2019b} to obtain the optimal control sequence $\Deltau^\star$. \REVIEW{The objective functions $J_0$ and $J_1$ assume that the PTO is ideal with no mechanical to electrical conversion losses. Thus, the conversion efficiency is taken to be 100\%, i.e., $\varepsilon = 1$. Readers are referred to Tona et al.~\cite{Tona2017}, who formulated a MPC problem with $\varepsilon < 1$ and investigated how a non-ideal PTO affects device dynamics and absorbed power~\footnote{\REVIEW{Although we have taken $\varepsilon = 1$ for all the cases in this work, our code (available at \url{https://github.com/IBAMR/cfd-mpc-wecs}) can also simulate the controlled dynamics of the WEC device with $\varepsilon < 1$. The non-ideal PTO problem is handled separately because it requires a sequential quadratic programming solver.}}.}

\end{enumerate}

\subsubsection{Including device/path constraints in MPC} \label{sec_path_constr}

In general, if the cost function $J_1$ is minimized as is, the device displacement, velocity, or actuator force will exceed the physical limits. An unconstrained control force could, for instance, cause the device to overshoot the free surface and slam into water with large impact forces. This can be avoided by using the following path/device constraints in MPC~\cite{Cretel2011, book_MPC_MATLAB, Faedo2017}: 
\begin{align}
z^\text{min}~\leq~&z(k)~\leq~z^\text{max} ,\nonumber \\
\dot{z}^\text{min}~\leq~&\dot{z}(k)~\leq~\dot{z}^\text{max},   \nonumber \\
u^\text{min}~\leq~&u(k)~\leq~u^\text{max}. 
\label{eq_path_constr}
\end{align}

Constraints written in Eq.~\eqref{eq_path_constr} are first expressed in the form $ \underline{\Zdmin} \le \underline{\Zd}  \le \underline{\Zdmax} $, which is then recast as $ \underline{\Zdmin} \le  \vcP \Xd{(k)} + \cJu \; \Deltau(k) + \cJv \; \Deltav(k)  \le \underline{\Zdmax} $ using Eq.~\eqref{eqn_output_pred_vector}. As both $\Xd(k)$ and $\Deltav(k)$ are known inputs to the quadratic program, the latter form of the inequality allows extraction of the constraint relationship for the variable of interest $\Deltau$.  

\subsubsection{Regularizing the MPC objective function} \label{sec_penalty_terms_obj_func}

The cost function $J_1$ of Eq.~\eqref{eq_J1_expn2} is further modified by adding two additional quadratic penalty terms:
 \begin{align}
J_2(k) &= J_1(k) + \lambda_1 \lVert \Deltau\rVert^2_2,  \label{eq_J2} \\
J_3(k) &= J_2(k) + \lambda_2 \lVert \underline{\boldsymbol{u}} \rVert^2_2. \label{eq_J3}
\end{align}
 
Adding the $ \lambda_1 \lVert \Deltau\rVert^2_2$ term to $J_1$  reduces the aggressiveness of the controller, i.e., $J_2$ results in smoother control force variation over time than the original cost function $J_1$~\cite{Cretel2011}.  The non-negative parameter $\lambda_1$ in Eq.~\eqref{eq_J2} has the dimensions of time. It is important to keep $\lambda_1$ positive in order to maintain or enhance $J_1$'s convexity. A smaller magnitude of $\lambda_1$ ensures that $J_1$ and $J_2$ are not too far apart.  
 
$J_2$ is further modified to $J_3$ by adding the quadratic penalty term $ \lambda_2 \lVert \underline{\boldsymbol{u}} \rVert^2_2$ (Eq.~\eqref{eq_J3}). The objective is to reduce the flow of power from the grid to the device, referred to as reactive power in wave energy literature~\cite{Korde2016,Faedo2017}. Even though reactive power aligns the device velocity with wave excitation forces to provide a higher overall energy output, it can lead to large instantaneous positive and negative powers in the PTO unit~\cite{Korde2016}. The two-way power flow complicates the design of a PTO system and increases its cost.   The goal of $J_3$ is to enforce the one-way power flow condition in the PTO machinery~\cite{Cretel2011}. As with $\lambda_1$, $\lambda_2$ should also be positive, smaller in magnitude, and has the dimensions of time. 



\subsection{Linear potential theory-based wave excitation/Froude-Krylov forces} \label{sec_Fexc_prediction}

The wave excitation forces acting on the body according to the LPT are the sum of effects coming from undisturbed incident waves (assuming that the body is removed from the path of the waves) and diffracted waves (which assumes the body is held stationary at its mean position). Wave excitation forces are also known as Froude-Krylov (FK) forces. FK forces can be computed using the undisturbed flow and diffracted wave potentials, $\phi_{I}$ and $\phi_{D}$, respectively, as
   
 \begin{equation}
\text{\bf{F}}_\text{FK}(t) = \text{\bf{F}}_{I}(t) + \text{\bf{F}}_{D}(t) = -\int_{{S_b}} (p_{I}(t) + p_D(t))~{\bf{n}}~\text{d}S_b, 
\label{eq_FK_force}
\end{equation}
in which ${S_b}$  is the wetted surface area of the body, $\text{\bf{n}}$ is the unit outward normal to the surface, $p_{I} = -\rho_w \D{\phi_I}{t}$ is the pressure due to incident waves, and  $p_{D} = -\rho_w \D{\phi_D}{t}$ is the pressure due to diffracted waves. It should be noted that the hydrostatic pressure $p_{H}(t) = -\rho_w g z(t)$ and the radiation pressure $p_{R}(t) = \rho_w \D{\phi_R}{t}$ are accounted for in the calculations of $\text{F}_h(t)$  and $\text{F}_r(t)$, respectively in Eq.~\eqref{eqn_Newton_2nd_law}.  Additionally, in Eq.~\eqref{eqn_Newton_2nd_law}, $\text{F}_{\text{exc}}$ is the $z$-component of $\text{\bf{F}}_{\text{FK}}$.


\subsubsection{Linear Froude-Krylov (LFK) forces: Up-wave measurements and future wave predictions}
\label{sec_LFK_calculation}

If the pressure integral of Eq.~\ref{eq_FK_force} is evaluated while the body is stationary at its mean equilibrium position, the Froude-Krylov forces are linear with respect to free surface elevation and are called linear Froude-Krylov forces (LFK).  The LFK forces can be computed more efficiently as a convolution integral between the wave impulse response function (WIRF) and water surface elevation at the device location $x_{B}$: $\text{F}_{\rm exc}(t) =  \int_{-\infty}^{\infty} \text{K}_e(\tau) \eta_\text{wave}(t-\tau; x_{B}) \text{d}\tau$ (repeated from Eq.~\eqref{eq_exc_force} for convenience). Assuming that the sea surface is calm prior to the start of the simulation at $t = 0$, i.e., $\eta_{\rm wave}(t < 0; \forall x) = 0$, the upper limit of the convolution integral $\text{K}_e * \eta_\text{wave}$ can be terminated at the current time $t$. 

\begin{figure}[]
 \centering 
  \includegraphics[scale = 0.38]{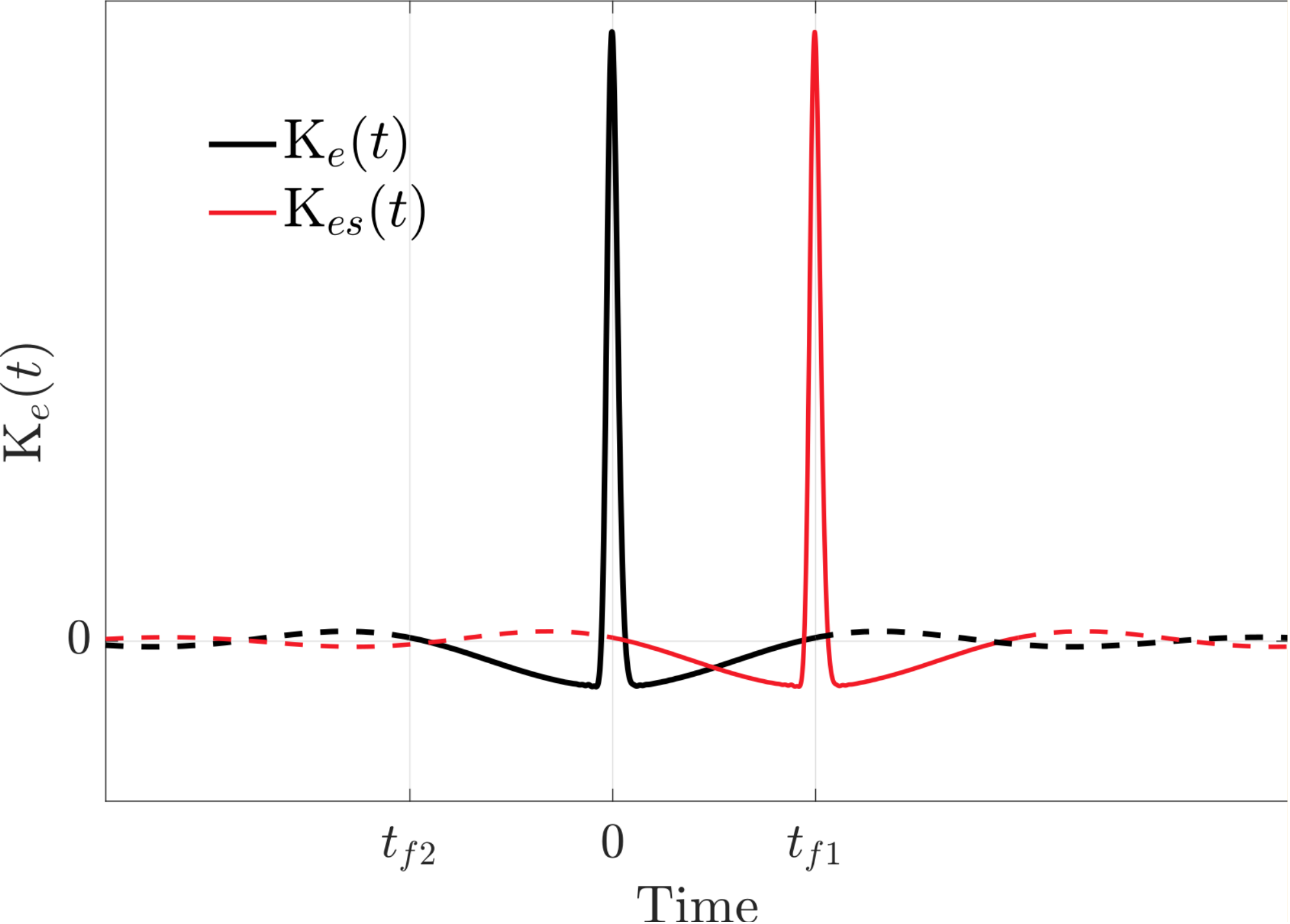}
 \caption{ Wave impulse response function (WIRF) for a vertical cylindrical in heave motion. The original WIRF K$_e(t)$ is shown in black and the right shifted WIRF K$_{es}(t)$ is shown in red. The dashed part of the curves represents the truncated region where the WIRF is close to zero. 
 }
 \label{fig_exc_IRF}
\end{figure}

In Fig.~\ref{fig_exc_IRF} we show the non-causal WIRF $\text{K}_\text{e}(t)$ as a black line. WIRF is the inverse Fourier transform of the frequency-domain excitation force $\widehat{\text{F}}(\omega) = \widehat{\text{F}}_I(\omega) + \widehat{\text{F}}_D(\omega) $ that we obtain using ANSYS AQWA software:
  \begin{equation}
  \text{K}_e(t) = \frac{1}{2\pi} \int_{-\infty}^{\infty} \widehat{\text{F}}(\omega)e^{(i\omega t)} \,\d{\omega}. 
\end{equation}
In practice, the incident wave forces $\widehat{\text{F}}_I$ and the diffracted wave forces $\widehat{\text{F}}_D$ can only be computed for discrete frequencies $\{\omega_i\}$, and a suitable numerical interpolation is required to evaluate the inverse Fourier transform.  From Fig.~\ref{fig_exc_IRF}, it can be seen that when $t > t_{f1}$ or  $ t < -|t_{f2}|$, $\text{K}_e(t) \rightarrow 0$. Truncated $\text{K}_e(t)$ is shown as a dashed line in Fig.~\ref{fig_exc_IRF}. The finite positive time interval where $\text{K}_\text{e}(t) \ne 0$ requires $\eta_\text{wave}$ data only until $ t - t_{f1}$ in the past to determine the convolution integral. Also, the finite negative time interval where $\text{K}_\text{e}(t) \ne 0$ implies that $\eta_\text{wave}$ data is only required up to $ t + |t_{f2}|$ into the future. The convolution integral of Eq.~\eqref{eq_exc_force} can therefore be performed efficiently as 
 \begin{equation}
\text{F}_\text{exc}(t)  = \int_{-t_f}^{t_f} \text{K}_e(\tau) \eta_\text{wave}(t-\tau; x_{B}) \;\text{d}\tau,
\label{eq_exc_force2}
\end{equation}
in which $t_f = \text{max}\left[t_{f1},|t_{f2}| \right]$. It follows that (with reasonable accuracy) $\text{F}_\text{exc}$ at the current time $t$ can be computed if the wave surface elevation data at the device location is available from $t - t_f$  to $t + t_f$. 

It is unrealistic to measure the undisturbed wave elevation at the device location since the incident waves cannot pass through the device. Furthermore, the waves near the body are altered by FSI and do not remain undisturbed in reality. Therefore, we need to find another way to estimate $\eta_\text{wave}$ at the device location $x_{B}$. We can take advantage of the fact that wave propagation is a hyperbolic phenomenon, which means that waves passing an up-wave location $x_{A}$ will arrive at the device at a later time. In order to locate a convenient up-wave location, we change the variable $\tau$ to $\tau' - t_f$ in Eq.~\eqref{eq_exc_force2}:
 \begin{align}
\text{F}_\text{exc}(t)  & = \int_{-t_f}^{t_f} \text{K}_e(\tau) \eta_\text{wave}(t-\tau; x_{B}) \;\text{d}\tau  \nonumber \\
                                  & = \int_{0}^{2 t_f} \text{K}_e(\tau' - t_f) \eta_\text{wave}(t + t_f - \tau'; x_{B}) \;\text{d}\tau'  \nonumber \\
                                  & = \int_{0}^{2 t_f} \text{K}_{es}(\tau') \eta_\text{wave}(t- \tau'; x_{A}) \;\text{d}\tau'.  \nonumber
\end{align}
Here, K$_{es}$ is the shifted WIRF obtained by shifting the original WIRF to the right side on the time-axis by an amount $t_f$. Symbolically, the time shift can be expressed by the relation $\text{K}_{es}(t)$ = $\text{K}_e(t-t_f)$. The shifted WIRF is shown as a red line in Fig.~\ref{fig_exc_IRF}. For the integral transformation above,  we defined the up-wave location $x_{A}$ so that the waves leaving this location reach the device after an additional time of $t_f$. Therefore, the water surface elevation at the device location at the present time $t$ is related to the up-wave elevation at the previous time $t-t_f$, i.e., $\eta(t + t_f; x_{B}) = \eta(t; x_{A})$. The distance of the up-wave point from the device is calculated by using the wave velocity $(\omega/\kappa)$ as 
 \begin{equation}
d_{f} = \frac{\omega}{\kappa}\cdot t_f,
\label{eq_df_for_xA}
\end{equation}
in which $\omega$ is the wave frequency and $\kappa$ is the wave number. In our CFD model, $x_{A}$ is chosen to be a point in the wave generation zone. See Fig.~\ref{fig_NWT_schematic} for a visual representation. In summary, the convolution integral of Eq.~\ref{eq_exc_force2} is equivalent to 
 \begin{equation}
\text{F}_\text{exc}(t) = \int_{0}^{2t_f} \text{K}_{es}(\tau) \eta_\text{wave}(t-\tau; x_{A}) \; \text{d}\tau.
\label{eq_exc_force3}
\end{equation}

It can be seen from Eq.~\ref{eq_exc_force3} that the wave excitation forces acting on the device at the present instant $t$ can be calculated from the $\eta_\text{wave}$ data recorded at the up-wave location between the period $[t-2t_f, t]$ for which no prediction or time-series estimation is needed. Wave forecasting is still necessary for MPC even if all the surface elevation data is obtained/measured at a nearby up-wave location. The reason is that for a prediction horizon of $\cT_h$, FK forces acting on the device are necessary between the period $[t, t + \cT_h]$ (to fill the entries of the vector $\Deltav$ in Eq.~\eqref{eqn_output_pred_vector} or Eq.~\eqref{eq_J1_expn2}). Accordingly, at the up-wave location $x_{A}$, $\eta_\text{wave}$ data is required in the interval $[t - 2t_f, t+\cT_h]$. In this study, we use the auto-regressive (AR) model for time series forecasting, one of the many techniques available to predict the future behavior of a time-series based on its past behavior. Detailed information about the implementation of an AR model for wave forecasting can be found in the thesis by Gieske~\cite{Gieske2007}.  A typical AR model is calibrated for a particular sea state and requires (manual) re-tuning to make accurate predictions in a different sea state. Sec.~\ref{subsec_mpc_adaptivity} describes the capability of MPC to adapt to changing sea states in which different AR models are used for different sea states. \REVIEW{Considering the importance of wave excitation force prediction, other methods of prediction are also described in the literature, including the recursive least squares filter~\cite{Bradley2015}, the Kalman and extended Kalman filters~\cite{Bonfanti2020, Shangyan2020, Garcia2017, Fusco2010, Hals2010}, and neural networks~\cite{Bonfanti2020, Fusco2010, Li2019}. In practice, some of these techniques may be easier to implement than AR. }


\subsubsection{Non-linear Froude-Krylov (NLFK) forces: A novel static grid approach based on implicit surfaces}
\label{sec_NLFK_calculation}

\begin{figure}[]
 \centering 
   \subfigure[]{
   	\includegraphics[scale= 0.4]{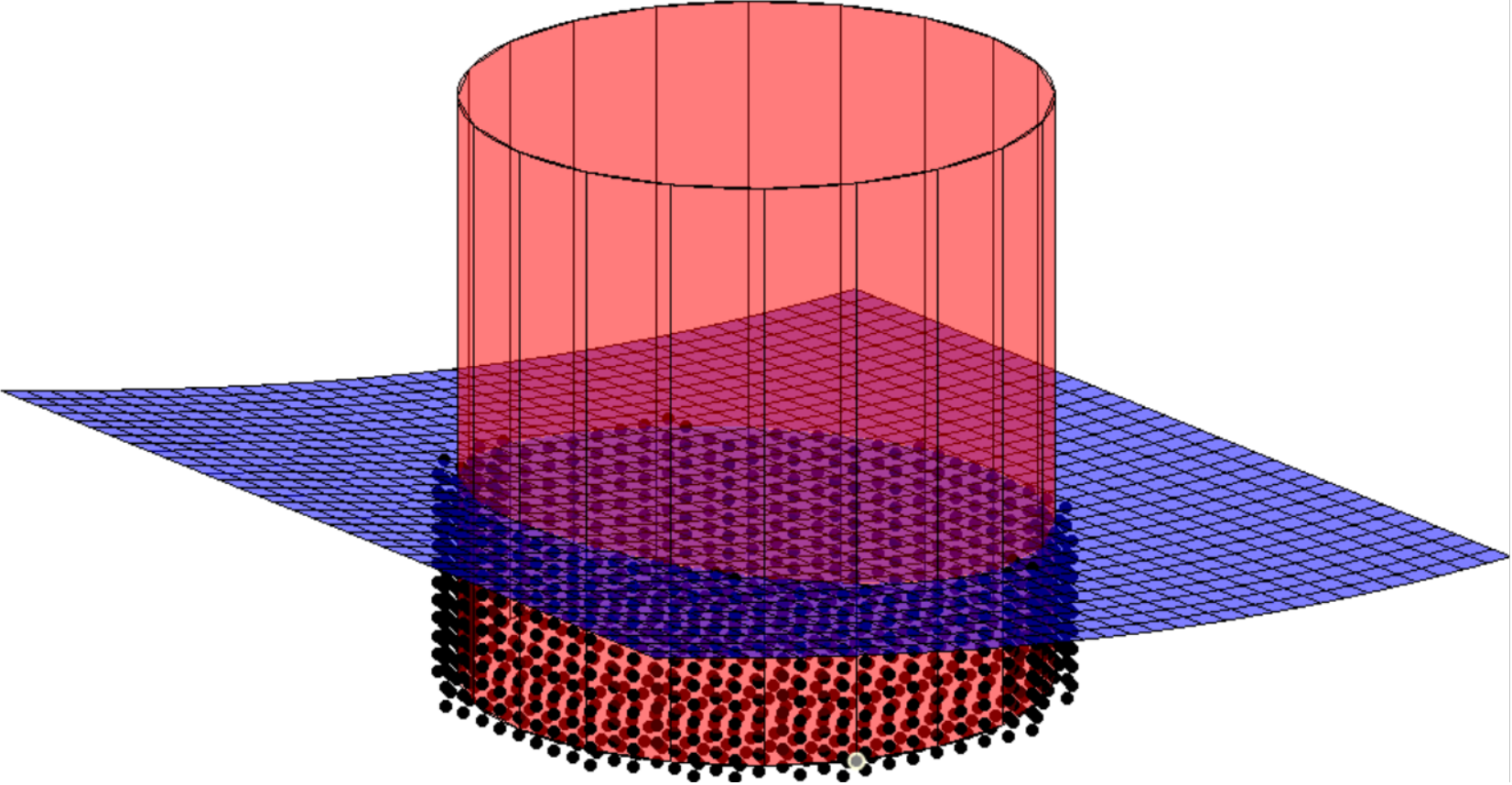}
	\label{fig_NLFK_schematic}
   }
      \subfigure[]{
   	\includegraphics[scale= 0.3]{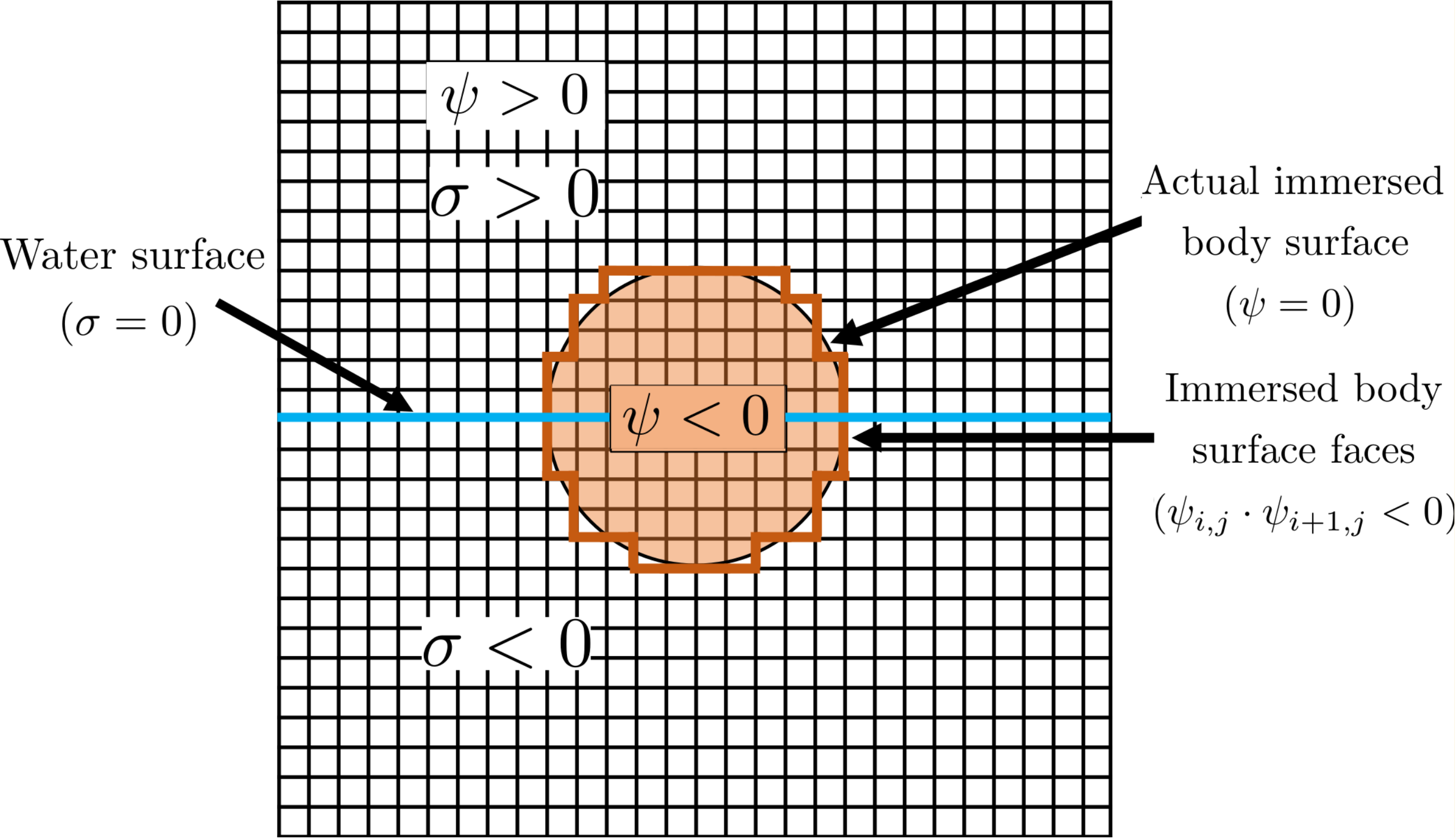}
	\label{fig_NLFK_integral}
   }
 \caption{NLFK force calculation using implicit signed distance functions $\sigma$ and $\psi$. \subref{fig_NLFK_schematic} A 3D schematic showing the instantaneous wetted surface area $S_b(t)$ of a vertical heaving cylinder interacting with the undulatory water surface. \subref{fig_NLFK_integral} A 2D schematic showing the stair-step representation of the immersed body on the Cartesian grid and the identification of the grid faces for evaluating the pressure integral using the body SDF $\psi$.  
 }
 \label{fig_NLFK}
\end{figure}

A significant amount of modeling accuracy can be achieved by considering the FK forces to be non-linear. The NLFK force is calculated by integrating the incident wave pressure $p_I(t)$ over the instantaneous wetted surface area $S_b(t)$ of the body; see Eq.~\eqref{eq_FK_force} and Fig.~\ref{fig_NLFK_schematic}. The computation of NLFK forces for practical control of WECs is considered prohibitively expensive in the wave energy literature. It is because such forces are typically computed using dynamic meshes, in which the computational domain is re-meshed to account for the relative motion between the body and the waves.  Nevertheless, computationally-efficient approaches have recently been developed for calculating NLFK forces. In~\cite{Giorgi2017}, Giorgi et al. presented an analytical method for evaluating the pressure integral for axisymmetric WECs. Though the method described in~\cite{Giorgi2017} is computationally attractive, it can only be applied to WEC devices that are geometrically solids of revolution. An alternative approach is presented in this section based upon the level set/signed distance function (SDF) that can effectively model the instantaneous wave-structure interaction of WECs on a static Cartesian grid. Moreover, the proposed technique can be applied to arbitrarily complex-shaped bodies because the SDF can be computed using efficient computational geometry algorithms within a narrow band of grid cells~\cite{Baerentzen2005}. Further, the level-set methodology is an embarrassingly parallel algorithm that is amenable to both distributed- and shared-memory parallelism.

First, we define a rectangular box region $\mathcal{R} = \mathcal{R}_w(t) \cup \mathcal{R}_a(t) $ around the WEC, which is discretized on a static Cartesian grid with rectangular cells. The grid cells are enumerated using the integer tuple $(i,j,k)$. The static region $\mathcal{R}$ should be a minimal one, covering only the wave amplitude and maximum displacement of the body expected in the simulation for computational efficiency. Next, define two level set functions $\psi(\x,t)$ and $\sigma(\x,t)$ over the entire box $\x \in \mathcal{R}$ that describe the signed distance to the WEC surface and the undulatory air-water interface, respectively. We take $\psi$ to be negative (positive) inside (outside) the body and $\sigma$ to be negative (positive) inside the water region $\mathcal{R}_w(t)$ (air region $\mathcal{R}_a(t)$). Zero-contours of $\psi$ and $\sigma$ implicitly define the WEC-fluid and the air-water interface, respectively. Sec.~\ref{sec_interface_track} provides more details on level set methodology, where we describe our multiphase CFD solver that is also based on the level set technique. The motion of the waves and the device is captured by redefining SDFs on the static grid~\footnote{SDF of a vertical cylinder can be prescribed analytically using constructive solid geometry operators, such as min/max acting on SDFs of primitive shapes. SDF of the air-water interface can also be prescribed analytically from the known surface elevation function $\eta_{\rm wave}(\x,t)$.}, which completely eliminates the need to re-mesh the computational domain $\mathcal{R}$.  The wave incident pressure $p_I$ is defined on the cell centers $\x_{i,j,k}$ of the static Cartesian grid in order to compute the NLFK force as
 \begin{align}
 p_I(\x_{i,j,k} ,t)  & = 0,   \hspace{18.5em}  \sigma(\x,t) > 0, \nonumber \\
  p_I(\x_{i,j,k} ,t)  & = \rho_w g \frac{\mathcal{H}}{2} \frac{\cosh(\kappa(d+\sigma))\cdot \cos(\kappa x - \omega t)}{\cosh(\kappa d)},   \hspace{3em}  \sigma(\x,t) \le 0, 
\label{eq_p_dynamic}
\end{align}
in which $\cH$ is the wave height, $\kappa$ is the wavenumber, $d$ is the depth of water above the sea floor, and $\omega$ is the wave frequency. The integral of $p_I$ over the wetted surface can be performed numerically as 
\begin{align}
	\text{\bf{F}}_{I}(t)  &= \sum_f   -p_I (\x_f,t) \, \n_f \, \Delta A_f.   \label{eqn_NLFK_p}
\end{align}
The discrete summation in Eq.~\eqref{eqn_NLFK_p} is carried over the Cartesian grid faces that provide a stair-step representation of the body on the Cartesian grid. This is shown in Fig.~\ref{fig_NLFK_integral}.  The set of the Cartesian grid faces $f$ can be easily identified by examining the sign change of $\sigma$.  The $\n_f $ and $\Delta A_f$ variables in the equation above represent the unit normal vector and the area of the cell face, respectively. The incident wave pressure $p_I(\x_f,t)$ on the cell face (where $\sigma$ is taken to be zero) is the weighted average of the neighboring cell center pressures, where the distance to the WEC surface $|\sigma(\x_{i,j,k}, t)|$ is used as the weights. In the heave direction, calculating F$_I(t)$ requires summing only over $z$-faces. 

The diffraction component of NLFK forces remains  linear. This is due to the assumption that the body is stationary when computing the diffraction forces. Similarly to LFK forces, the $z$-component of $\text{\bf{F}}_{D}(t)$ can be computed as a convolution integral between the diffraction impulse response function (DIRF) K$_d(t)$ in the heave direction and the water surface elevation as
 \begin{equation}
\text{F}_D(t) = K_d * \eta_\text{wave} = \int_{-\infty}^{\infty} \text{K}_d(\tau) \eta_\text{wave}(t-\tau) \; \text{d}\tau = \int_{-t_f}^{t_f} \text{K}_d(\tau) \eta_\text{wave}(t-\tau; x_{B}) \; \text{d}\tau .
\label{eq_diff_force}
\end{equation}
DIRF is the inverse Fourier transform of frequency-domain diffraction force data $\widehat{F}_D(\omega)$ that we obtain using ANSYS AQWA. 

We remark that the technique described in this section can be easily modified to model nonlinear buoyancy forces for varying cross-sectional WEC devices. This is achieved by replacing $p_I$ by $p_H = - \rho_w g z(t)$ in Eq.~\eqref{eqn_NLFK_p}.

%% file: Wave_dynamics.tex
This section describes the Stokes theory of regular and irregular water waves.

\subsection{Regular waves} \label{sec_regular_waves}

First-order Stokes waves, or regular waves, are simple harmonic waves of height $\cH$, time period $\cT$, and wavelength $\lambda$~\cite{book_Offshore, book_Waves_in_oceanic_and_coastal_waters}. Assuming that the waves travel in the positive $x$-direction, the wave elevation $\eta(x,t)$ from the still water surface at a depth of $d$ above the sea floor is
\begin{equation}
	\eta(x, t) = \frac{\cH}{2} \cos(\kappa x - \omega t),
\label{eq_first_order_wave_elev}
\end{equation} 
in which $\kappa = 2 \pi/\lambda$ is the wavenumber and $\omega = 2 \pi/\cT$ is the angular wave frequency. The first-order Stokes wave satisfies the dispersion relation given by
\begin{equation}
	\omega^2 =  g \kappa \tanh{(\kappa d)}, 
\label{eq_dispersion_relation}
\end{equation}
which relates the wave frequency $\omega$ to wavenumber $\kappa$ and water depth $d$. Eq.~\ref{eq_dispersion_relation} is a transcendental equation that requires an iterative procedure to calculate $\kappa$ for given $\omega$, or vice versa. Instead, we use an explicit relationship between these quantities that is accurate enough for practical purposes at all water depths~\cite{Fenton1988}:
\begin{equation}
	\label{eq_explicit_dispersion_relation}
   \kappa d \approx \frac{\Gamma+ \beta^{2}\left(\cosh\beta\right)^{-2}}{\tanh\beta + \beta\left(\cosh\beta\right)^{-2}},
\end{equation}
in which $\beta = \Gamma \left(\tanh \Gamma \right)^{-\half}$ and $\Gamma = \omega^2 d / g$.

As the waves travel along the ocean or sea surface, they carry kinetic and potential energy---this energy is partially absorbed by the WEC device. The time-averaged wave power per unit crest width carried by the regular waves in the direction of propagation is given by~\cite{book_Offshore}
\begin{equation}
	\widebar{P}_{\rm wave} = \frac{1}{8} \rhow g \cH^2 c_g,
\label{eq_regular_wave_power}
\end{equation}
in which $c_g$ is the group velocity of the waves, i.e., the velocity with which wave energy is transported and it is given by the relation
\begin{equation}
	c_g = \half \frac{\lambda}{\cT} \left(1 + \frac{2 \kappa d}{\sinh (2 \kappa d )} \right).
\label{eq_group_velocity}
\end{equation}
In the deep water limit, where $d > \lambda/2$ and $\kappa d\rightarrow \infty$, Eqs.~\ref{eq_dispersion_relation} and \ref{eq_group_velocity} become
\begin{equation}
	\label{eq_deep_water_limit}
	\omega^2 = g \kappa  \quad \text{or} \quad \lambda = \frac{g\mathcal{T}^2}{2\pi} \qquad \text{and} \qquad c_g = \frac{\lambda}{2 \cT}.  \qquad \text{(deep water limit)}
\end{equation}
Substituting Eq.~\eqref{eq_deep_water_limit} into Eq.~\eqref{eq_regular_wave_power}, the wave power per unit crest width in the deep water limit is expressed as
\begin{equation}
		\label{eq_regular_wave_power_deepwater}
		\widebar{P}_{\rm wave} = \frac{\rho_w g^2 \cH^2 \cT}{32 \pi} \approx \cH^2 \cT \;\; \text{kW/m}, \qquad \text{(deep water limit)}
\end{equation}
in which the constant numerical factor $\rho_w g^2/32\pi \approx 10^3$ when all quantities are evaluated in SI units.  

\subsection{Irregular waves} \label{sec_irregular_waves}

A realistic sea state consists of irregular waves. Mathematically, an irregular wave can be described as a linear superposition of a large number of (first-order) regular wave components. Using the superposition principle, the sea surface elevation can be expressed as
\begin{equation}
     \label{eq_irregular_elevation}
     \eta(x,t) = \sum_{i=1}^{N_w} a_i \cos (\kappa_i x - \omega_i t + \theta_i ),
\end{equation}
in which $N_w$ is the number of (regular) wave components. Each wave component has its own amplitude $a_i = \cH_i/2$, angular frequency $\omega_i$, wavenumber $\kappa_i$, and a random phase $\theta_i$. Each component also satisfies the dispersion relation between $\kappa_i$ and $\omega_i$ given by Eq.~\ref{eq_dispersion_relation}. The random phase $\theta_i$ follows the uniform distribution in the interval $[0, 2 \pi]$.

The linear superposition of first-order waves implies that the total energy carried by the irregular wave is the sum of wave energy carried by the individual wave components. To describe the energy content of irregular waves, a continuous wave spectral density function $S(\omega)$ is used, wherein the number of wave components $N_w$ tend to infinity and an infinitesimal small frequency bandwidth $\text{d}\omega$ separates the wave components. The area under the $S(\omega)$ versus $\omega$ curve gives the total energy of the irregular wave, modulo the factor $\rho_w g$. Discretely, the wave frequencies are chosen at an equal interval of $\Delta \omega$ and the wave spectral density function $S(\omega)$ approaches zero for frequencies outside the narrow bandwidth. In this work, we consider only singly-peaked wave spectra with $S(\omega)$ peaking at a particular frequency $\omega_p$. Each wave component of an irregular wave has a wave amplitude that is related to the spectral density function by
\begin{equation}
		\label{eq_wave_spectrum_amplitude}
		a_{i} = \sqrt{ 2 \cdot S (\omega_i) \cdot \Delta{\omega} } \; .
\end{equation}

\begin{figure}[]
 \centering
 \includegraphics[scale = 0.45]{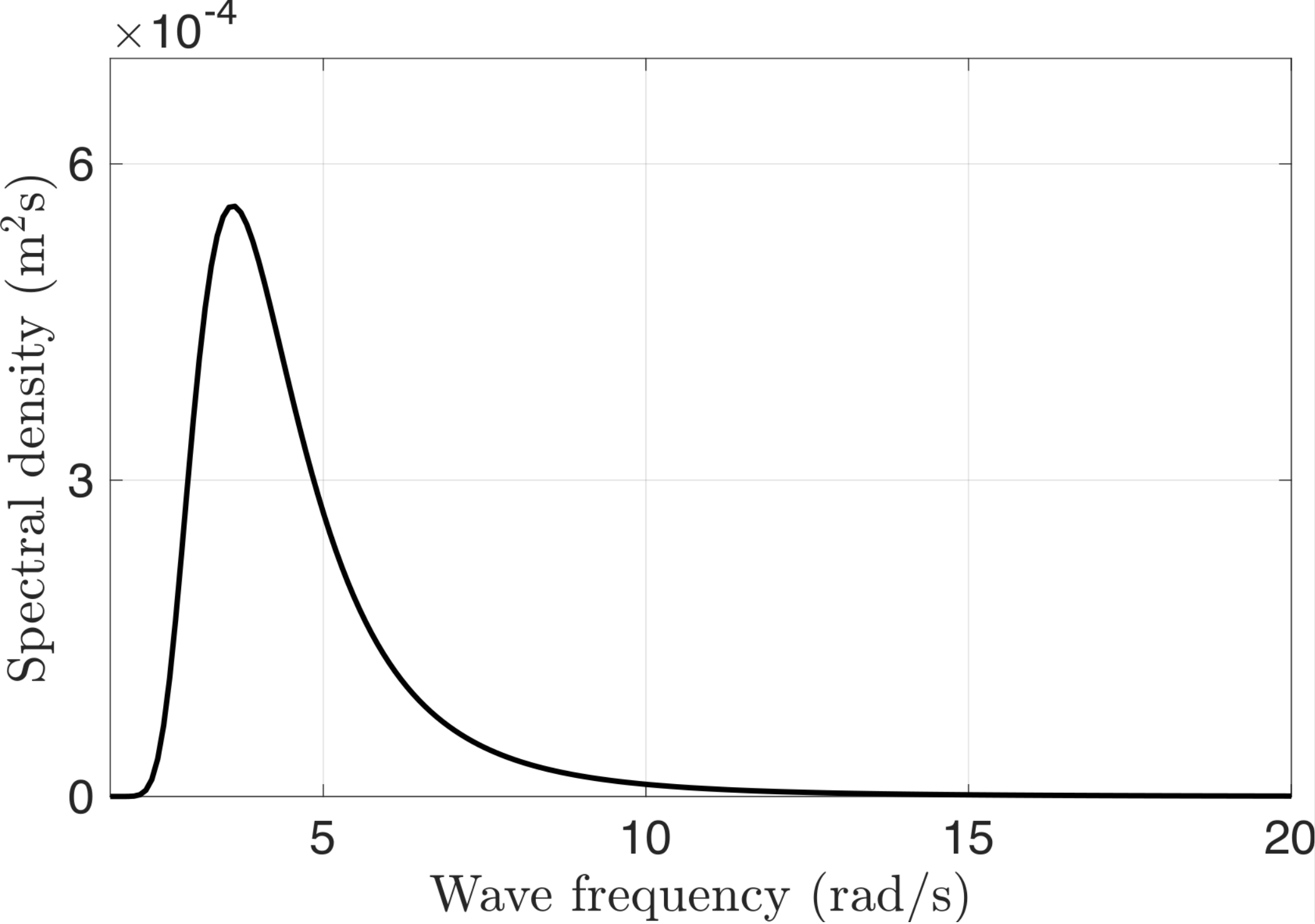}
 \caption{The Bretschneider wave spectrum obtained using $\cH_s = 0.15$ m and $\cT_p = 1.7475$ s ($\omega_p = 2\pi / \cT_p = 3.5955 $ rad/s).}
 \label{fig_Bretschneider_spectrum}
\end{figure}

We consider the two-parameter Bretschneider spectrum~\cite{book_Offshore}, which is suited for open seas where our WEC device is considered to be located. Specifically, the Bretschneider spectrum $S(\omega)$ is based on the significant wave height $\cH_s$ and the peak wave time period $\cT_p$ and it reads as
\begin{equation}
	S(\omega) = \frac{173 \cdot \cH^2_s}{\cT^4_p} \cdot \omega^{-5} \cdot \exp \left(\frac{-692}{\cT^4_p} \cdot \omega^{-4}\right). 
\label{eq_Bretshneider_wave_spec}
\end{equation}
The peak wave time period $\cT_p$  is the time period with the highest spectral density; see Fig.~\ref{fig_Bretschneider_spectrum}.

For irregular waves the mean wave power per unit crest width is calculated as
\begin{equation}
	\widebar{P}_{\rm wave} = \rho_w g \left(\int_{0}^{\infty} S(\omega) \; \text{d}\omega \right) c_g \approx   \rho_w g \left(\sum_{i=1}^{N_w}\frac{1}{2}a^2_{i}\right) c_g
\label{eq_irreg_wave_power}
\end{equation}
in which the group velocity $c_g$ is calculated from Eq.~\ref{eq_group_velocity} using the significant wavelength and peak time period of the spectrum. In the deep water limit, Eq.~\ref{eq_irreg_wave_power} becomes
\begin{equation}
	\label{eq_irreg_wave_power_deep}
	\widebar{P}_{\rm wave} \approx 0.49 \cH_\text{s}^2 \cT_p \;\; \text{kW/m}. \qquad \text{(deep water limit)}
\end{equation}

%% file: IBAMR.tex

This section begins with a description of continuous equations of motion solved by the \emph{fictitious domain} Brinkman penalization (FD/BP) method~\cite{BhallaBP2019,Khedkar2021}. Following this, we discuss the multiphase interface tracking technique. Afterwards, the spatiotemporal discretization, the overall solution methodology, and the time-stepping scheme are briefly discussed. Next, the numerical wave tank setup for performing fully-resolved and control-informed multiphase WSI simulations is discussed. A time-averaged kinetic energy equation is also derived to describe how power transfers from waves to the PTO system. 


\subsection{Continuous equations of motion} 
\label{sec_cont_eqs}

Let $\Omega \subset \mathbb{R}^d$ with $d = 3$ represent a fixed three-dimensional region in space. The incompressible Navier-Stokes (INS) equations govern the dynamics of the coupled multiphase fluid-structure system occupying this domain:
\begin{align}
  \D{\rho \u(\x,t)}{t} + \div (\rho \, \u(\x,t) \otimes \u(\x,t)) &= -\grad p(\x,t) + \div \left[\mu \left(\grad \u(\x,t) + \grad \u(\x,t)^T\right) \right]+ \rho\g + \f_c(\x,t), \label{eqn_momentum}\\
  \div \u(\x,t) &= 0, \label{eqn_continuity} 
\end{align}
which describe the momentum and incompressibility of a fluid with velocity $\u(\x,t)$ and pressure $p(\x,t)$ in an Eulerian coordinate system $\x = (x,y,z) \in \Omega$.  Eqs.~\ref{eqn_momentum} and~\ref{eqn_continuity} are written for the entire computational domain $\Omega$. The domain $\Omega$ is further decomposed into two non-overlapping regions, one occupied by the fluid $\Omega_f(t) \subset \Omega$ and the other by an immersed body $\Omega_b(t) \subset \Omega$, so that $\Omega = \Omega_f(t) \cup \Omega_b(t)$. The term $\f_c(\x,t)$ is the constraint force (density) that vanishes outside $\Omega_b(t)$ and ensures a rigid body velocity $\u_b(\x,t)$ within the solid. The density and viscosity fields vary spatiotemporally and are denoted $\rho(\x,t)$ and $\mu(\x,t)$, respectively. The location of the solid body is tracked using an indicator function $\chi(\x,t)$, which is non-zero only within $\Omega_b(t)$. The acceleration due to gravity is directed towards the negative $z$-direction: $\g = (0, 0, -g)$. Fig.~\ref{fig_cfd_domains} shows the schematic representation of the domain occupied by the three (air, water, and solid) phases. 

The immersed body is treated as a porous region with vanishing permeability $\kappa_p \ll 1$. The Brinkman penalization constraint force is given by
\begin{align}  
	\f_c(\x,t)  = \frac{\chi(\x,t)}{\kappa_p}\left(\u_b(\x,t) - \u(\x,t)\right).
\label{eqn_brinkman_force}
\end{align}
The rigid body velocity $\u_b(\x,t)$ in the solid region $\Omega_b(t)$ is determined by the combined actions of the hydrodynamic force (estimated by the multiphase flow solver) and the control force $\text{F}_\text{PTO}$ (estimated by the MPC). Sec.~\ref{sec_fsi_coupling} explains this. 

\subsection{Interface tracking}
\label{sec_interface_track}

\begin{figure}
  \centering
  \subfigure[Continuous domain]{
    \includegraphics[scale = 0.45]{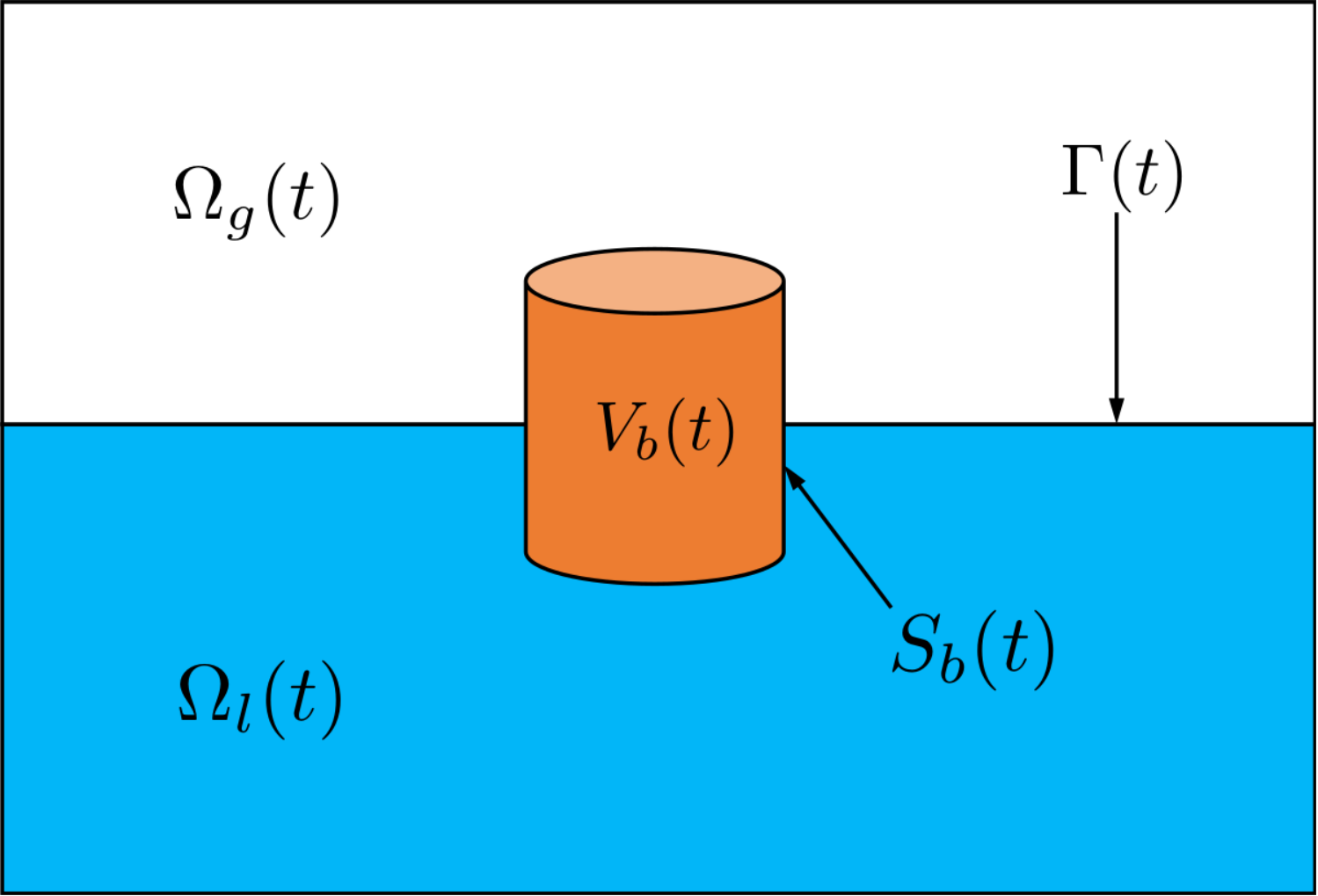} 
    \label{fig_cont_domain}
  }
   \subfigure[FD/BP discretized domain]{
    \includegraphics[scale = 0.45]{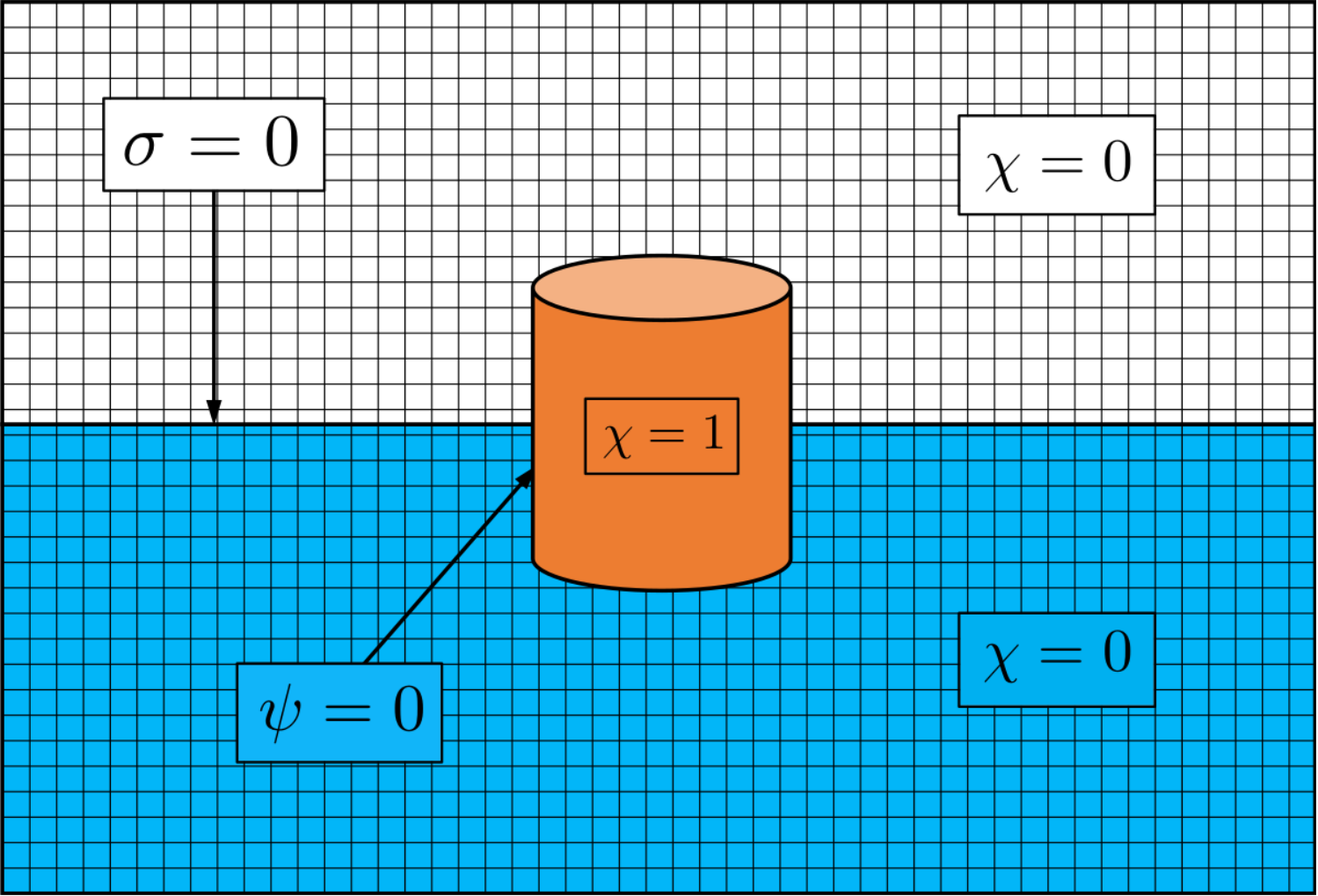}
    \label{fig_discrete_domain}
  }
  \caption{\subref{fig_cont_domain} Schematic of a two-dimensional slice through the computational domain $\Omega$ showing an immersed body interacting with an air-water interface.~\subref{fig_discrete_domain} Discretization of the domain $\Omega$ on a Cartesian mesh and values of the indicator function $\chi(\x,t)$ used to differentiate the fluid and solid regions in the FD/BP method. Here, $\chi(\x,t)= 1$ inside the solid domain and $\chi(\x,t) = 0$ in air and water domains. The air-water interface $\Gamma(t)$ is tracked by the zero-contour of $\sigma(\x,t)$, while the zero-contour of $\psi(\x,t)$ tracks the solid-fluid interface $\Sb(t)$.
  }
\label{fig_cfd_domains}
\end{figure}

Here, we briefly describe the interface tracking method for capturing the air-water and fluid-solid interfaces; details on the implementation of the technique can be found in our prior works~\cite{Nangia2019WSI, Nangia2019MF}. A scalar level set/signed distance function (SDF)  $\sigma(\x,t)$ is used to demarcate the liquid (water) and the gas (air) regions, $\Omega_l \subset \Omega$ and $\Omega_g \subset \Omega$, respectively, in the computational domain. The zero-contour of $\sigma$ defines the air-water interface $\Gamma(t) = \Omega_l \cap \Omegag$. Similarly, the surface of the immersed body $\Sb(t) = \partial V_b(t)$ is tracked using the zero-contour of the level set function $\psi(\x,t)$; see Fig.~\ref{fig_discrete_domain}. The indicator function $\chi(\x,t)$ for the solid domain is computed based on the level set function $\psi$. The two SDFs are advected using the local fluid velocity:
\begin{align}
	\D{\sigma}{t} + \u \cdot \grad \sigma &= 0, \label{eq_ls_fluid_advection} \\
	\D{\psi}{t} +  \u \cdot \grad \psi  &= 0. \label{eq_ls_solid_advection}\\
\end{align}
The density and viscosity in the entire computational domain is expressed as a function of $\sigma(\x,t)$ and $\psi(\x,t)$ using the signed distance property: 
\begin{align}
\rho (\x,t) &= \rho(\sigma(\x,t), \psi(\x,t)), \label{eq_rho_ls} \\
\mu (\x,t) &= \mu(\sigma(\x,t), \psi(\x,t)). \label{eq_mu_ls}
\end{align}
To maintain their signed distance property, both level set functions need to be reinitialized after each time step. To reinitialize $\sigma$, we use the relaxation approach of Sussman et al.~\cite{Sussman1994} to compute the steady-state solution to the Hamilton-Jacobi equation.  $\psi$, on the other hand, is reinitialized directly because the SDF of a vertical cylinder can be constructed analytically by using constructive solid geometry operators (i.e., the min/max operators) on primitive shapes~\cite{Zhang2019}.  

\subsection{Spatial discretization} 
\label{sec_spatial_discretization}

\begin{figure}
  \centering
  \subfigure[\REVIEW{2D staggered Cartesian grid}]{
    \includegraphics[scale = 0.45]{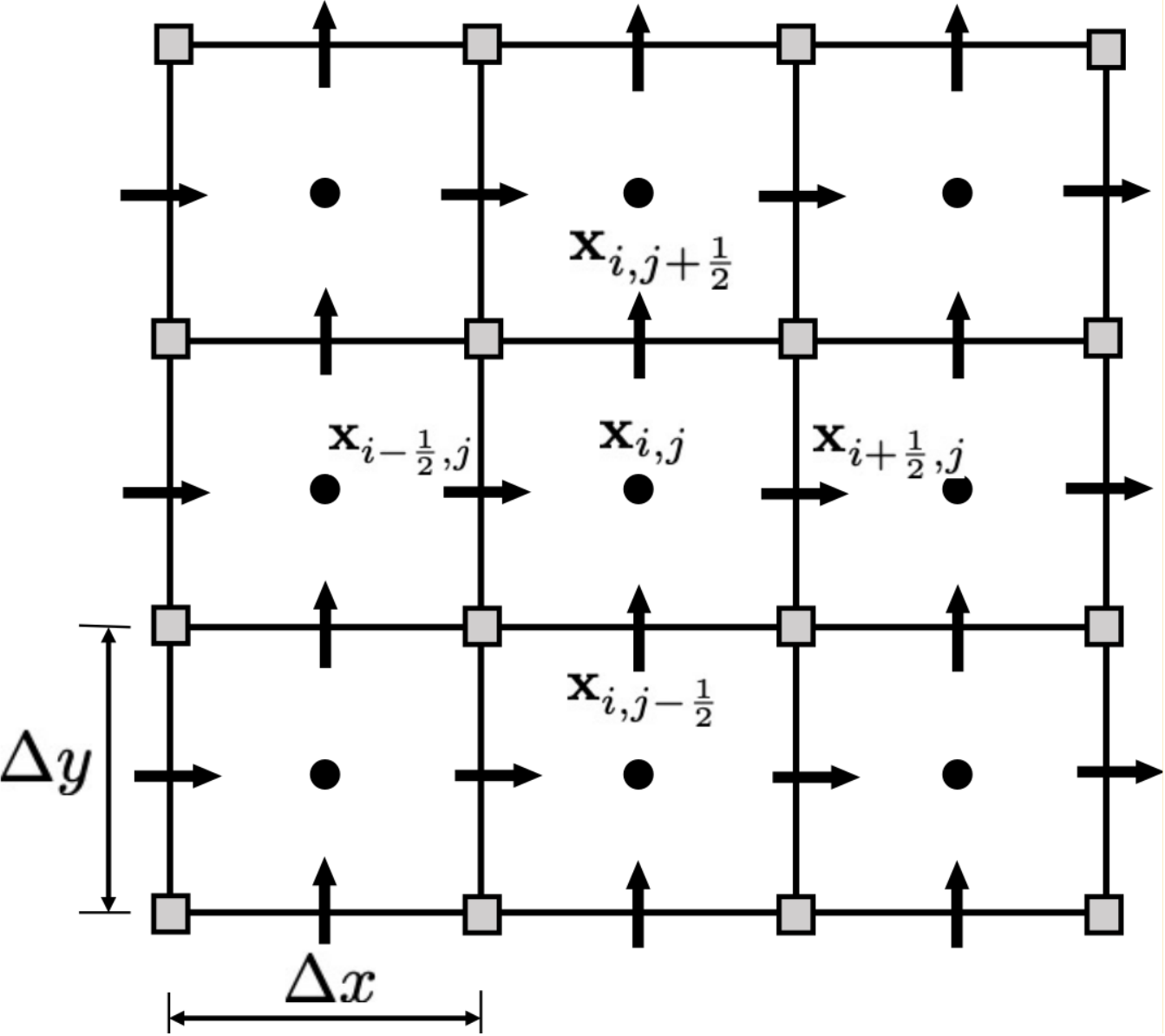} 
    \label{fig_discretized_staggered_grid}
  }
   \subfigure[\REVIEW{Single grid cell}]{
    \includegraphics[scale = 0.35]{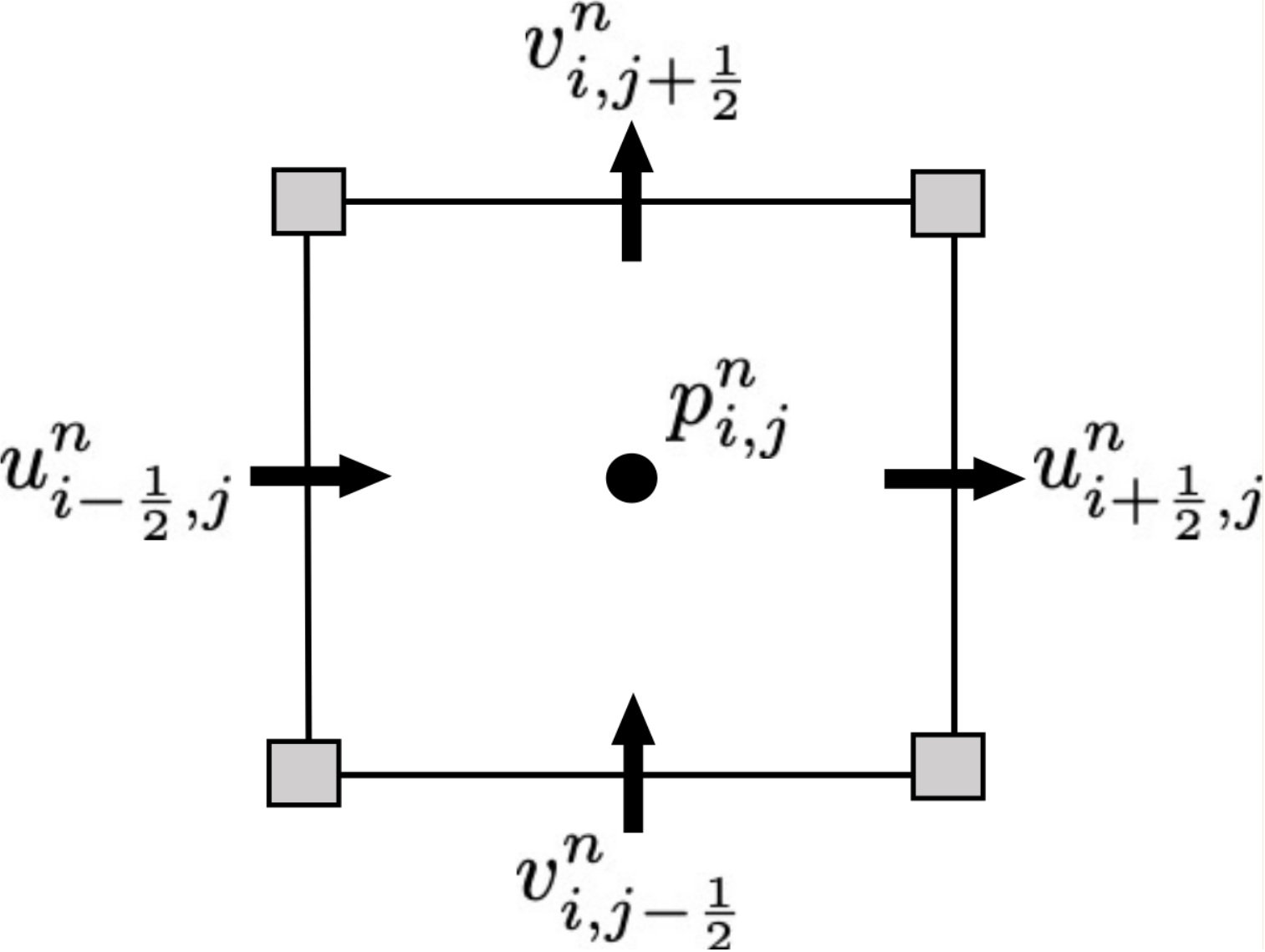}
    \label{fig_single_cell}
  }
  \caption{\REVIEW{Schematic representation of a 2D staggered Cartesian grid. \subref{fig_discretized_staggered_grid} shows the coordinate system for the staggered grid. \subref{fig_single_cell} shows a single grid cell with velocity components $u$ and $v$ approximated at the cell faces (${\bf{\rightarrow}}$) and scalar variable pressure $p$ approximated at the cell center ($\bullet$) at $n^\text{th}$ time step.}}
\label{fig_cfd_domains}
\end{figure}

The continuous equations of motion given by Eqs.~\ref{eqn_momentum}-\ref{eqn_continuity} are discretized on a locally-refined staggered Cartesian grid. The grid covers the domain $\Omega$ with $\Nx \times \Ny \times \Nz$ rectangular cells. The grid spacing in the three spatial directions are $\dx$, $\dy$, and $\dz$ respectively. Without any loss of generality, the lower left corner of the domain is considered the origin $(0,0,0)$ of the coordinate system such that each cell center of the grid has a position $\x_{i,j,k} = \left((i + \half)\dx,(j + \half)\dy,(k + \half)\dz\right)$ for $i = 0, \ldots, \Nx - 1$, $j = 0, \ldots, \Ny - 1$, and $k = 0, \ldots, \Nz - 1$. The location of a cell face which is half a grid cell away from $\x_{i,j,k}$ in the $x$-direction is at $\x_{i-\half,j,k} = \left(i\dx,(j + \half)\dy,(k + \half)\dz\right)$. Similarly, the location of a cell face that is half a grid cell away from $\x_{i,j,k}$ in the $y$-directions is $\x_{i,j-\half,k} =\left((i + \half)\dx,j\dy,(k + \half)\dz\right)$ and in the $z$-direction it is $\x_{i,j,k-\half} =\left((i + \half)\dx,(j + \half)\dy,k\dz\right)$. \REVIEW{See Fig.~\ref{fig_discretized_staggered_grid}}. The time at time step $n$ is denoted by $t^n$. The scalar quantities: level set functions, pressure, and the material properties (density and viscosity) are all approximated at cell centers and are denoted $\sigma_{i,j,k}^{n} \approx \sigma \left(\x_{i,j,k}, t^n\right)$, $\psi_{i,j,k}^{n} \approx \psi\left(\x_{i,j,k}, t^n\right)$, $p_{i,j,k}^{n} \approx p\left(\x_{i,j,k},t^{n}\right)$, $\rho_{i,j,k}^{n} \approx \rho\left(\x_{i,j,k},t^{n}\right)$ and  $\mu_{i,j,k}^{n} \approx \mu\left(\x_{i,j,k},t^{n}\right)$, respectively. \REVIEW{See Fig.~\ref{fig_single_cell}}. Some of these scalar quantities need to be interpolated onto the required degrees of freedom; see Nangia et al.~\cite{Nangia2019MF} for further details. The velocity degrees of freedom are approximated on the cell faces as $u_{i-\half,j,k}^{n} \approx u\left(\x_{i-\half,j,k}, t^{n}\right)$, $v_{i,j-\half,k}^{n} \approx v\left(\x_{i,j-\half,k}, t^{n}\right)$, and $w_{i,j,k-\half}^{n} \approx w\left(\x_{i,j,k-\half}, t^{n}\right)$. The constraint force density and the gravitational body force in the momentum Eq.~\ref{eqn_momentum} are also approximated on cell faces. For all spatial derivatives, second-order finite differences are used.  A uniform grid spacing  $\Delta x = \Delta y = \Delta z = h$ is used for all simulations in this work, unless stated otherwise. For readability, the discretized version of the differential operators are denoted with a $h$ subscript, e.g., $\grad \approx \grad_h$. For further details on the spatial discretization on a hierarchy of adaptively refined meshes, see our prior works~\cite{Nangia2019MF, Cai2014, Griffith2009, Bhalla13}.

\subsection{Numerical wave tank (NWT)} \label{sec_NWT}

\begin{figure}[]
 \centering
 \includegraphics[scale = 0.45]{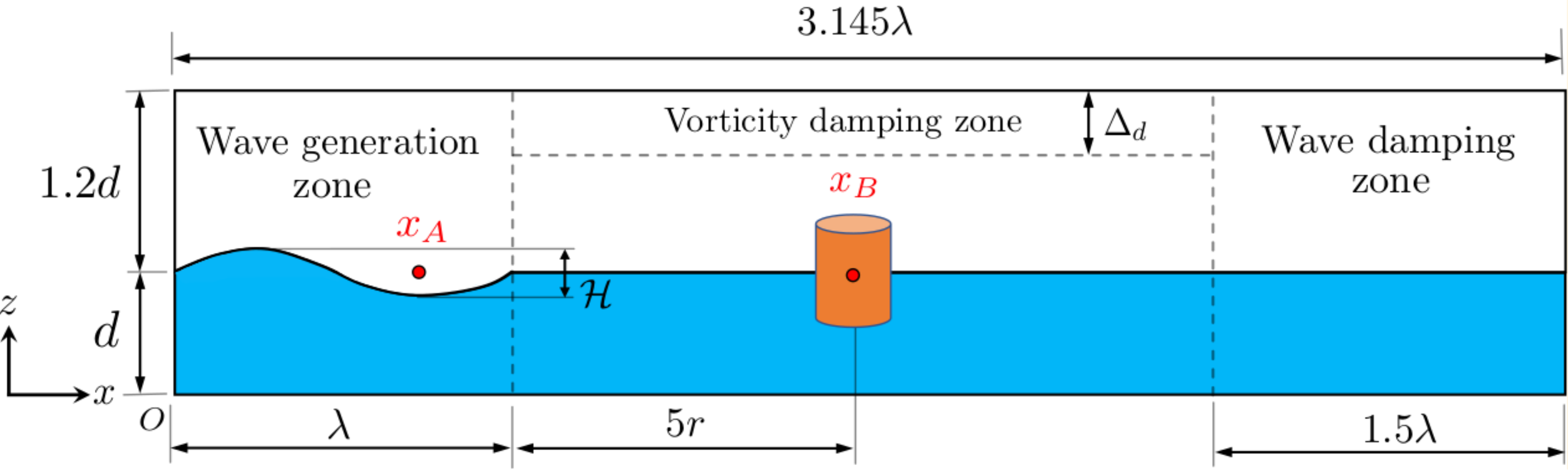}
 \caption{Numerical wave tank (NWT) schematic showing wave generation, wave damping, and vorticity damping zones. The WEC device is placed in the working zone of length $3.145\lambda$.
 }
 \label{fig_NWT_schematic}
\end{figure}

The fully-resolved and control-informed WSI of the device is simulated using a NWT, depicted in Fig.~\ref{fig_NWT_schematic}.  In the tank, the converter is located at position $x_B$. Dirichlet boundary condition for the velocity components is used to generate regular and irregular water waves at the left boundary of the domain. The waves travel in the positive $x$-direction and are reflected back from the right boundary of the domain and the device surface.  A reflected wave can cause wave distortion and interference phenomena and reduce the quality of waves reaching the device if it is not handled properly.  A variety of numerical techniques have been proposed in the literature to mitigate these effects~\cite{Miquel2018,Windt2018,Windt2019}, such as the relaxation zone method~\cite{Jacobsen2012}, the active wave absorption method~\cite{Higuera2013,Frigaard1995,Schaffer2000}, the momentum damping method~\cite{Choi2009,Ha2013}, the viscous beach method~\cite{Ghasemi2014}, the porous media method~\cite{Dong2009,Jacobsen2015}, and the mass-balance PDE method~\cite{Hu2016}. The relaxation zone method is used in this paper because of its simplicity and effectiveness. To smoothly extend the Dirichlet velocity boundary conditions into the wave tank, a relaxation zone called the wave generation zone is added near the inlet boundary. The wave generation zone reduces the interaction of the reflected waves (from the device) with the inlet boundary. The wave generation zone being relatively free of reflected waves, the up-wave point $x_A$ is also placed inside this zone, which accurately records the wave elevation data $\eta_{\rm wave}(t; x_A)$ and sends it to the MPC. Also sent to the controller are the device's displacement and velocity, $z$ and $\dot{z}$, computed from the fully-resolved WSI. Near the outlet boundary, a second relaxation zone, known as the wave damping zone, is located to smoothly dampen out waves that reach the right end of the NWT. The length of the wave damping zone is set to $1.5 \lambda$ in all simulations. 

The top boundary of the NWT is a zero pressure boundary. To dissipate the vortical structures reaching the top boundary, a vorticity damping zone is implemented (see Fig.~\ref{fig_NWT_schematic}). The vortical structures shed by the device (as a result of FSI) move freely in the air region (which is taken to be small in order to reduce the computational cost of the simulations) and interfere with the top boundary. In order to implement the vorticity damping zone, a damping force $\f_d$ is added to the momentum equation, which reads as
\begin{equation}
	\f_d =  -g(\tilde{z}) \u.
\label{eq_spongelayer}
\end{equation}
Here, $g(\tilde{z})$ = $\rho_a(\cos(\pi \tilde{z})+1)/(4\Delta t)$ is the smoothed damping coefficient, $\rho_a$ is the density of the air phase, $\Delta t$ is the time step size of the multiphase flow solver, $\tilde{z} = (z - z_\text{max})/{\Delta_d}$ is the normalized $z$ coordinate, $z_\text{max} = 2.2d$, and $\Delta_d$ is the vorticity damping zone width. In all our simulations, $\Delta_d$ is taken to be four (coarsest grid) cell size wide. More details on the implementation of the relaxation zone method and the level set-based NWT can be found in  our prior work~\cite{Nangia2019WSI}.

\subsection{Solution methodology} 
\label{sec_sol_method}

The methodology to solve the discretized equations of motion involves three major steps:
\begin{enumerate}
\item Specify the material properties, density $\rho(\x, t)$ and viscosity $\mu(\x, t)$ in the entire computational domain.
\item Calculate the Brinkman penalization rigidity constraint force density $\f_c(\x,t)$ based on the vertical cylinder WEC dynamics.
\item Update the solutions for $\sigma$, $\psi$, $\u$, and $p$.
\end{enumerate}

We briefly review the computations described above for a vertical cylinder device with a single degree of freedom (in the $z$-direction). We refer readers to Bhalla et al.~\cite{BhallaBP2019} and references therein for a general FSI treatment. 

\subsubsection{Density and viscosity specification}

To transition between the air-water interface $\Gamma(t)$ and the fluid-solid interface $S_b(t)$, a smoothed Heaviside function is used. $\ncells$ grid cells are used on either side of the interface to smoothly vary the material properties in the transition region. For example, a given material property $\Im$, say density or viscosity, is prescribed throughout the computational domain by first calculating the \emph{flowing} phase (i.e., air and water) property as
\begin{equation}
	\Im^{\text{flow}}_{i,j,k} = \Im_l + (\Im_g - \Im_l) \widetilde{H}^{\text{flow}}_{i,j,k},
\label{eqn_ls_flow}
\end{equation}
and later correcting $\Im^{\text{flow}}$ to account for the solid body by 
\begin{equation}
	\Im_{i,j,k}^{\text{full}} = \Im_s + (\Im^{\text{flow}}_{i,j,k} - \Im_s) \widetilde{H}^{\text{body}}_{i,j,k}.
\label{eqn_ls_solid}
\end{equation}
Here, $\Im^{\text{full}}$ is the final scalar material property field throughout $\Omega$. To specify the transition specified by Eqs.~\ref{eqn_ls_flow} and~\ref{eqn_ls_solid}, the standard numerical Heaviside functions are used:
\begin{align}
\widetilde{H}^{\text{flow}}_{i,j,k} &= 
\begin{cases} 
       0,  & \sigma_{i,j,k} < -\ncells \, h,\\
        \frac{1}{2}\left(1 + \frac{1}{\ncells \, h} \sigma_{i,j,k} + \frac{1}{\pi} \sin\left(\frac{\pi}{ \ncells \, h} \sigma_{i,j,k}\right)\right) ,  & |\sigma_{i,j,k}| \le \ncells \, h,\\
        1,  & \textrm{otherwise}.
\end{cases}       \label{eqn_Hflow} \\
\widetilde{H}^{\text{body}}_{i,j,k} &= 
\begin{cases} 
       0,  & \psi_{i,j,k} < -\ncells \, h,\\
        \frac{1}{2}\left(1 + \frac{1}{\ncells \, h} \psi_{i,j,k} + \frac{1}{\pi} \sin\left(\frac{\pi}{ \ncells \, h} \psi_{i,j,k}\right)\right) ,  & |\psi_{i,j,k}| \le \ncells \, h,\\
        1,  & \textrm{otherwise}.  \label{eqn_Hbody}
\end{cases}
\end{align}

In all simulations performed in this study, the number of transition cells $\ncells = 1$ for both air-water and fluid-solid interfaces. 

\subsubsection{Time stepping scheme}

The time stepping scheme employs a fixed-point iteration with $\ncycles$ cycles per time step to evolve quantities from time level $t^n$ to time level $t^{n+1} = t^n + \Delta t$. To denote the cycle number of a fixed-point iteration, a $k$ superscript is used. At the beginning of every time step, the solutions from the previous time step are used to initialize cycle $k = 0$: $\u^{n+1,0} = \u^{n}$, $p^{n+\half,0} = p^{n-\half}$, $\sigma^{n+1,0} = \sigma^{n}$, and $\psi^{n+1,0} = \psi^{n}$. The physical quantities at the initial time $n = 0$ are prescribed via initial conditions.  A larger number of cycles in the simulation allows a larger, more stable time step size. In this work, we limit $\ncycles$ to 1 so that the number of linear solves per time step is reduced for the computationally expensive 3D simulations.
 
\subsubsection{Level set advection}

To evolve the two level set/signed distance functions $\sigma$ and $\psi$, we use a standard explicit advection scheme as follows
\begin{align}
\frac{\sigma^{n+1,k+1} - \sigma^{n}}{\dt} + Q\left(\u^{n+\half,k}, \sigma^{n+\half,k}\right) &= 0, \\
\frac{\psi^{n+1,k+1} - \psi^{n}}{\dt} + Q\left(\u^{n+\half,k}, \psi^{n+\half,k}\right) &= 0,
\end{align}
in which $Q(\cdot,\cdot)$ represents an explicit piecewise parabolic method (xsPPM7-limited) approximation to the linear advection terms on cell centers~\cite{Griffith2009, Rider2007}.

\subsubsection{Multiphase incompressible Navier-Stokes solution}

The discretized form of the multiphase incompressible Navier-Stokes Eqs.~\ref{eqn_momentum}-\ref{eqn_continuity} in conservative form reads as
\begin{align}
	\frac{\breve{\V \rho}^{n+1,k+1} \u^{n+1,k+1} - { \V \rho}^{n} \u^n}{\dt} + \C^{n+1,k} &= -\grad_h \, p^{n+\half, k+1}
	+ \left(\L_{\mu} \u\right)^{n+\half, k+1}
	+  \V \wp^{n+1,k+1}\g +  \f_c^{n+1,k+1}, \label{eqn_c_discrete_momentum}\\
	 \grad \cdot \u^{n+1,k+1} &= \V{0}, \label{eqn_c_discrete_continuity}
\end{align}
in which $\C^{n+1,k}$ is the discretized version of the convective term $\div (\rho \, \u \otimes \u)$ and the density approximation $\breve{\V \rho}^{n+1,k+1}$ is computed using a consistent mass/momentum transport scheme. The consistent mass/momentum transport scheme ensures the numerical stability of cases involving high density contrast between various phases, such as air, water, and the solid device. See previous works by Nangia et al.~\cite{Nangia2019MF} and Bhalla et al.~\cite{BhallaBP2019} for a detailed exposition of the consistent mass/momentum transport scheme employed in this study.

The viscous strain rate  in Eq.~\eqref{eqn_c_discrete_momentum} is handled using the Crank-Nicolson approximation: $\left(\L_{\mu} \u\right)^{n+\half, k+1} =  \half\left[\left(\L_{\mu} \u\right)^{n+1,k+1} + \left(\L_{\mu} \u\right)^n\right]$, in which $\left(\L_{\mu}\right)^{n+1} = \grad_h \cdot \left[\mu^{n+1} \left(\grad \u + \grad \u^T\right)^{n+1}\right]$.
The newest approximation to the viscosity $\mu^{n+1,k+1}$ is obtained using the two-stage process described by Eqs.~\ref{eqn_ls_flow} and~\ref{eqn_ls_solid}. The gravitational body force term $\V \wp \g = \V{\rho}^{\text{flow}} \g$ is calculated using the flow density field to avoid spurious currents generated due to large density variations near the fluid-solid interface~\cite{Nangia2019WSI}.

\subsubsection{Fluid-structure coupling}
\label{sec_fsi_coupling}

The Brinkman penalization term $\f_c$ given by Eq.~\ref{eqn_brinkman_force} is treated implicitly in the discretized version of the momentum Eq.~\eqref{eqn_c_discrete_momentum} and computed as
\begin{align}
	\f_c^{n+1,k+1} = \frac{\widetilde{\chi}}{\kappa_p}\left(\u_b^{n+1,k+1} - \u^{n+1,k+1}\right),  \label{eqn_bp_discrete}
\end{align}
in which the discretized indicator function is defined using the body Heaviside function (see Eq.~\ref{eqn_Hbody}) as $\widetilde{\chi} = 1 - \widetilde{H}^{\text{body}}$; $\widetilde{\chi} = 1$ inside the solid region. A sufficiently small value of the permeability coefficient $\kappa_p \sim \cO(10^{-8})$ is shown to be effective in enforcing the rigidity constraint~\cite{BhallaBP2019, Gazzola2011}.

The rigid body velocity $\u_b$ in Eq.~\ref{eqn_bp_discrete} can be expressed as the sum of translational $\U_r$ and rotational $\W_r$ velocities:
\begin{equation}
	\u_b = \U_r + \W_r \times \left(\x-\Xcom\right),
\label{eq_ub_velocity}
\end{equation}
in which $\Xcom$ is the position of the center of mass of the body. In this study, the WEC device is allowed to move only in the heave direction. Hence, $\U_r = (0, 0, \dot{z}(t))$ and $\W_r = \V{0}$. The rigid body velocity is simplified to
\begin{equation}
	\u_b^{n+1,k+1} = \dot{z}^{n+1,k+1} \; \widehat{\V z}.
\end{equation}
The heave velocity $\dot{z}$ resulting from the WSI can be computed using Newton's second law of motion as
\begin{align}
		m {\frac{{\rm d} \dot{z} }{{\rm d} t}} = m \frac{\dot{z}^{n+1,k+1} - \dot{z}^n}{\dt} &=  \mathcal{F}_{\text{hydro}}^{n+1,k} - m g + \text{F}_{\text{PTO}}^{n+1,k+1},  \label{eqn_newton_u} 
\end{align}
in which $\text{F}_{\text{hydro}}$ is the net hydrodynamic force (pressure and viscous) in the heave direction and $m$ is the mass of the cylinder (same as the one used in Eq.~\eqref{eqn_Newton_2nd_law}). The method that was previously described in Sec.~\ref{sec_NLFK_calculation} to compute NLFK forces using the SDF $\psi$ (see Eq.~\eqref{eqn_NLFK_p} can be easily extended to include both pressure and viscous force contributions. Following the SDF approach, the net hydrodynamic force acting on the body is computed as
\begin{align}
	\V{\mathcal{F}}_{\text{hydro}}^{n+1,k}  &= \sum_f    \left(-p^{n+1,k} \n_f + \mu_f \left(\grad_h \, \u^{n+1,k} +  \left(\grad_h \, \u^{n+1,k}\right)^T \right)\cdot \n_f \right) \Delta A_f.   \label{eqn_int_fh}
\end{align}
We remark that whereas Eq.~\eqref{eqn_NLFK_p} is evaluated using a simple and a minimal box region $\mathcal{R}$ surrounding the device and the waves near it, Eq.~\eqref{eqn_int_fh} is evaluated using the actual CFD grid that is distributed across multiple processors. Lastly, the $\text{F}_\text{PTO}$ term of Eq.~\eqref{eqn_newton_u} is computed by the MPC algorithm as discussed in Sec.~\ref{sec_wec_model_mpc_form}. 

\subsection{Power transfer from waves to the PTO} \label{sec_power_transfer_pathway}

Here, we mathematically describe the pathway of power transfer from the sea waves to the PTO system. The relationships derived in this section can also be used to quickly verify the accuracy of the CFD simulations. 

To begin, multiply the dynamical Eq.~\eqref{eqn_newton_u} by the heave velocity $\dot{z}$ and rearrange the terms to obtain: 
\begin{equation}
m\frac{\text{d}}{\text{d}t}\left(\frac{\dot{z}(t)^2}{2}\right) =  \mathcal{F}_{\text{hydro}}(t)\dot{z}(t) - mg\dot{z}(t) + \text{F}_\text{PTO}(t)\dot{z}(t).
\label{eq_KE_eq1}
\end{equation}
Taking the time average of the above equation over one wave period $\mathcal{T}$ and rearranging terms, we get

\begin{equation}
\langle{\mathcal{F}_{\text{hydro}}(t)\dot{z}(t)}\rangle = \langle{m\frac{\dot{z}(t)^2}{2}}\rangle  + \langle{mg\dot{z}(t)}\rangle - \langle{\text{F}_\text{PTO}(t)\dot{z}(t)}\rangle,
\label{eq_KE_eq2}
\end{equation}
in which $\langle (\cdot) \rangle = \int_t^{t + \mathcal{T}} (\cdot) \, \d{\tau}$ represents the time-averaging operator. For regular waves, contributions from the inertial and the gravity terms are zero due to the time periodicity of the heave velocity. Hence, we have:
\begin{equation}
   \langle{\mathcal{F}_{\text{hydro}}(t)\dot{z}(t)}\rangle = - \langle{\text{F}_\text{PTO}(t)\dot{z}(t)}\rangle. \label{eq_KE_eq3}
\end{equation}   
    
The term $\langle{\mathcal{F}_{\text{hydro}}(t)\dot{z}(t)}\rangle$ describes the mechanical work done by the waves to oscillate the converter and the term -$\langle{\text{F}_\text{PTO}(t)\dot{z}(t)}\rangle$ describes the power absorbed by the device.  For irregular waves, the inertial and gravity terms may not equal zero when averaged over an exact wave period. Nevertheless, Eq.~\ref{eq_KE_eq2} remains valid. The power transfer relationships are verified in Sec.~\ref{subsec_power_transfer_verification}.


\section{WSI and MPC solvers}

Secs.~\ref{sec_wec_model_mpc_form}  and~\ref{sec_wsi_eqs} describe methods naturally suited to different types of WSI and MPC solvers. There are two types of WSI solvers that can be derived from  Sec.~\ref{sec_wec_model_mpc_form}: (\textbf{1}) BEM-LFK and  (\textbf{2}) BEM-NLFK. Here, BEM implies a WSI solver that solves Eq.~\eqref{eqn_Cummins_eq}, LFK implies the excitation force is calculated using Eq.~\ref{eq_exc_force2} (or Eq.~\eqref{eq_exc_force3}), and NLFK implies the excitation force is calculated using Eqs.~\eqref{eqn_NLFK_p} and~\eqref{eq_diff_force}.  MPC solvers can also be divided into two types: 
(\textbf{1}) MPC-LFK and  (\textbf{2}) MPC-NLFK, where the excitation force vector $\Deltav$ is computed linearly and non-linearly, respectively. Lastly, based upon Sec.~\ref{sec_wsi_eqs}, we have a multiphase IB/CFD solver that solves Eqs.~\eqref{eqn_momentum}-\eqref{eqn_continuity}. Table~\ref{tab_solvers} shows six possible WSI/MPC combinations.  Note that it is computationally unfeasible (if not impossible) to implement MPC using a CFD-based solver. Moreover, results of Sec.~\ref{subsec_LFK_NLFK_CFD_MPC_comparison} suggest that a higher fidelity hydrodynamical model within MPC does not necessarily improve accuracy. 
\begin{table}[]
 \centering
 \caption{Various WSI/MPC solver combinations considered in this work.}
 \rowcolors{3}{}{gray!10}
\begin{tabular}{ccc}
\toprule
 & Solver                      & MPC             \\
 \midrule
1   &   BEM-LFK           & LFK               \\
2   &   BEM-LFK           & NLFK             \\
3   &   BEM-NLFK         & LFK                \\
4   &   BEM-NLFK         & NLFK             \\
5   &   CFD                    & LFK                \\
6    &  CFD                    & NLFK             \\
 \bottomrule
\end{tabular}
\label{tab_solvers}
\end{table}

\section{Software implementation} \label{sec_software}

\subsection{CFD solver} \label{sec_ibamr_software}
The FD/BP and the numerical wave tank methods presented here are implemented within the IBAMR library~\cite{IBAMR-web-page}, which is an open-source C++ simulation software focused on immersed boundary methods with adaptive mesh refinement; the code is publicly hosted at \url{https://github.com/IBAMR/IBAMR}. The C++ application/driver code (\texttt{main.cpp}) and the MATLAB MPC routines link directly against the compiled IBAMR library and are publicly hosted in a separate GitHub repository at \url{https://github.com/IBAMR/cfd-mpc-wecs}. IBAMR relies on SAMRAI~\cite{HornungKohn02, samrai-web-page} for Cartesian grid management and the AMR framework. Linear and nonlinear solver support in IBAMR is provided by the PETSc library~\cite{petsc-efficient, petsc-user-ref, petsc-web-page}. All of the CFD cases in the present work made use of distributed-memory parallelism using the Message Passing Interface (MPI) library.  Simulations were carried out on both the XSEDE Comet cluster at the San Diego Supercomputer Center (SDSC) and the Fermi cluster at San Diego State University (SDSU). 

\subsection{Communication layer between the CFD and MPC solvers} 
\label{sec_cfd_mpc_interface}

\begin{figure}[]
   \centering
   \includegraphics[scale= 0.45]{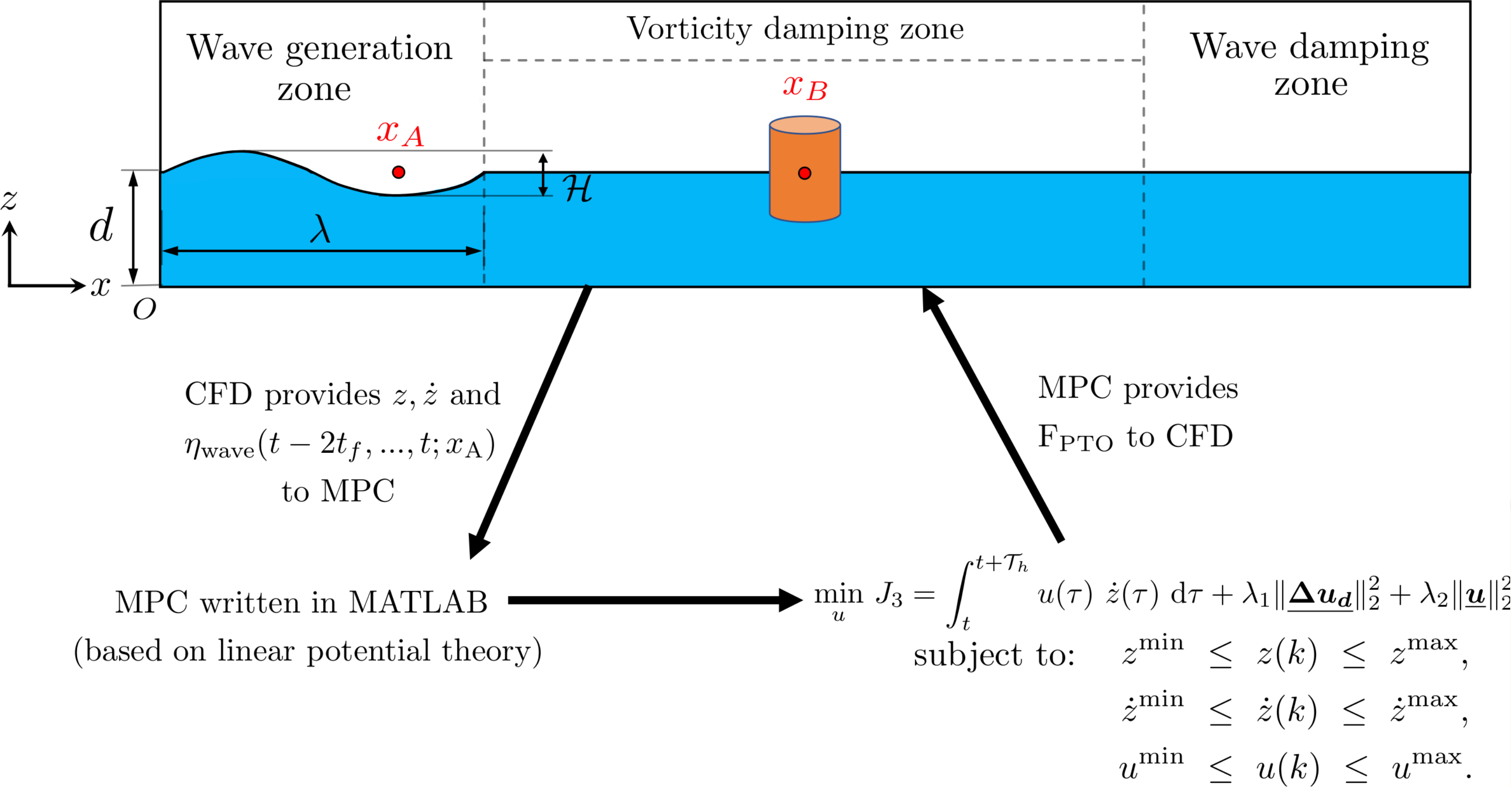}
   \caption{Schematic representation of the dynamic interaction between the MPC algorithm and multiphase IB solver.}
   \label{fig_CFD_integrated_MPC}
\end{figure}

In this section, we present the custom communication layer between the CFD and MPC solvers. The ``glue code" is written using PETSc~\cite{petsc-web-page}, which provides a high-level communication channel between IBAMR~\cite{IBAMR-web-page} and MATLAB~\cite{MATLAB2019b}.  As discussed in Sec.~\ref{sec_wec_model_mpc_form}, MPC requires quadratic programming (QP) and autoregressive models (AR). Although there are several compiled language implementations of QP (e.g., QuadProg++~\cite{QuadProgpp-web-page}) and AR (e.g., Cronos~\cite{Cronos-web-page}) techniques, we implement the MPC algorithm in MATLAB, which has built-in support for QP and AR techniques. MATLAB is probably the most widely used programming environment for dynamical systems modeling and control in academia and industry, so our current implementation can easily be adapted to integrate other optimal control strategies for WECs into a different CFD code of choice, e.g., OpenFOAM.     

In the following, we describe the interaction between the CFD and MPC solver codes as a three-part algorithm. Fig.~\ref{fig_CFD_integrated_MPC} shows this interaction pictorially. Communication between the CFD and MPC codes is handled by the PETScMatlabEngine object provided by the PETSc library. Details on the PETSc functions and objects can be found in its user manual~\cite{petsc-user-ref}.  

\begin{enumerate}
 

\item {\bf{Accessing the MATLAB workspace}}: Algorithm~\ref{alg_algorithm_1} is called towards the beginning of the driver code to create the PETScMatlabEngine object `mengine' on MPI (Message Passing Interface) rank 0. 
 This is achieved by calling the PETSc function \texttt{PetscMatlabEngineCreate()} on line~\ref{algo_1_line_2} of the algorithm, in which  `PETSC\_COMM\_SELF' is the MPI communicator containing the single MPI rank 0.
 Next, the MATLAB workspace is cleared for any data already present and the path to the `MPC\_matlab\_code' directory is added to MATLAB's standard search path. The directory `MPC\_matlab\_code' contains all the MPC code scripts and related functions. The PETSc function to achieve this is called on line \ref{algo_1_line_3} of Algorithm \ref{alg_algorithm_1}. Then, various wave ($\cH$, $\cT_p$, $\omega$, $\kappa$, $d$) and device parameters ($m$, $R_\text{cyl}$, $\text{L}_\text{cyl}$) are loaded into the workspace by calling the PETSc function \texttt{PetscMatlabEngineEvaluate()} on line~\ref{algo_1_line_4}, wherein a MATLAB variable `var' is created with the numerical value of var\_value. Next, the BEM data is read and loaded into the workspace by executing the MATLAB script `load\_mpc\_parameters.m'.  This includes the added mass of the cylinder $m_\infty$ and the impulse response functions $\text{K}_e(t)$ and $\text{K}_d(t)$. The script also sets various MPC parameters ($\Delta t_p$, $\cT_h$, $N_p$, $n_r$, $t_f$), device constraints ($z_\text{min/max}$, $\dot{z}_\text{min/max}$, $u_\text{min/max}$), wave type (regular or irregular), and the method of wave excitation force calculation (LFK or NLFK). The coefficients of the quadratic penalty terms $\lambda_1$ and $\lambda_2$ and the MPC solver options (maximum iterations,  solver tolerance, etc.) are also set by the same script `load\_mpc\_parameters.m'. Since the CFD solver sends the device and wave elevation data to the MPC code, it needs to know the MPC time step size $\Delta t_p$ and the next time to synchronize data with the controller $t_\text{next-sync}$. The values from the MATLAB workspace are obtained by calling the function \texttt{PetscMatlabEngineGetArray()}. Finally, the remaining CFD parameters and variable values (${\Delta t}$, ${\bf{\text{X}}}_\text{com}$, $x_\text{B}$, and $x_\text{A}$) that could not be added to the workspace earlier (on line~\ref{algo_1_line_4}) are loaded to the workspace on line~\ref{algo_1_line_7}.

\begin{algorithm}
\caption{Creating and initializing the MATLAB workspace.}
\label{alg_algorithm_1}
 \If{(\text{MPI}\_\text{rank} == 0)}{
 	\tt{PetscMatlabEngineCreate}(PETSC\_COMM\_SELF, NULL, \&(mengine)); \tcp{Create a MATLAB engine on MPI rank 0.} \label{algo_1_line_2}
	
	\tt{PetscMatlabEngineEvaluate}(mengine, ``clc;  clear all;  close all;  addpath(`./MPC\_matlab\_code')"); \tcp{Execute MATLAB commands and add the MPC code directory path to the standard search path.} \label{algo_1_line_3}
	 
 	\tt{PetscMatlabEngineEvaluate}(mengine,``var = \%f", var\_value); \tcp{Load the wave and device parameters into the MATLAB workspace.} \label{algo_1_line_4}
	
	\tt{PetscMatlabEngineEvaluate}(mengine, ``load\_mpc\_parameters"); \tcp{Execute the script to read and load the BEM data and MPC parameters into the MATLAB workspace.}
 
	\tt{PetscMatlabEngineGetArray}(mengine, ...); \tcp{Load the values of the MATLAB variables into the CFD code.} 
	
	$\ldots \ldots \ldots$ \tcp{Code to do CFD related setup and calculations.} 
	
	\tt{PetscMatlabEngineEvaluate}(mengine, ...); \tcp{Load the remaining CFD variables into the workspace that were not available/calculated earlier.} \label{algo_1_line_7}
}
\end{algorithm}

\item {\bf{The main time loop}}: Algorithm~\ref{alg_algorithm_2} describes the time-loop interaction between the CFD and MPC solvers. First, the MPI rank 0 updates the MATLAB workspace with the CFD solver time $t_\text{CFD} = t^{n+1}$ and the device displacement and velocity data, as shown on line \ref{algo_2_line_4} of the algorithm. Next, the algorithm checks if the CFD solver time is greater than or equal to the controller synchronization time $t_\text{next-sync}$. If the statement evaluates to true, then a new set of MPC matrices $\V{\cP}$, $\cJu$, and $\cJv$ are calculated and the radiation damping vector ${\bm{x_r}}$ is advanced in time using the MATLAB scripts `calculate\_mpc\_matrices.m' and `calculate\_radiation\_damping\_xr.m', respectively. To enable the calculation of wave excitation forces over a prediction horizon of $\cT_h$,  the CFD solver sends the past up-wave surface elevation data (from the NWT) to the MATLAB workspace. Using the updated matrices, vectors, and FK forces, the MPC solver predicts the optimal control sequence for the entire prediction horizon on line~\ref{algo_2_line_mpc}. The first signal of the optimal control sequence is sent to the CFD solver, which is then interpolated to time   $t_\text{CFD}$ using Eq.~\ref{eq_uc_interpolation}. Note that since the CFD solver time step size $\Delta t$ is typically smaller than the MPC solver time step size $\Delta_p$, line~\ref{algo_2_line_time_interp} of Algorithm~\ref{alg_algorithm_2} ensures that $\text{F}_\text{PTO}$ is computed at the correct time level in the case the if block is not executed.  Lastly, both $t_\text{next-sync}$ and $t_\text{CFD}$ are updated and the time level is moved to $n+2$.

\begin{algorithm}
\caption{Time-loop interaction between the CFD and MPC solvers.}
\label{alg_algorithm_2}
Initialize the MATLAB workspace and load the BEM data and MPC parameters. \tcp{See Algorithm~\ref{alg_algorithm_1}.}
\While{($t_\text{CFD}  \leq t_\text{end}$)}{

     \If{ (MPI\_rank == 0)}{
   
          \tt{PetscMatlabEngineEvaluate} (mengine, ``$t_\text{CFD} = \%f$; $\Delta t = \%f$; $z = \%f$; $\dot{z} = \%f$;" , $t_\text{CFD}$, ${\Delta t}$, $z$, $\dot{z}$); \tcp{Send the latest CFD and device data to MATLAB workspace.} \label{algo_2_line_4}
       
  \If{($t_\text{CFD} \geq t_\text{next-sync}$)}{
  
          \tt{PetscMatlabEngineEvaluate}(mengine, ``calculate\_mpc\_matrices; calculate\_radiation\_damping\_xr;"); \tcp{Execute the MATLAB scripts to update the discrete-time dynamical matrices.}
          
           \BlankLine
           \BlankLine
           
          \tcp{Send the past up-wave surface elevation data to the MATLAB workspace.}
          \tt{PetscMatlabEnginePutArray}(mengine, $t_\text{past}$.size(), 1, \&($t_\text{past}$[0]),``$t_\text{past}$");           
          
          \tt{PetscMatlabEnginePutArray}(mengine, $\eta_\text{A}$.size(), 1, \&($\eta_\text{A}$[0]),``$\eta_{\text{A}_\text{past}}$");    
           
          \BlankLine
           \BlankLine
           
          \tt{PetscMatlabEngineEvaluate}(mengine, ``calculate\_control\_force;");  \tcp{Compute the optimal control sequence using Algorithm~\ref{alg_algorithm_3}.} \label{algo_2_line_mpc}
          
          \tt{PetscMatlabEngineGetArray}(mengine, 1, 1, \&$({u})$,``$ u$"); \tcp{Get the first signal of the optimal control sequence from MPC for the CFD solver.}

   	    $t_\text{next-sync} \gets t_\text{next-sync}$ + $\Delta t_p$ \tcp{Update the synchronization time.}
   }
      Interpolate $\text{F}_\text{PTO} \gets (m+ m_\infty)  u$ to $ t_\text{CFD}$ using Eq.~\ref{eq_uc_interpolation}. \label{algo_2_line_time_interp}
   }
   \tt{MPI\_Bcast}($\text{F}_\text{PTO}$); \tcp{Broadcast the value of the PTO force to all processors.} 
   Solve the FSI problem using the multiphase IB solver. 
   
   $t_{\rm CFD} \gets  t_{\rm CFD} + \Delta t$
  }
\end{algorithm}

\item {\bf{The MPC routine}}: Algorithm~\ref{alg_algorithm_3} describes the AR predictions and the LFK and NLFK force calculations required by the MPC to compute an optimal control force sequence. This algorithm is executed by the MATLAB script `calculate\_control\_force.m'. First, line \ref{algo_3_line_1} calculates the discrete time instants over the prediction horizon at a uniform interval ${\Delta t}_p$. Next, the algorithm checks if AR predictions are to be used or not. If the value of the variable $AR\_start\_time$ is set to a large number (larger than the simulation end time), then the if condition on line \ref{algo_3_line_2} always evaluates to true. In this case, the algorithm computes the LFK or the NLFK force based on the analytical expression of the wave elevation. In the case $AR\_start\_time$ is set to the controller/MPC start time, the if condition on line \ref{algo_3_line_2} evaluates to false when $t_\text{CFD}$ becomes equal or larger than the MPC start time. In that case, the wave elevation data over the prediction horizon is calculated using AR predictions; see line \ref{algo_3_line_16}.  The wave excitation force is computed using the convolution integral given by Eq.~\ref{eq_exc_force3} based on the AR predictions of wave elevation. Next, other necessary terms like the viscous force, the state vector $\X_{\boldsymbol{d}}$, etc., are calculated on lines~\ref{algo_3_line_19}-\ref{algo_3_line_21}. Finally, the QP functionality of MATLAB is used to compute the optimal control force sequence $\Deltau$ considering the necessary device constraints and penalty terms.

\begin{algorithm}
\caption{MATLAB-based MPC routine.}
\label{alg_algorithm_3}

time\_horizon = $t_\text{CFD}$ + (0:$N_p$) $\times {\Delta t}_p$;  \tcp{Calculate the discrete time horizon.} \label{algo_3_line_1}

\eIf{($t_\text{CFD} \leq AR\_start\_time$)} {\label{algo_3_line_2}

	\eIf{(strcmp($\text{F}_{\text{exc-type}}$, `LINEAR\_FK'))}
	{
	
        $\text{F}_\text{exc}$ $\gets$ \tt{calculate\_excitation\_force}(time\_horizon);   \tcp{Calculate $\text{F}_\text{exc}$ using Eq.~\ref{eq_exc_force2}. $\eta_\text{wave}$ is calculated using Eq.~\ref{eq_first_order_wave_elev} for regular  and Eq.~\ref{eq_irregular_elevation} for irregular waves.}
		
	}
	{
	
		$z_\text{predicted}$ $\gets$ \tt{$\text{AR\_prediction}$}($z_\text{past}$, $t_\text{past}$, $N_p$, $\text{AR\_order}$, ${\Delta t}_p$); \tcp{Predict the device displacement using the AR model based on past data.}
		
	\BlankLine
		
		$\text{F}_D$ $\gets$ \tt{calculate\_diffraction\_force}(time\_horizon); \tcp{Calculate the wave diffraction force using Eq.~\ref{eq_diff_force}. $\eta_\text{wave}$ is calculated using Eq.~\ref{eq_first_order_wave_elev} for regular and Eq.~\ref{eq_irregular_elevation} for irregular waves.}
		
	 \BlankLine
		           
		\For{$(m = 1$ $\KwTo$ $(N_p+1))$}
		{   
			$\psi$ $\gets$ \tt{calculate\_level\_set\_for\_cylinder}($z_\text{predicted}(m)$, $R_\text{cyl}$, $L_\text{cyl}$); \tcp{Compute the level set for the cylinder on a static grid region $\mathcal{R}$.}
			
            		$\sigma$ $\gets$ \tt{calculate\_level\_set\_for\_wave}($\text{time\_horizon}(m)$); \tcp{Compute the level set for the undulatory air-wave interface.}
		
			$\text{F}_I $(m) $\gets$ \tt{calculate\_NLFK\_force}($\psi$, $\sigma$, $\text{time\_horizon}(m)$); \tcp{Calculate the incident wave force using Eq.~\eqref{eqn_NLFK_p}.}
		
	      }
	      $\text{F}_\text{exc} \gets \text{F}_D + \text{F}_I$  \tcp{Compute $\text{F}_\text{exc}$ for  $ \forall t \in \text{time\_horizon}$.}
    }
}
{
	$\eta_{\text{A}_\text{predicted}}$ $\gets$ \tt{AR\_prediction}($\eta_{\text{A}_\text{past}}$, $t_\text{past}$, $N_p$, $\text{AR}\_\text{order}$, ${\Delta t}_p$);\label{algo_3_line_16}
	
	Calculate the future $N_p$ values of $\text{F}_\text{exc}$ for  $ \forall t \in \text{time\_horizon}$ using Eq~\ref{eq_exc_force3}.
	
}

Calculate the first term in the linearized form of the viscous force $\text{F}_v$ given in Eq.~\ref{eqn_linear_Fv}. \label{algo_3_line_19}

Calculate  the vectors $\bf{\text{X}_d}$ and $\Deltav$. 

Calculate $\cJu^T Q \cJu$ and $\cJu^T Q (\vcP \Xd + \cJv \Deltav)$ terms of Eq.~\ref{eq_J1_expn1}. \label{algo_3_line_21}
 
Minimize the cost function $J_3$ (with constraints) using the QP functionality of MATLAB to obtain the optimal control sequence $\Deltau$.
 
\end{algorithm}

\end{enumerate}

%% file: Validation_and_motivation.tex

While MPC has been used in the process industries (chemical plants and oil refineries) since the 1980s, its formulation for the wave energy conversion application was first suggested by Gieske~\cite{Gieske2007} in 2007. The study involved optimizing the control of the Archimedes wave swing (AWS) device modeled as a second-order linear system. In 2010, Cretel et al.~\cite{Cretel2010} implemented a zero-order hold (ZOH) method based MPC for a half-submerged heaving vertical cylinder A later study published by Cretel et al.~\cite{Cretel2011} suggested using the first-order hold (FOH) method, which yielded better results than ZOH-based MPC. The BEM-LFK solver was used in all the aforementioned studies. 

\begin{figure}[]
  \centering
  \subfigure[Heave velocity]{
  	\includegraphics[scale= 0.35]{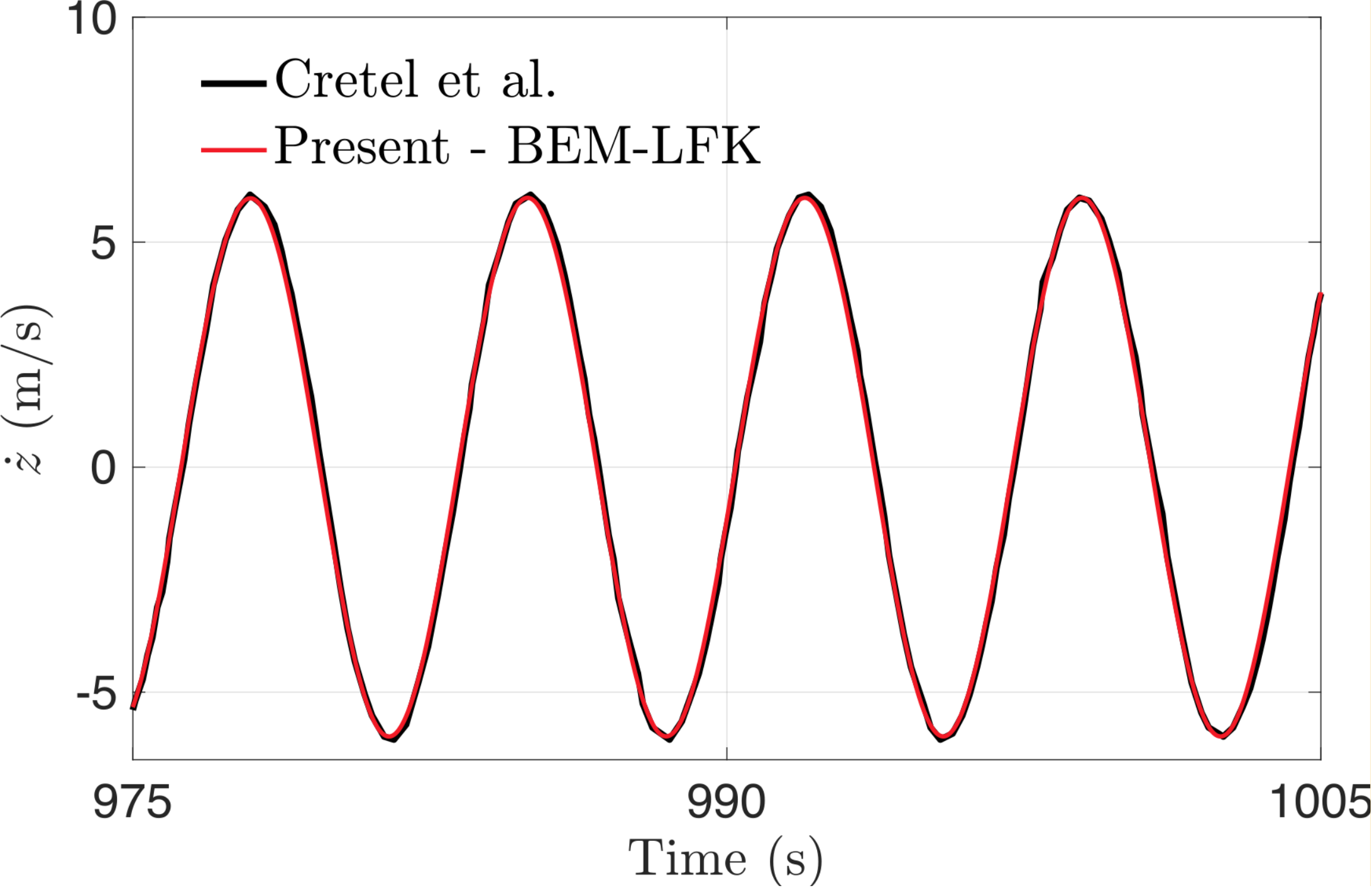}
	\label{fig_velocity_Cretel_case}
  }
  \subfigure[\REVIEW{Wave excitation force}]{
  	\includegraphics[scale = 0.34]{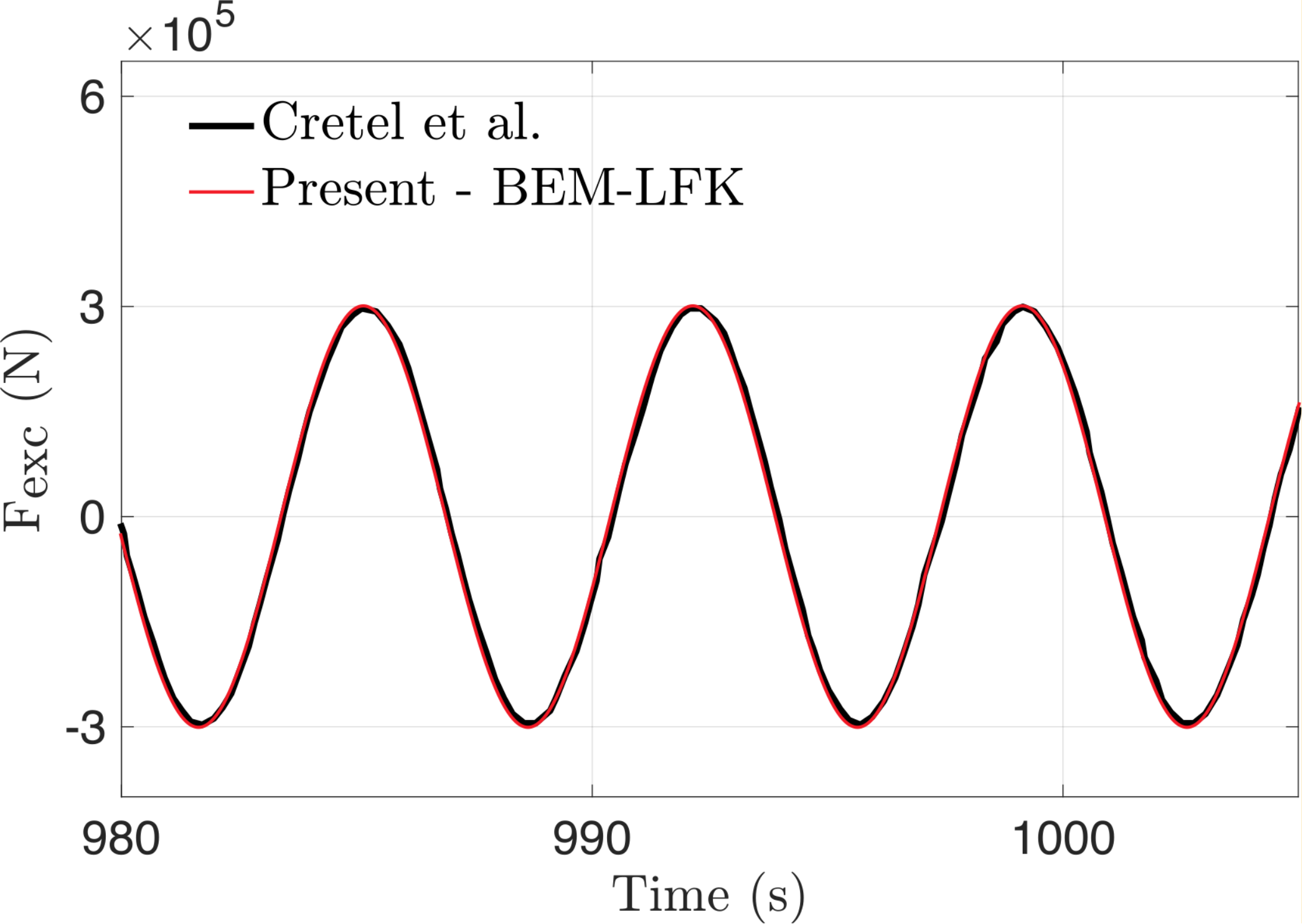}
	\label{fig_Fexc_Cretel_case}
  }
  \caption{Temporal evolution of~\subref{fig_velocity_Cretel_case} the heave velocity and \subref{fig_Fexc_Cretel_case} wave excitation forces acting on the heaving vertical cylinder. Results are compared against Cretel et al.~\cite{Cretel2011} for first-order regular waves of height $\cH = 2$ m and time period $\cT = 7$ s. The MPC parameters are  $\Delta t_p = 0.1$ s, $\cT_h = 6$ s, $\lambda_1 = 2$ s, and $\lambda_2 = 0$ s. }
  \label{fig_validation_with_Cretel_case}
\end{figure}

In order to validate our (FOH-based) BEM-LFK solver and MPC implementations, we consider the same half-submerged vertical cylinder case as Cretel et al.~\cite{Cretel2011}. The cylinder has a radius of $R_{\rm cyl} = 5$ m and an upright length of $L_{\rm cyl} = 16$ m. Regular waves of height $\cH = 2$ m and time period $\cT = 7$ s are used. This corresponds to a small have height case and the BEM solvers are expected to be accurate in this wave regime. The BEM parameters $m_\infty$ and $\text{K}_e(t)$ are obtained using ANSYS AQWA by performing frequency domain WSI simulations. The MPC parameters are taken to be  $\Delta t_p = 0.1$ s, $N_p = 60$ (and consequently a prediction horizon of $\cT_h = 6$ s), $\lambda_1 = 2$ s, and $\lambda_2 = 0$ s.  There are no device constraints included, and $J_2$ cost function is used in the MPC to match Cretel et al.'s setup. Fig.~\ref{fig_validation_with_Cretel_case} shows the temporal evolution of the heave velocity and excitation forces and compares it against the steady-state results of Cretel et al.~\cite{Cretel2011}. Both studies agree very well. The steady-state time-averaged power $\widebar{P}_\text{PTO}$ absorbed by the device is 353.5301 kW, which is also close to the value of 395.08 kW reported in~\cite{Cretel2011}. We conclude from these results that our BEM-LFK  solver and MPC implementations are correct.

\begin{figure}[]
  \centering
  \subfigure[Wave excitation force]{
  	\includegraphics[scale= 0.35]{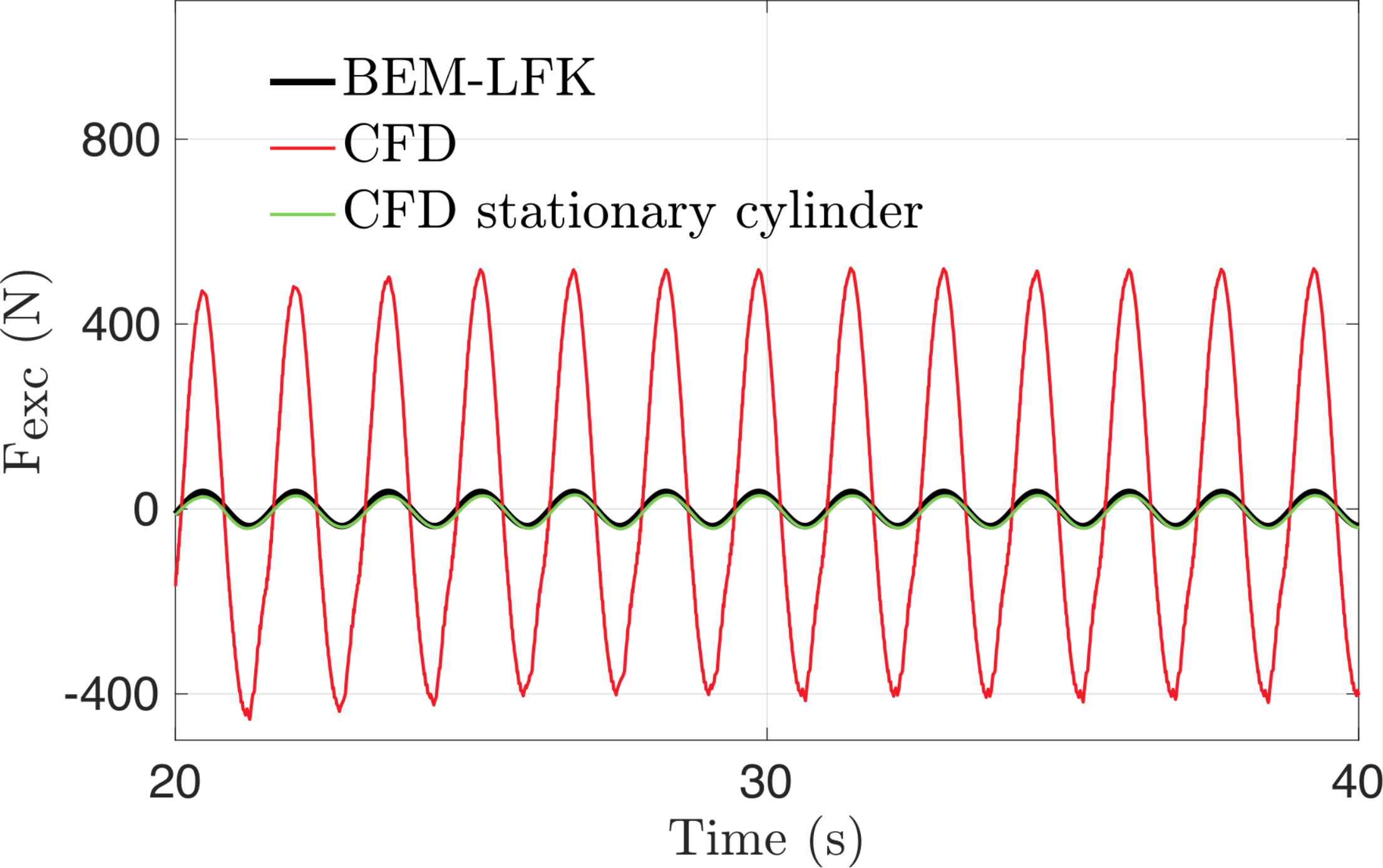}
	\label{fig_Fexc_Cretel_case_CFD}
  }
  \subfigure[Control force]{
  	\includegraphics[scale = 0.35]{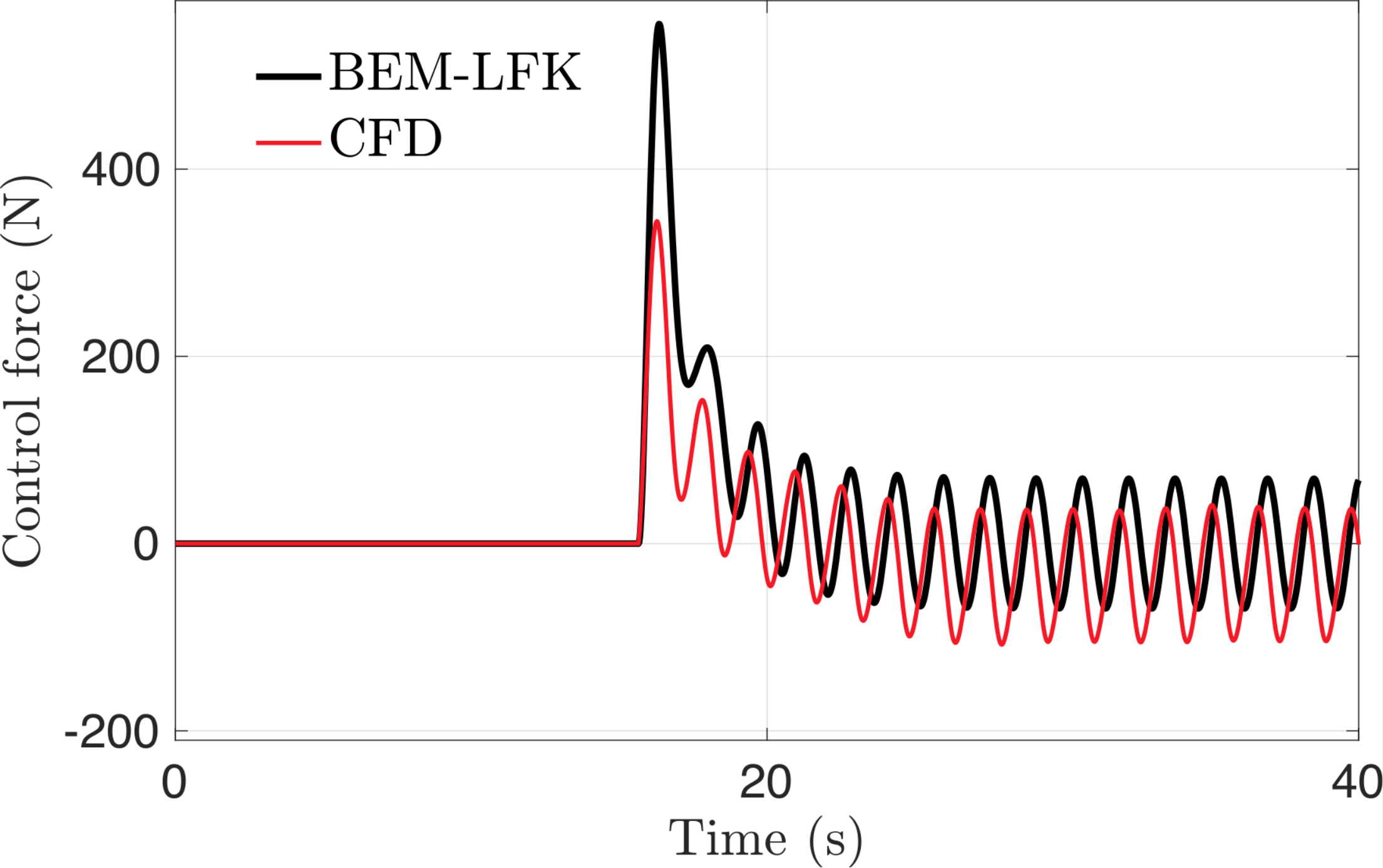}
	\label{fig_Fpto_Cretel_case_CFD}
  }  
  \subfigure[Instantaneous and steady-state powers]{
  	\includegraphics[scale = 0.35]{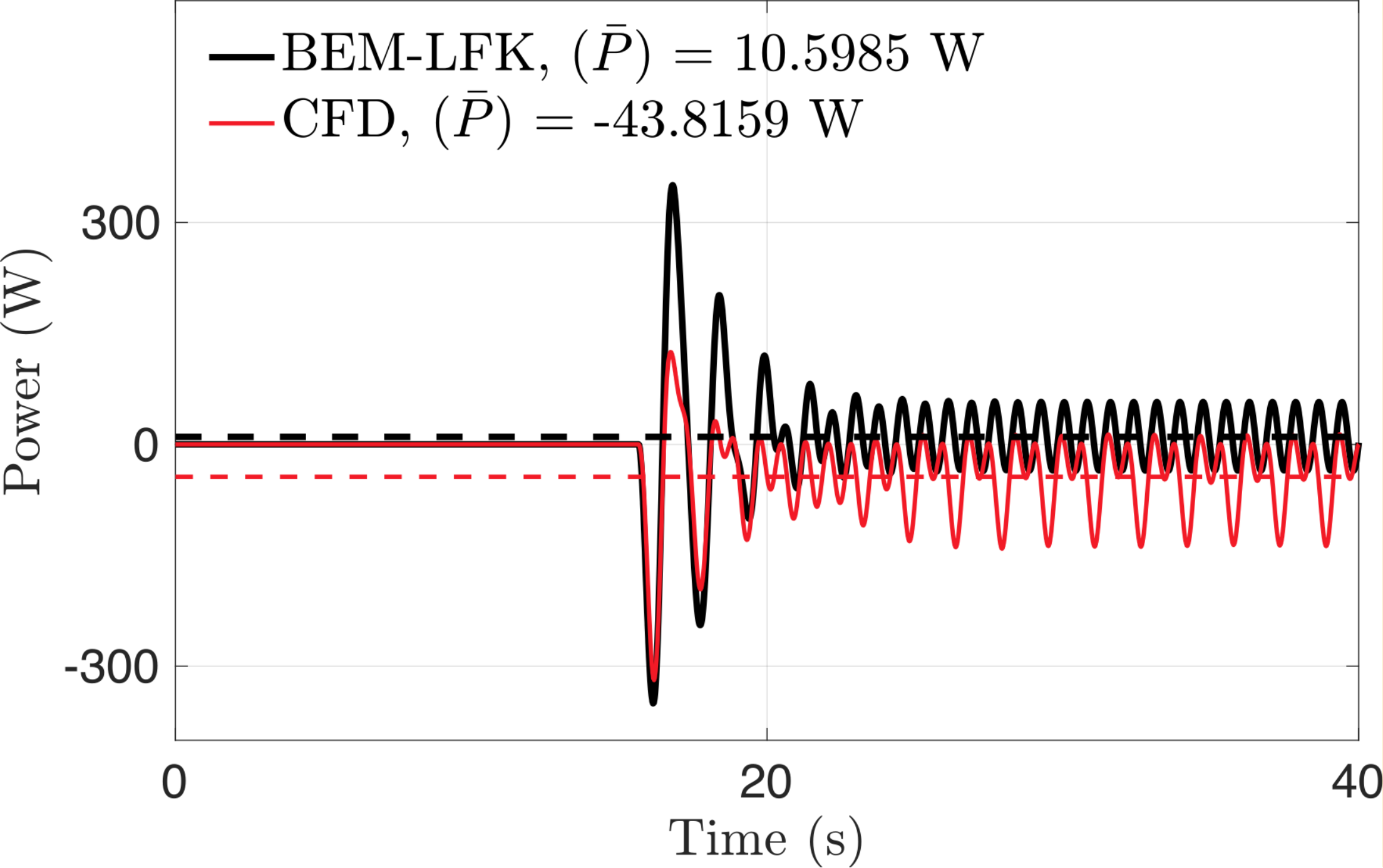}
	\label{fig_power_Cretel_case_CFD}
  }
  \caption{Temporal evolution of \subref{fig_Fexc_Cretel_case_CFD} wave excitation force, \subref{fig_Fpto_Cretel_case_CFD} control force, and \subref{fig_power_Cretel_case_CFD} instantaneous power absorbed for vertical cylinder WEC device heaving on the sea surface for first order regular wave of $\cH = 0.1$ m and $\cT = 1.5652$ s. The MPC parameters are  $N_p = 60$,  $\Delta t_p = 0.0223$ s, $\cT_h = 1.3415$ s, $\lambda_1 = 2$ s, and $\lambda_2 = 0$ s.}
  \label{fig_Cretel_case_CFD}
\end{figure}

Next, we compare the predictions of the BEM and CFD solvers for a 1:20 scaled-down version of the device (using Froude scaling). We do this to reduce the computational cost of CFD simulations, as the full-scale WEC device requires a larger computational domain and a higher mesh resolution to resolve the high Reynolds number flow.  For further details on the Froude scaling of the device and wave characteristics ($\cH$ and $\cT$), the readers are referred to Khedkar et al.~\cite{Khedkar2021}. The size of the domain, grid resolution, and time step size of the CFD simulation are determined by the spatial-temporal simulation performed in the next section~\ref{sec_spatiotemporal_tests}. Both solvers use regular waves of height $\cH = 0.1$ m and time period $\cT = 1.5652$ s and the MPC parameters are  $N_p = 60$,  $\Delta t_p = 0.0223$ s, $\cT_h = 1.3415$ s, $\lambda_1 = 2$ s, and $\lambda_2 = 0$ s. Fig.~\ref{fig_Cretel_case_CFD} compares the predictions of the two solvers. Fig.~\ref{fig_Fexc_Cretel_case_CFD} clearly shows that the wave excitation force of the CFD simulation is much larger than that of the BEM-LFK simulation. A similar discrepancy is observed using the BEM-NLFK solver whose results are closer to the BEM-LFK solver (data not shown for brevity). Since this is a low wave amplitude case, we attribute the discrepancy between the CFD and BEM solvers to the non-linear WSI  caused by the controller. To confirm this hypothesis  an additional CFD simulation is conducted, in which hydrodynamic loads are calculated on a vertical cylinder that has the same dimensions, but is fixed at equilibrium. The case is represented by the green curve in Fig.~\ref{fig_Fexc_Cretel_case_CFD}.  It is clear that both solvers (CFD and BEM-LFK) estimate the same hydrodynamic force on the stationary cylinder. Furthermore, an uncontrolled dynamics case is simulated in the next Sec.~\ref{sec_spatiotemporal_tests}, where the BEM and CFD solvers' predictions match for the same wave conditions of this section. These additional tests confirm our hypothesis that even in calm sea conditions, the controller can cause a mismatch between the solvers' predictions.

Figs.~\ref{fig_Fpto_Cretel_case_CFD} and~\ref{fig_power_Cretel_case_CFD} compare the MPC control force and the instantaneous power absorbed by the heaving device (respectively) using the BEM-LFK and CFD solvers. The comparison shows that, while the BEM-LFK solver estimates the power produced by the device at 10.6 W~\footnote{Using Froude scaling, this value corresponds to $10.6 \times (20)^{\frac{7}{2}} = 379.2$ kW for the full-scale device, which is close what is predicted earlier in this section.}  during its steady-state operation, the CFD solver predicts a large withdrawal of power from the grid (-43.8 W). The  power results of the BEM-NLFK solver are close to those of the BEM-LFK solver (data not presented). There was only a small effect of changing the penalty term $\lambda_1$ on the power results of the two solvers. The results presented in this section, therefore, suggest that the BEM solvers may not always provide a reliable estimate of the power production capability of the WEC device under certain operating/controlled conditions. Furthermore, it can also be appreciated that it is necessary to include the $\lambda_2$ term in the objective function to eliminate or mitigate the large negative powers. This section summarizes the motivation for the work conducted here, which is to investigate why the performance of various types of solvers differs and to compare them under different operating conditions. Due to the reasons noted above, we compare the performance of various solvers using $J_3$ instead of $J_2$ in the results and discussion section~\ref{sec_results_and_discussion}. The case of this section is also repeated (Case 2 of Table~\ref{tab_wave_cases}) using the $J_3$ cost function because it is more suitable for the model predictive control of WECs. 

Before proceeding to the main results Sec.~\ref{sec_results_and_discussion}, we first perform a grid convergence study for the CFD solver in the next section. 

%% file: Spatial_temporal_resolution_tests.tex

In this section, we perform a grid convergence study on the heaving WEC device using the CFD solver. Convergence tests are performed without the MPC. In WSI simulations, both regular and irregular waves are considered. The spatial resolution study is based on three spatial resolutions listed in Table~\ref{tab_grid_convergence_study}, while the temporal resolution study is based on three values of the time step size $\Delta t$ for irregular waves. In all tests, the maximum Courant-Friedrichs-Levy (CFL) number is less than or equal to 0.5. Simulations are performed on locally refined grids in order to reduce computational costs.

\begin{figure}
   \centering
   \subfigure[Locally refined Cartesian mesh]{
   	\includegraphics[scale= 0.185]{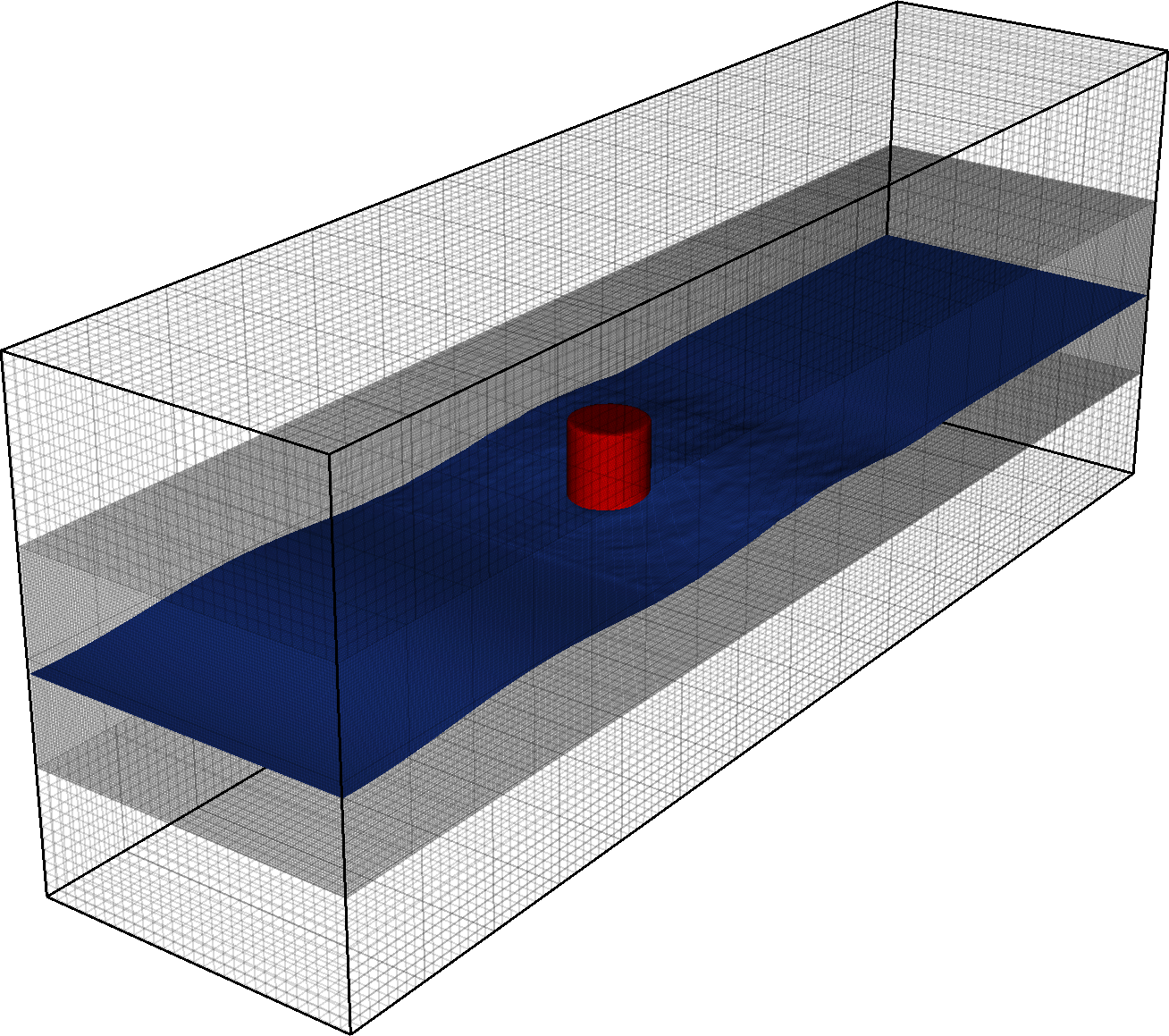}
	\label{fig_3d_vcyl_mesh}
   }
   \subfigure[Regular waves]{
   	\includegraphics[scale = 0.168]{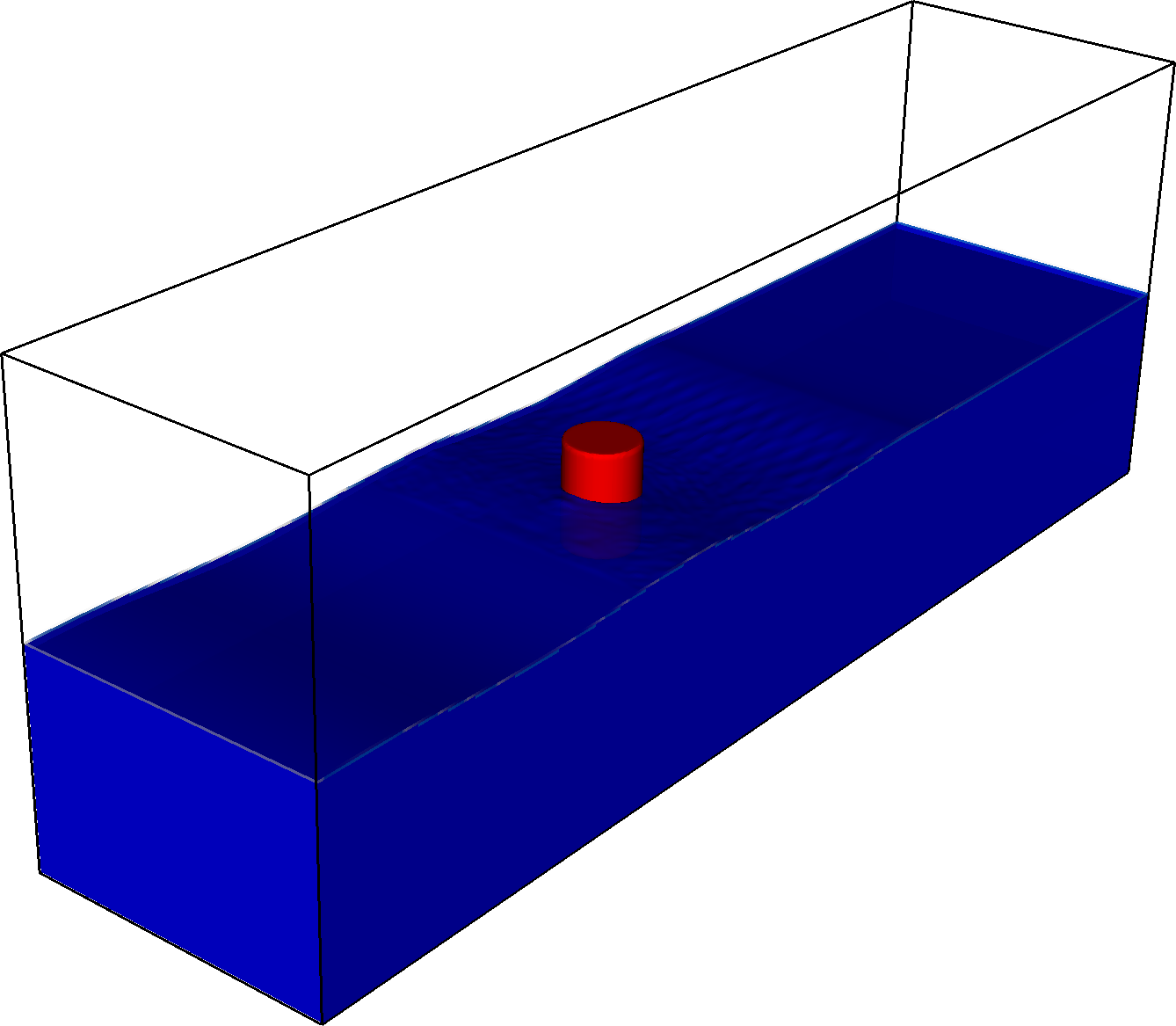}
	\label{fig_3d_vcyl_regular}
   }
   \subfigure[Irregular waves]{
   	\includegraphics[scale= 0.185]{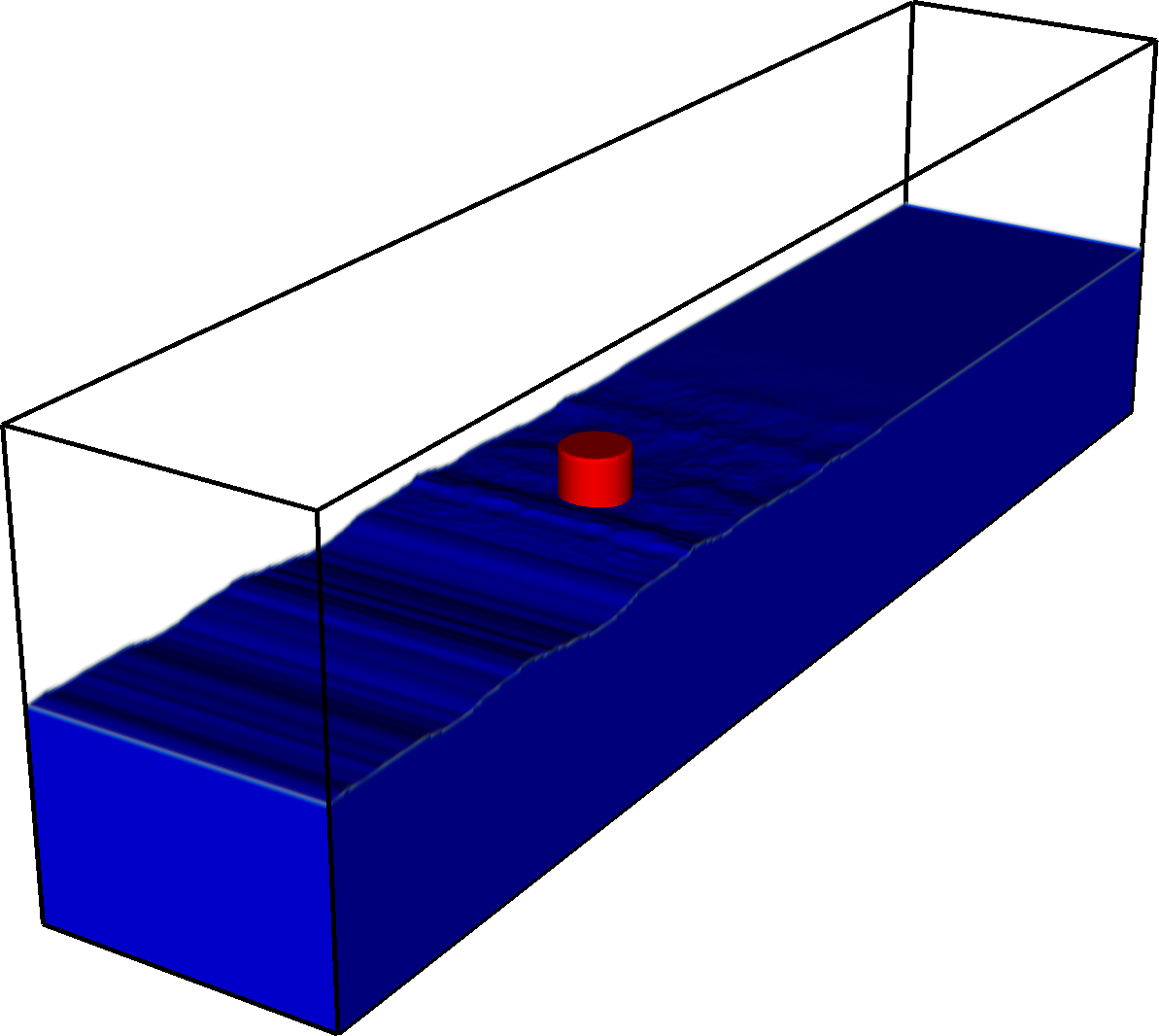}
	\label{fig_3d_vcyl_irregular}
   }
   \caption{\subref{fig_3d_vcyl_mesh} Locally refined Cartesian mesh with two levels of mesh refinement for the 3D NWT. Representative WSI of the 3D WEC model at $t$ = 37.5 s:~\subref{fig_3d_vcyl_regular} for regular waves and~\subref{fig_3d_vcyl_irregular} for irregular waves.}
   \label{fig_3d_regular_irregular_medium}
\end{figure}

The computational domain for regular waves is $\Omega$ = [0, 3.145$\lambda$] $\times$ [0, 12$R_\text{cyl}$] $\times$ [0, 2.2$d$], whereas for irregular waves it is $\Omega$ = [0, 3.176$\lambda$] $\times$ [0, 12$R_\text{cyl}$] $\times$ [0, 2.2$d$].  The domain size is large enough to eliminate boundary effects. This is based on our previous experience modeling WSI of WEC devices~\cite{Dafnakis2020,Khedkar2021}. The origin of the domain is located at the bottom left corner; see Fig.~\ref{fig_NWT_schematic}. The initial center of mass of the device is located at $\X_{\rm com}$ = ($\lambda + 5 R_\text{cyl}$, $6 R_\text{cyl}$, $d$).  $R_\text{cyl} = 0.25$ m and $L_\text{cyl} = 0.8$ m,  which is a 1:20 scaled-down version of the one presented in~\cite{Cretel2011}. The cylinder is half-submerged in its equilibrium position.The quiescent water depth is $d$ = 2 m, acceleration due to gravity is g = 9.81 m/s$^2$ (directed in the negative z-direction), density of water is $\rho_w$ = 1025 kg/m$^3$, density of air is $\rho_a = 1.225$ kg/m$^3$, viscosity of water is $\mu_w = 10^{-3}$ Pa$\cdot$s, and viscosity of air is $\mu_a = 1.8 \times 10^{-5}$ Pa$\cdot$s. At this scale, surface tension at the air-water interface has no effect on WEC dynamics and is therefore ignored. All of the CFD simulations in this work, including those of the previous Sec.~\ref{sec_validation_and_motivation} use the same material properties and computational domain setup.   Fig.~\ref{fig_3d_regular_irregular_medium} shows the grid layout and typical wave-structure interactions of the device in the NWT. 

\subsection{Grid convergence study} \label{subsec_grid_convergence}

Here, a grid convergence study is performed to determine the optimal mesh spacing for the CFD simulations.  Three grid sizes  are used to conduct the grid convergence test: coarse, medium, and fine (see also Table~\ref{tab_grid_convergence_study}). The coarse mesh size corresponds to 5 cells per radius of the cylinder (CPR), the medium mesh size is 10 CPR, and the fine mesh size is 15 CPR. The computational mesh consists of a hierarchy of $\ell$ grid levels. The coarsest grid level is discretized into $N_x \times N_y \times N_z$ grid cells and covers the entire computational domain $\Omega$. A sub-region of the coarsest level is then locally refined $(\ell - 1)$ times by an integer refinement ratio of $n_\text{ref}$. The local refining is done in such a way that the device and the air-water interface remains embedded on the finest grid level throughout the simulation. The grid spacing on the finest grid level is calculated as: $\Delta x = \Delta x_0 / n_\text{ref}^{\ell-1}$, $\Delta y = \Delta y_0 / n_\text{ref}^{\ell-1}$, and $\Delta z = \Delta z_0 / n_\text{ref}^{\ell-1}$, in which $\Delta x_0$, $\Delta y_0$, and $\Delta z_0$ are the grid spacings on the coarsest grid level.

\begin{figure}[]
 \centering
 \subfigure[Heave displacement]{
  \includegraphics[scale = 0.33]{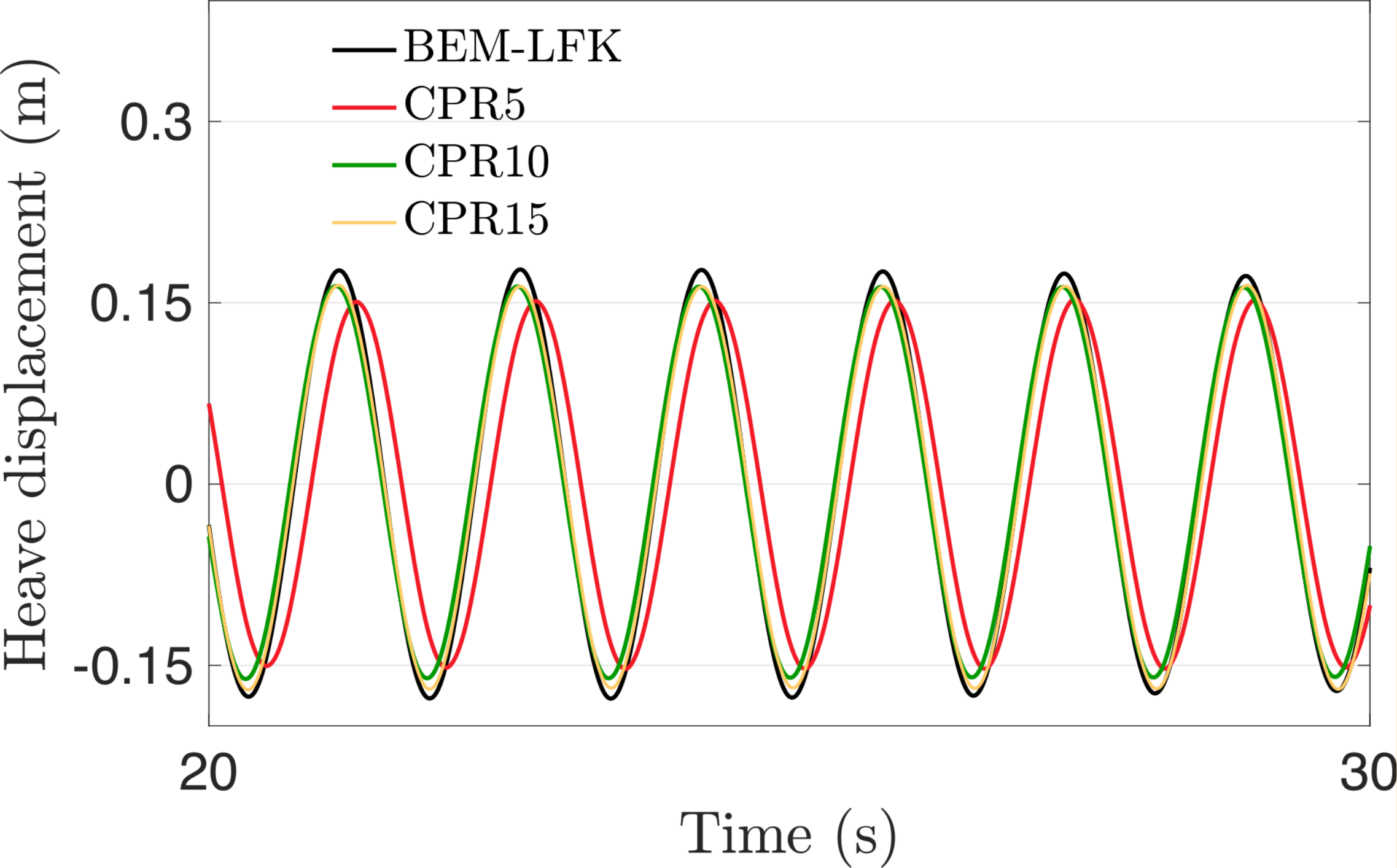}
  \label{heave_vs_time_grid_conv}
 }
   \subfigure[Heave velocity]{
  \includegraphics[scale = 0.33]{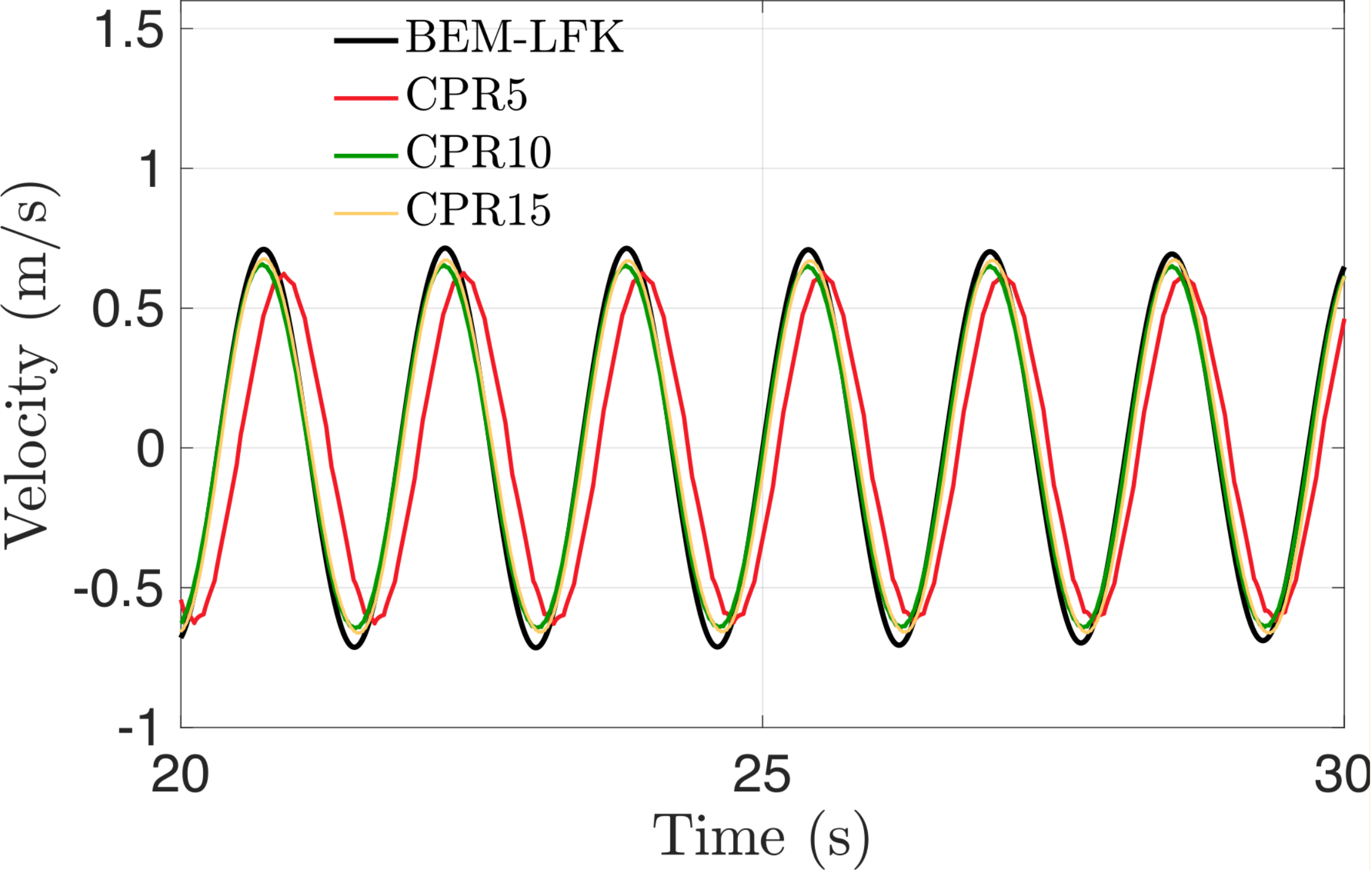}
  \label{velocity_vs_time_grid_conv}
 }
 \caption{Temporal evolution of the heave~\subref{heave_vs_time_grid_conv} displacement 
and~\subref{velocity_vs_time_grid_conv} velocity of the uncontrolled WEC device using BEM-LFK  (\textcolor{black}{\textbf{-----}}, black) and CFD solvers. Three grid resolutions of CPR5 (\textcolor{red}{\textbf{-----}}, red), CPR10 (\textcolor{ForestGreen}{\textbf{-----}}, green), and CPR15 (\textcolor{yellow}{\textbf{-----}}, yellow) are used for the CFD solver. The first-order regular wave characteristics are: $\cH$ = 0.1 m, $\cT$ = 1.5652 s, and $\lambda$ = 3.8144 m.
}
 \label{fig_grid_convergence_plots}
\end{figure}

\begin{figure}[]
   \centering
   	\includegraphics[scale = 0.18]{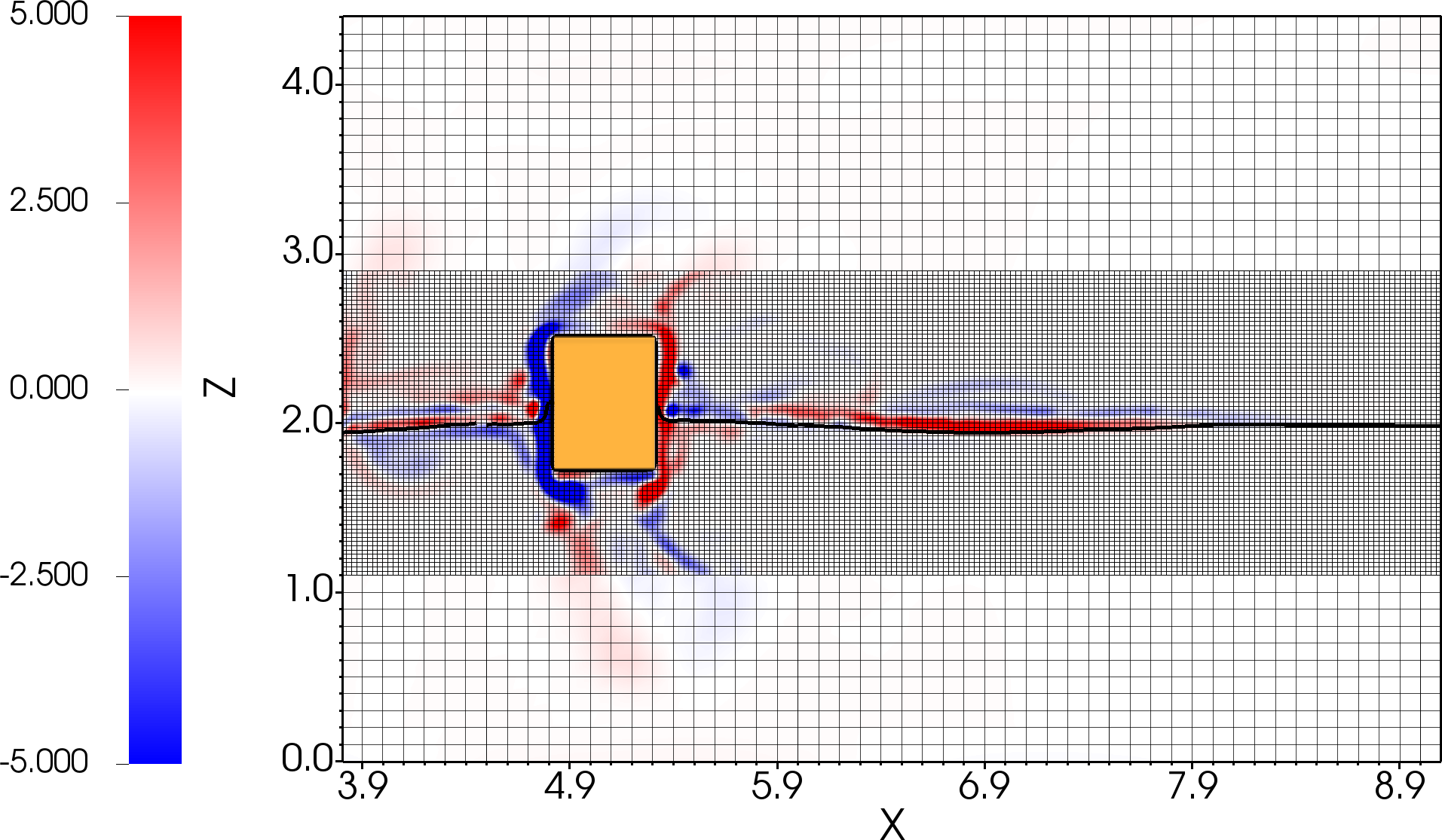}
   \caption{Wave-structure interaction of the 3D vertical cylinder WEC device (here shown in the $x-z$ plane) at $t = 22$ s using the medium grid resolution (CPR10). A locally refined mesh with $\ell = 2$ and $\nref = 4$ is used. The air-water interface and the vortical structures resulting from the WSI are plotted.
}
\label{fig_vcyl_nwt_regular_medium}
\end{figure}

First-order regular waves of height $\cH$ = 0.1 m and time period $\cT$ = 1.5652 s enter from the left side of the domain and interact with the 3D vertical cylinder. The temporal evolution of the device displacement and velocity using three mesh resolutions are shown in Figs.~\ref{heave_vs_time_grid_conv} and~\ref{velocity_vs_time_grid_conv}, respectively. The average percentage change in the peak values of the heave displacement between two consecutive grid resolutions is calculated from $t$ = 20 s to 30 s. The average percentage change between the coarse and medium grids 
is 6 $\%$, and between the medium and fine grids is 2.7 $\%$. For heave velocity these values are 3.6 $\%$ and 2.5 $\%$, respectively. Fig.~\ref{fig_vcyl_nwt_regular_medium} shows the air-water interface and the vortical structures arising from the WSI using the medium grid (CPR10) resolution. It can be observed that both these fluid dynamical quantities are adequately resolved by the CPR10 grid. From Figs.~\ref{fig_grid_convergence_plots} and~\ref{fig_vcyl_nwt_regular_medium}, it can be concluded that the medium grid resolution is able to capture the WSI dynamics with good accuracy and hence is used for the rest of the CFD simulations. 

The device dynamics are also simulated using the BEM-LFK solver, which solves Eqs.~\ref{eqn_cont_SS_sysa}-\eqref{eqn_cont_SS_sysb} of Sec.~\eqref{sec_wec_model_mpc_form}. Since the present test simulates the WSI without MPC, the device undergoes a small motion from its mean equilibrium position under the action of first-order Stokes waves. Therefore, the CFD results are expected to match the BEM results in this situation. Indeed, this can be confirmed from the results of Figs.~\ref{heave_vs_time_grid_conv} and~\ref{velocity_vs_time_grid_conv}.  

\begin{table}[]
 \centering
 \caption{Grid refinement parameters used for the grid convergence study.}
 \rowcolors{2}{}{gray!10}
 \begin{tabular}{*6c}
 \toprule
 Parameters & Coarse & Medium & Fine \\
 \midrule
  $n_{\text{ref}}$ & 4   & 4   & 4  \\
  $\ell$         & 2   & 2   & 2         \\
 $N_{x}$       & 60 & 120 & 180   \\
 $N_{y}$       & 15 & 30 & 45   \\
 $N_{z}$       & 22  & 44  & 66   \\
  \REVIEW{$\Delta x_0 = \Delta y_0 = \Delta z_0$ (m)}       & \REVIEW{0.2}  & \REVIEW{0.1}  & \REVIEW{0.0667}   \\
  \REVIEW{$\Delta x = \Delta y = \Delta z$ (m)}       & \REVIEW{0.05}  & \REVIEW{0.025}  & \REVIEW{0.0166}   \\
 $\Delta t$ (s)   & $5\times 10^{-3}$ & $2.5\times 10^{-3}$ & $1.5\times 10^{-3}$ \\
 \bottomrule
 \end{tabular}
 \label{tab_grid_convergence_study}
\end{table}

\subsection{Temporal resolution study} \label{temporal_resol_study}

\begin{figure}[!h]
   \centering
   \subfigure[Heave displacement]{
   	\includegraphics[scale = 0.33]{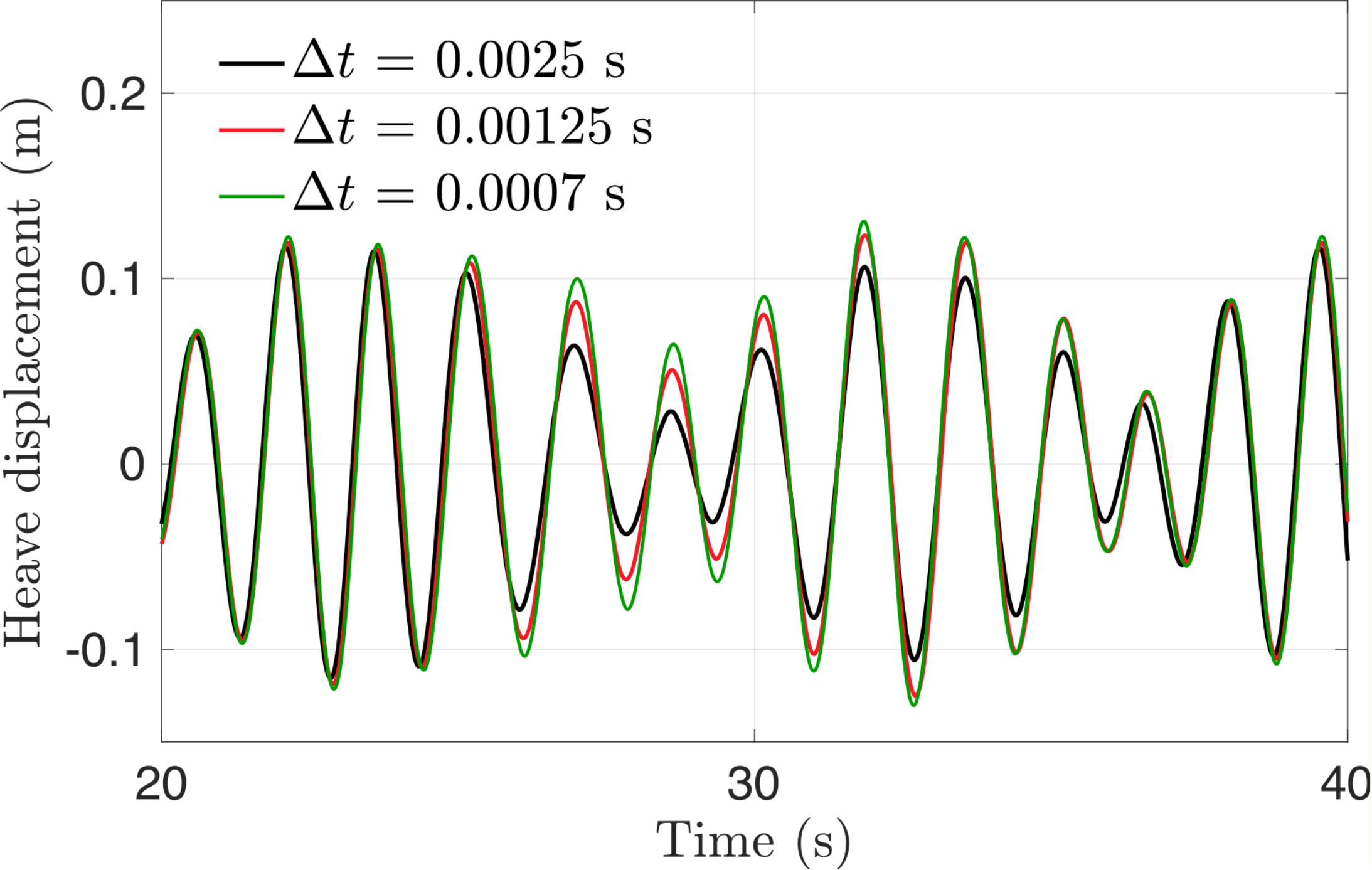}
	\label{heave_vs_time_temporal_res}
   }
   \subfigure[Heave velocity]{
   	\includegraphics[scale= 0.33]{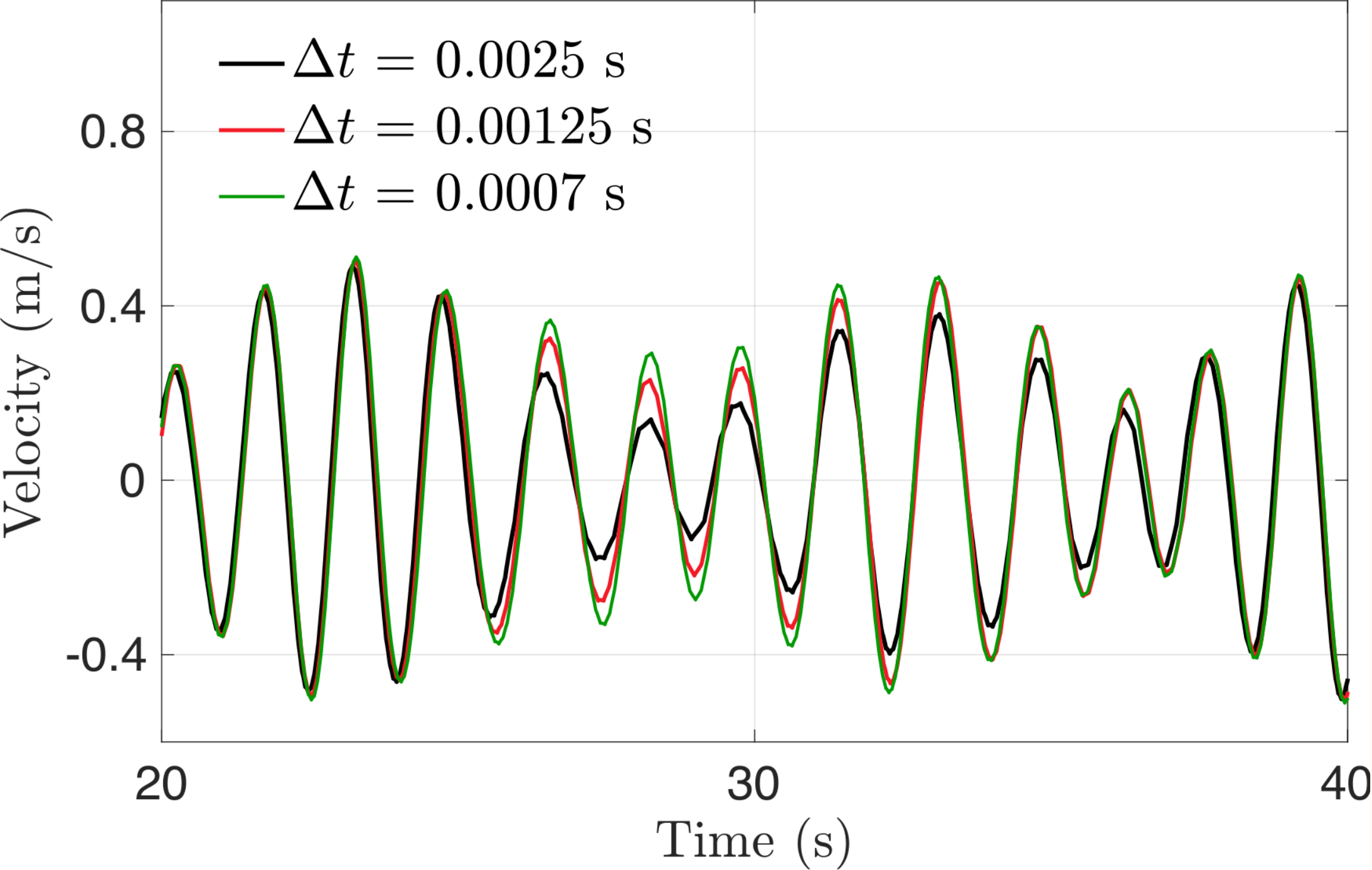}
	\label{velocity_vs_time_temporal_res}
   }
   \caption{Temporal evolution of~\subref{heave_vs_time_temporal_res} the heave displacement and~\subref{velocity_vs_time_temporal_res} heave velocity for three different time step sizes: $\Delta t = 2.5 \times 10^{-3}$ s (\textcolor{black}{\textbf{-----}}, black), $\Delta t = 1.25 \times 10^{-3}$ s (\textcolor{red}{\textbf{-----}}, red), and $\Delta t= 7 \times 10^{-4}$ s (\textcolor{ForestGreen}{\textbf{-----}}, green).
Irregular water waves are generated with $\cH_\text{s}$ = 0.15 m, $\cT_p$ = 1.7475 s, and 
 $N$ = 50 wave components, with wave component frequencies in the range 1.6 rad/s to 20 rad/s distributed uniformly.}
   \label{fig_irregwave_temporal_resol}
\end{figure}

In this section,  we conduct a time step size study to find the step size $\Delta t$  that adequately resolves the energy content of irregular waves. Specifically,  $\Delta t$ should be such that the high-frequency wave components that carry a considerable amount of energy are adequately represented in the simulation. Irregular waves of  height $\cH$ = 0.15 m, peak time period $\cT_p$ = 1.7475 s, and $N$ = 50 wave components are generated at the left end of the NWT. We use three different time step sizes for the temporal convergence study: $\Delta t = 2.5 \times 10^{-3} $ s, $1.25 \times 10^{-3} $ s, and $7 \times 10^{-4} $ s. The medium grid resolution (CPR10) of the previous section is used here.  The temporal evolution of the heave displacement and velocity of the device are compared in Fig.~\ref{fig_irregwave_temporal_resol}.  With smaller $\Delta t$ values, we are able to resolve the amplitudes of the heave displacement and velocity more accurately, as seen in Fig.~\ref{heave_vs_time_temporal_res} and Fig.~\ref{velocity_vs_time_temporal_res}, respectively. The average percentage change in the peak values of the heave displacement and velocity between two consecutive time step sizes is calculated from $t$ = 20 s to 40 s. The average percentage change for the heave displacement between $\Delta t$ = $2.5 \times 10^{-3}$ s and $\Delta t$ = $1.25 \times 10^{-3}$ s is 15.06 \% and betwen $\Delta t$ = $1.25 \times 10^{-3}$ s and $\Delta t$ = $7 \times 10^{-4}$ s is 9.89 \%. For velocity, the percentage changes are 14.68 \% and 5.45 \%, respectively. According to these results, $\Delta t$ = $1.25 \times 10^{-3}$ s is sufficient to model WSI with irregular waves.

Based on the tests of this section, we hereafter use the medium grid spatial resolution with  $\Delta t$ = $2.5 \times 10^{-3}$ s for regular waves and $\Delta t$ = $1.25 \times 10^{-3}$ s for irregular waves.

%% file: Results_and_discussion.tex
Sec.~\ref{sec_validation_and_motivation} motivates us to investigate the following questions:

\begin{enumerate}

\item At various sea states, how do the predictions of different WSI and MPC solvers compare?

\item  In the case of the predictions of the solvers differing widely, what is the main reason for this?

\item How do AR predictions affect MPC performance?

\item By using CFD simulations, can the wave-to-PTO power transfer relationships be adequately captured? 

\item How well does the MPC adapt to changing sea states?

%
%
\end{enumerate}

We perform MPC-integrated WSI simulations of the cylindrical WEC device operating in different sea states to answer these questions. CFD simulations are conducted in a computational domain described in Sec.~\ref{sec_spatiotemporal_tests}. The following MPC parameters are used in all simulations, unless stated otherwise:  $\Delta t_p = 0.05$ s, $\cT_h = \cT$ (or $\cT_p$),  $N_p = \lceil \frac{\cT_h}{\Delta t_p} \rceil$, $\lambda_1 = 2$ s, and $\lambda_2 = 0.2$ s. Here, $\lceil \cdot \rceil$ is the nearest-integer/ceil function. The controller is activated at $t = 10 \; \cT$ (or $10 \; \cT_p$), i.e., when the device starts oscillating steadily. We do this to avoid the possibility of creating a large PTO force at the start of the simulation, which could destabilize it.   

\subsection{Comparing the predictions of different solvers}
 \label{subsec_LFK_NLFK_CFD_MPC_comparison}


This section compares the predictions of various WSI and MPC solvers listed in Table~\ref{tab_solvers}. The results presented here are not based on the AR model, but on analytical expressions to predict the wave elevation data. We discuss the effect of AR predictions on MPC performance separately in Sec.~\ref{subsec_real_time_control}.  Table~\ref{tab_wave_cases} lists the sea states and the PTO force limits. In order to simplify the discussion,  constraints on the device displacement and velocity are not included in the MPC.  Furthermore, preliminary testing showed that adding the displacement and velocity constraints (along with the PTO force constraint) did not significantly alter the results of this section (data not shown for brevity).

\begin{table}[]
\centering
  \caption{Cases considered for comparing results for various solvers and MPC methodologies.}
   \rowcolors{2}{}{gray!10}
\begin{tabular}{cccc}
 \toprule
Case & Wave type           & Wave height (m) & Control force ($\text{F}_\text{PTO}$) constraint (N) \\
 \midrule
1    & First-order regular & 0.1             & $\pm$ 25                           \\
2    & First-order regular & 0.1             & $\pm$ 100                          \\
3    & First-order regular & 0.5             & $\pm$ 25                           \\
4    & First-order regular & 0.5             & $\pm$ 100                          \\
5    & First-order regular & 0.5             & $\pm$ 300                          \\
6    & Irregular           & 0.15            & $\pm$ 25                           \\
7    & Irregular           & 0.15            & $\pm$ 100                          \\
8    & Irregular           & 0.3             & $\pm$ 25                           \\
9    & Irregular           & 0.3             & $\pm$ 100                        \\
\bottomrule          
\end{tabular}
\label{tab_wave_cases}
\end{table}

\subsubsection{Comparing the predictions with regular waves}
\label{subsec_regwave_solver_comp}

Here, the controlled heave dynamics of the WEC device operating in regular sea conditions are compared. As listed in Table~\ref{tab_wave_cases}, Cases 1 and 2 consider regular waves of small height $\cH = 0.1$ m and time period $\cT = 1.5652$ s, with  control force limits of $\pm 25$ N and $\pm 100$ N, respectively. Cases 3,  4, and 5 consider regular waves of large height $\cH = 0.5$ m and (the same) time period $\cT = 1.5652$ s, with control force limits of $\pm 25$ N, $\pm 100$ N, and $\pm 300$ N, respectively.  Allowing a larger control force in MPC leads to a higher heave amplitude of the device. However, this puts more strain on the actuator system, which can damage the hardware or negatively impact the actuator efficiency (actuator efficiency is not considered in this work).

\begin{table}[]
 \centering
 \caption{Time-averaged power output using different WSI and MPC solvers for Cases 2, 5, 7, and 9 of Table~\ref{tab_wave_cases}.}
 \rowcolors{3}{}{gray!10}
\begin{tabular}{c c c c c c c}
\toprule
\multirow{2}{*}{} & \multirow{2}{*}{Solver} & \multirow{2}{*}{MPC} & \multicolumn{4}{c}{Time-averaged power (W)}        \\
\cline{4-7}
                         &                         &                      & Case 2 & Case 5    & Case 7 & Case 9  \\
                         \midrule
1                        & BEM-LFK                 & LFK                  & 5.4458 & 138.9282 & 3.9463 & 12.6718 \\
2                        & BEM-LFK                 & NLFK                 & 5.674  & 142.7581 & 3.8786 & 12.1793 \\
3                        & BEM-NLFK                & LFK                  & 5.4766 & 40.9436  & 3.726  & 13.0561 \\
4                        & BEM-NLFK                & NLFK                 & 5.5401 & 36.9532  & 3.7235 & 12.8456 \\
5                        & CFD                     & LFK                  & 3.7216 & 34.2936  & 2.4871 & 7.9284  \\
6                        & CFD                     & NLFK                 & 3.7407 & 34.4263  & 2.7513 & 9.0553  \\
\bottomrule
\end{tabular}
\label{tab_power_case2579}
\end{table}

\begin{figure}[]
   \centering
   \subfigure[Heave displacement for Case 2]{
   	\includegraphics[scale= 0.33]{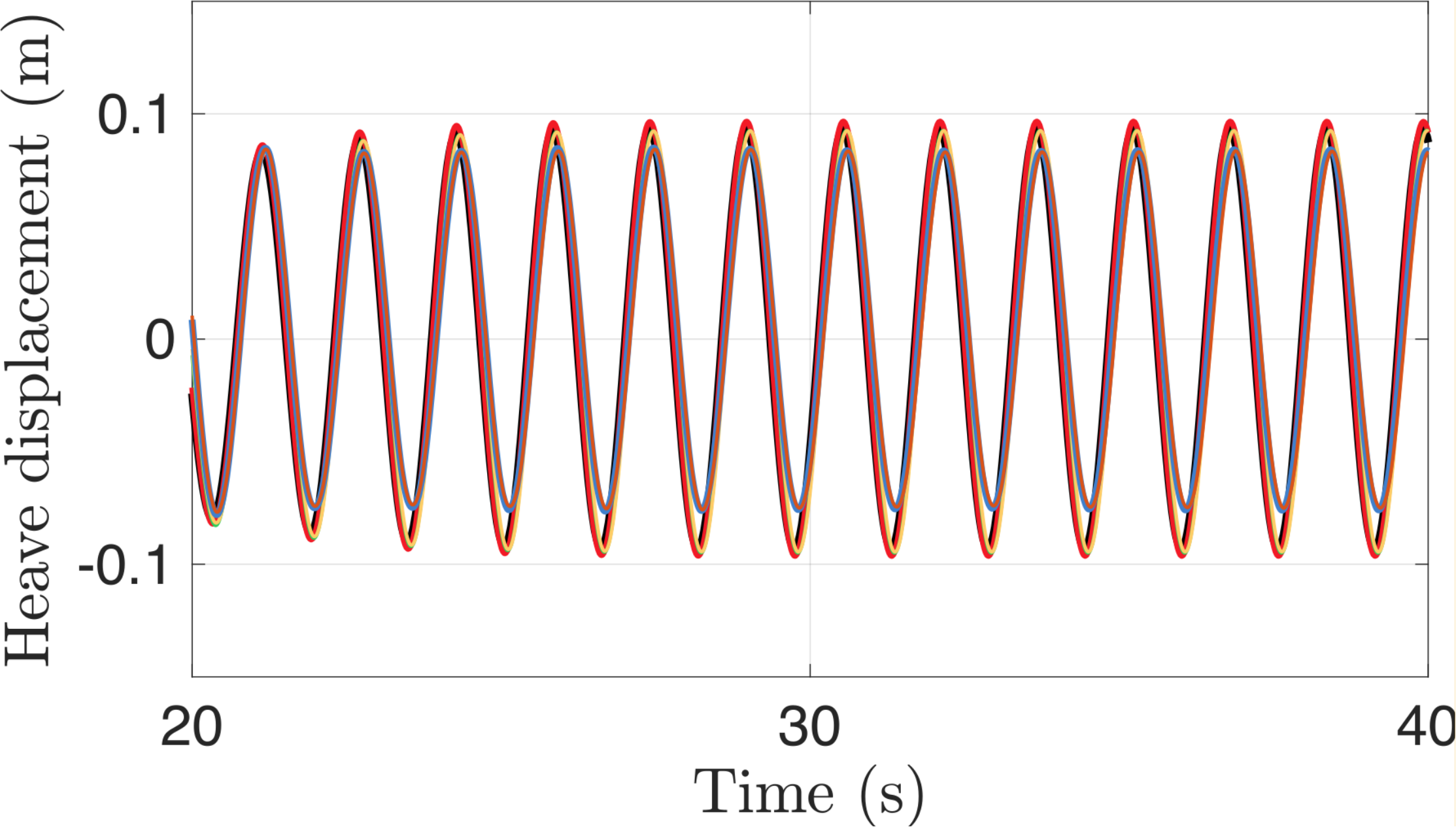}
	\label{fig_heave_CF100_1st_H1_Tp2}
   }
      \subfigure[Heave displacement for Case 5]{
   	\includegraphics[scale= 0.33]{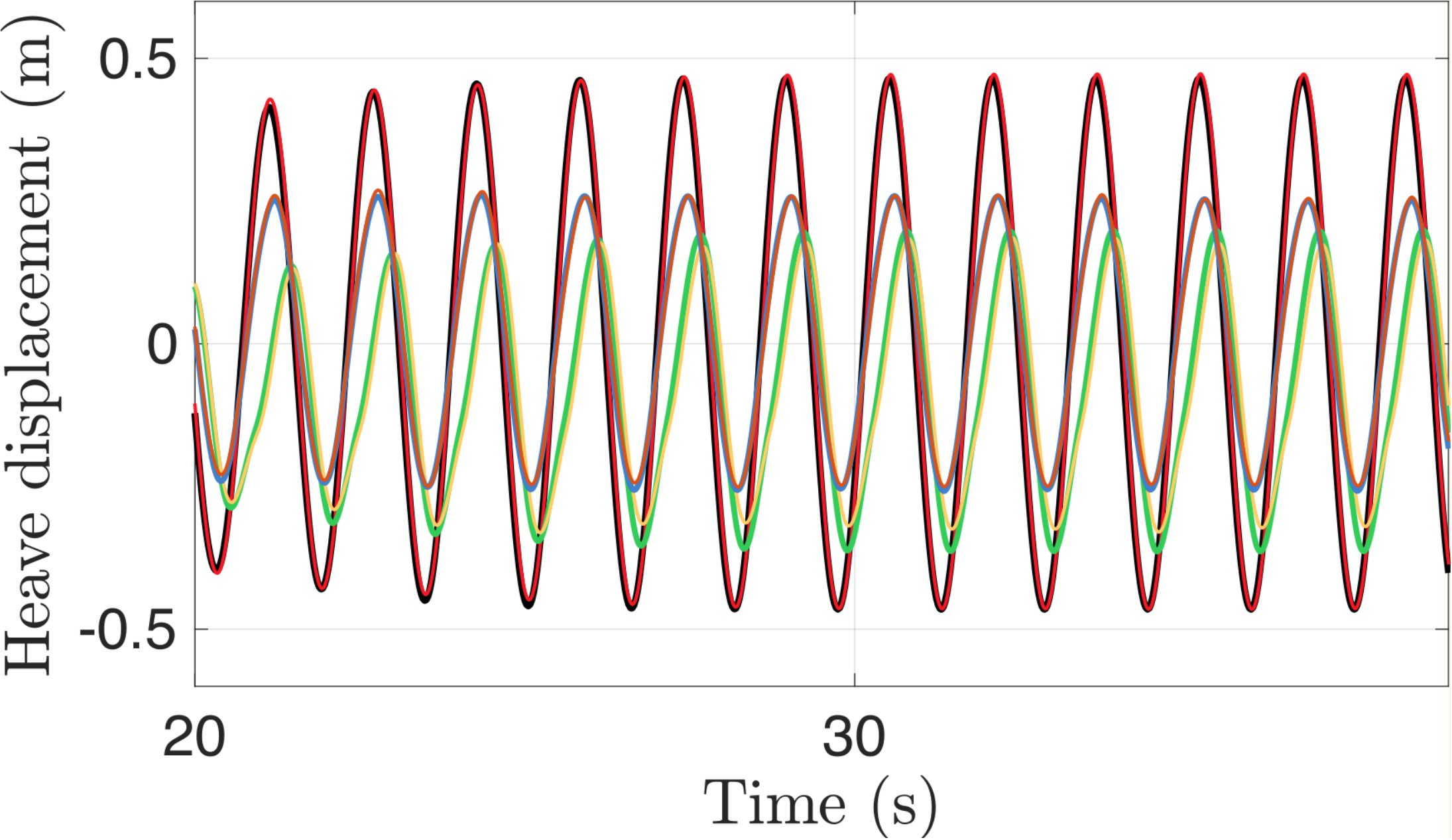}
	\label{fig_heave_CF300_1st_H5_Tp2}
   }
   \subfigure[Control force for Case 2]{
   	\includegraphics[scale = 0.33]{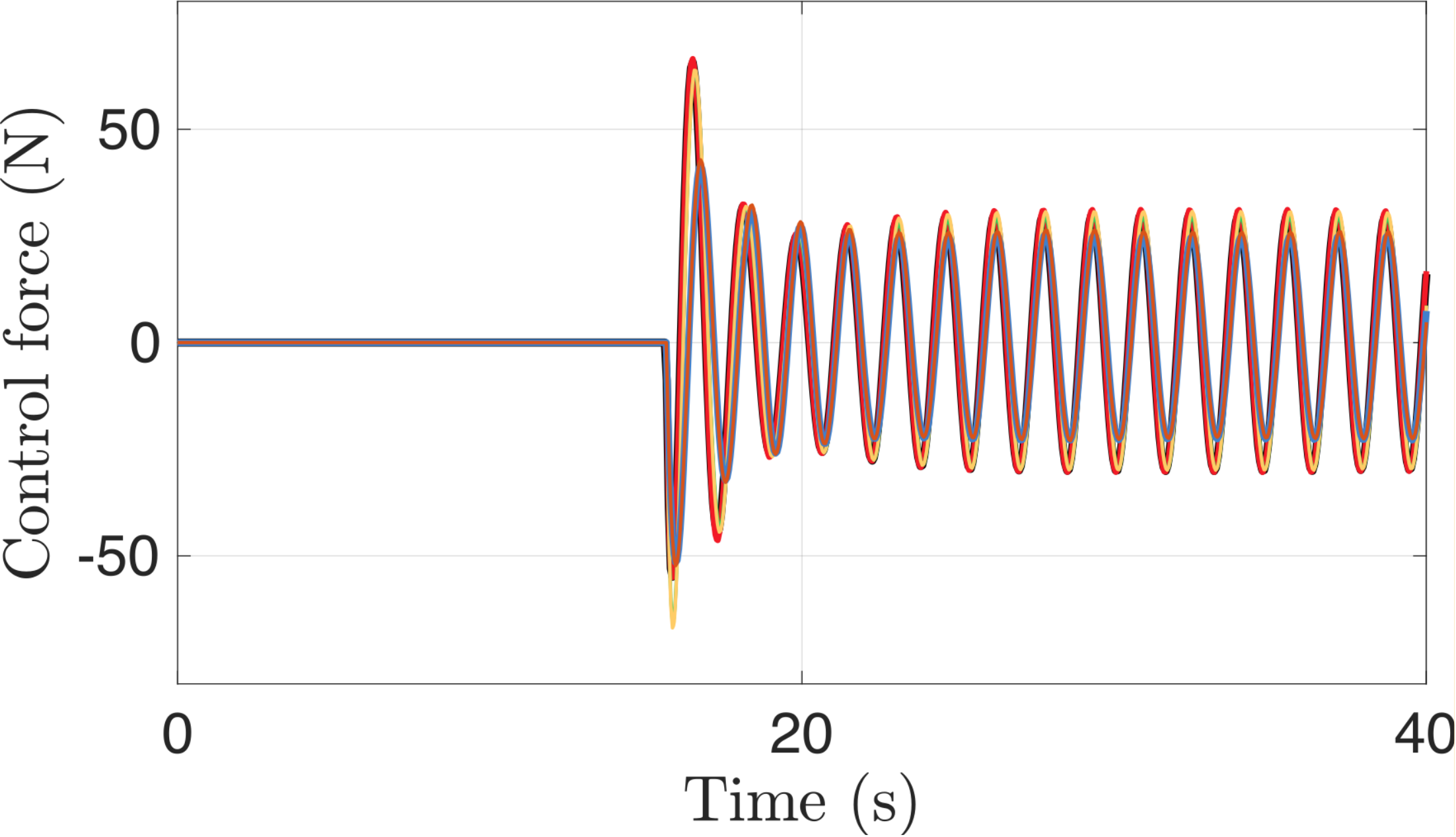}
	\label{fig_CF_CF100_1st_H1_Tp2}
   }
    \subfigure[Control force for Case 5]{
   	\includegraphics[scale = 0.33]{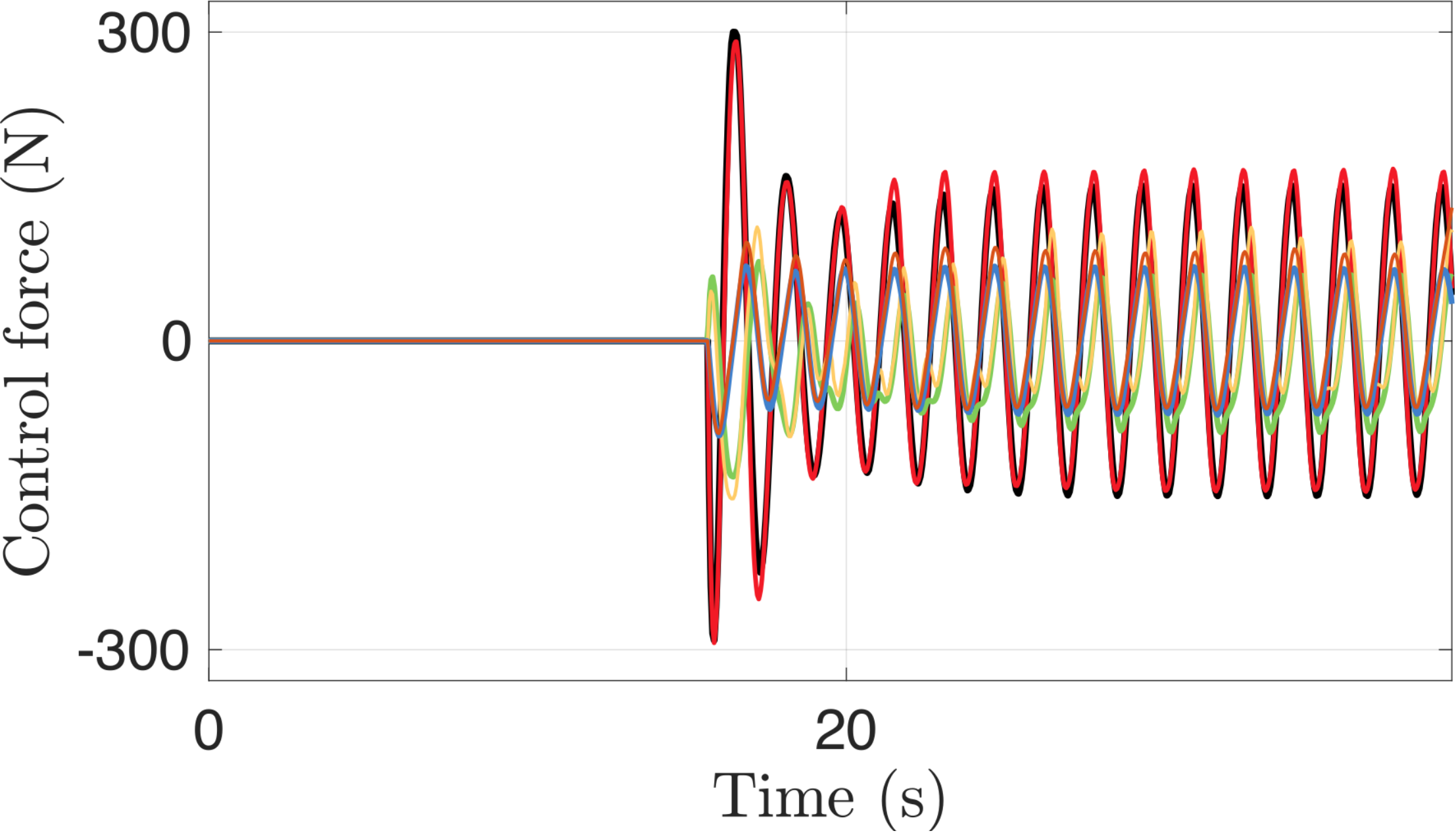}
	\label{fig_CF_CF300_1st_H5_Tp2}
   }
   \subfigure[Power for Case 2]{
   	\includegraphics[scale= 0.33]{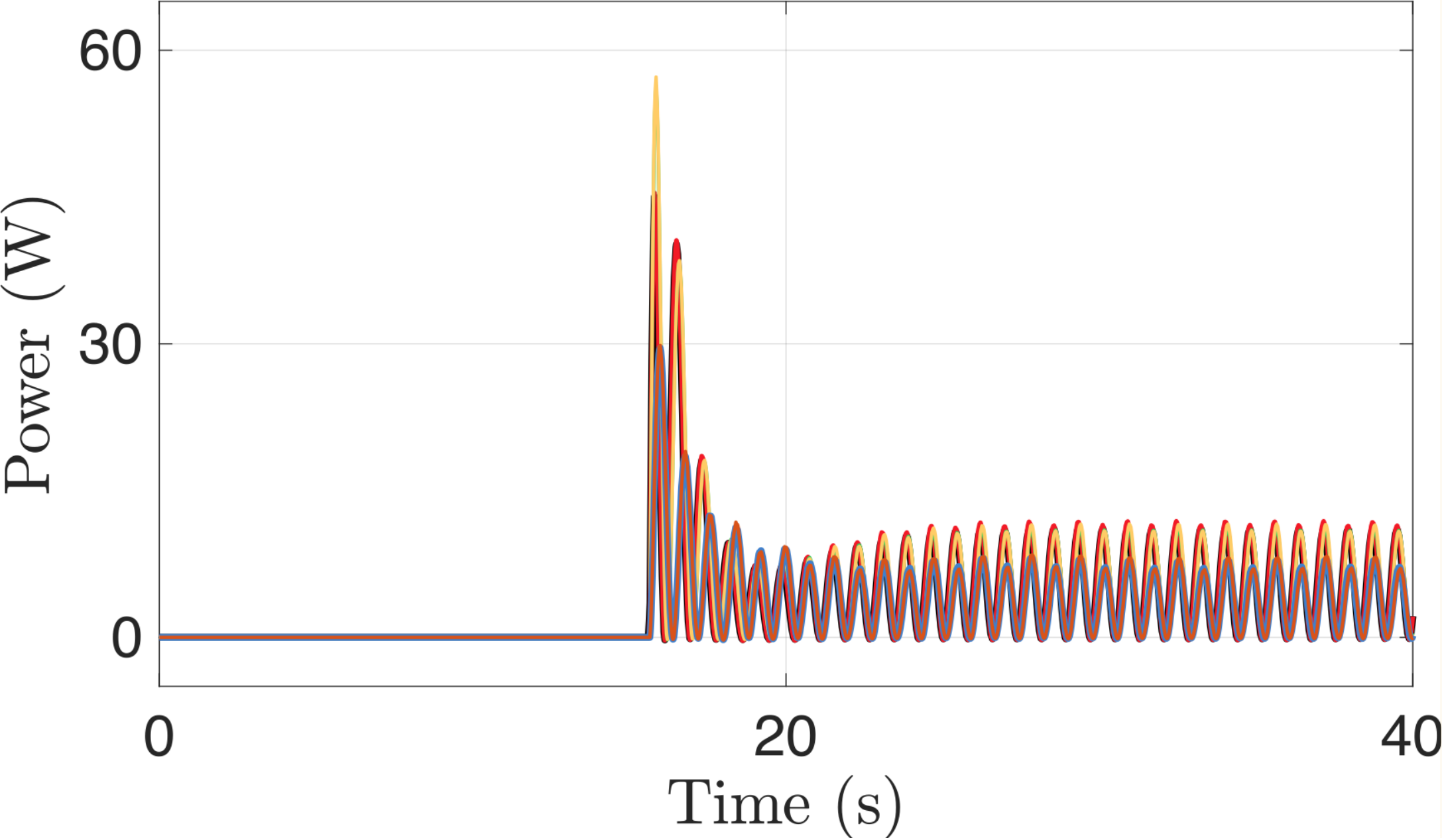}
	\label{fig_power_CF100_1st_H1_Tp2}
   }
    \subfigure[Power for Case 5]{
   	\includegraphics[scale= 0.33]{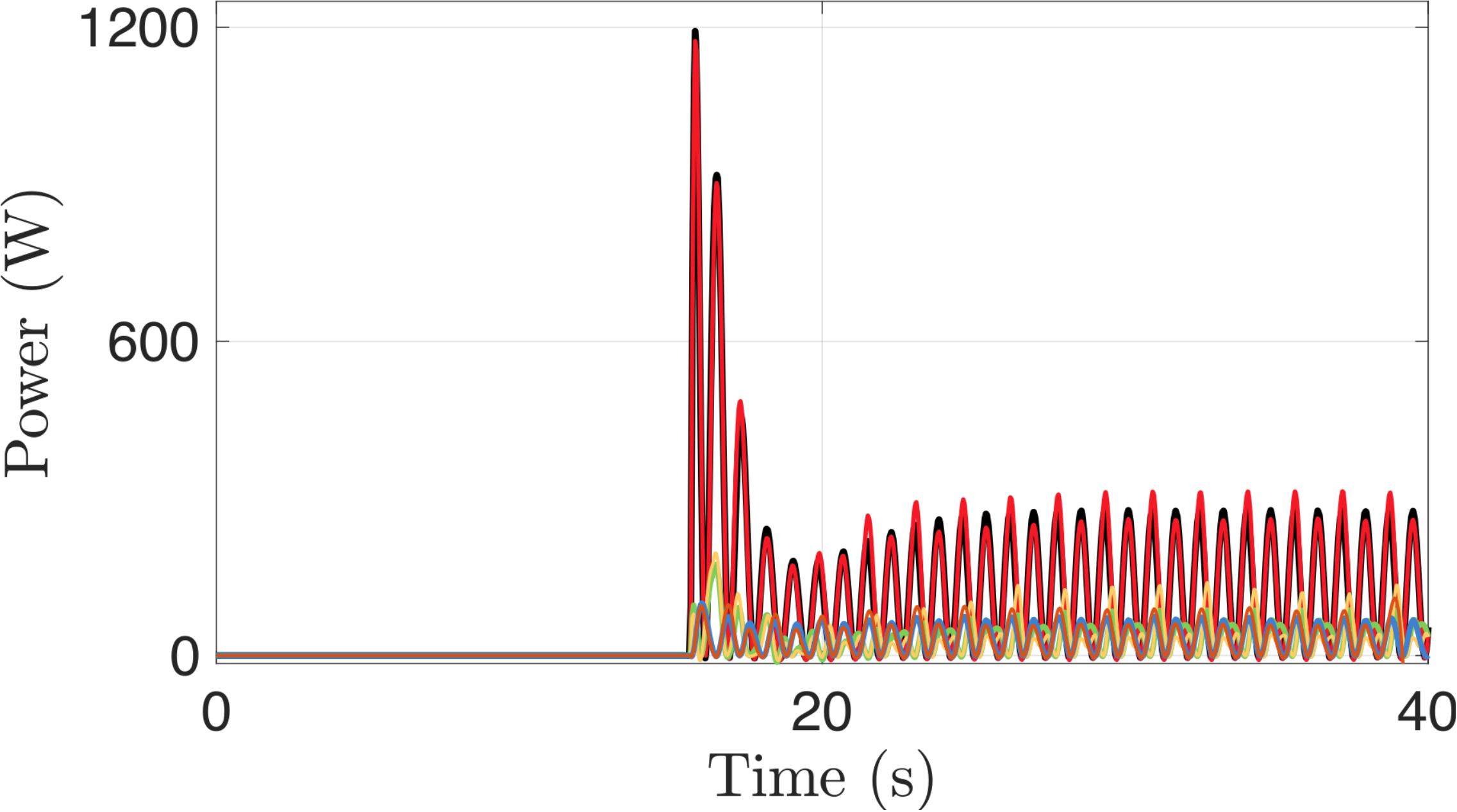}
	\label{fig_power_CF300_1st_H5_Tp2}
   }
   \caption{Comparison of the controlled heave dynamics of the 3D vertical cylinder WEC device with regular waves.  Case 2 and Case 5 of Table~\ref{tab_wave_cases} are considered here. The WSI and MPC solver combinations are: BEM-LFK and MPC-LFK  (\textcolor{Black}{\textbf{-----}}, black);  BEM-LFK and MPC-NLFK  (\textcolor{Red}{\textbf{-----}}, red), BEM-NLFK and MPC-LFK  (\textcolor{Green}{\textbf{-----}}, green), BEM-NLFK and MPC-NLFK  (\textcolor{Goldenrod}{\textbf{-----}}, mustard), CFD and MPC-LFK  (\textcolor{Blue}{\textbf{-----}}, blue), and CFD and MPC-NLFK  (\textcolor{Orange}{\textbf{-----}}, orange).}
   \label{fig_Case2_5_results}
\end{figure}

Figs.~\ref{fig_heave_CF100_1st_H1_Tp2} and~\ref{fig_heave_CF300_1st_H5_Tp2} compare the heave displacement, \ref{fig_CF_CF100_1st_H1_Tp2} and \ref{fig_CF_CF300_1st_H5_Tp2} compare the optimal control force, and \ref{fig_power_CF100_1st_H1_Tp2} and \ref{fig_power_CF300_1st_H5_Tp2} compare the instantaneous power absorbed by the device using different WSI and MPC solvers for Case 2 and 5, respectively. The time-averaged power of the device for Cases 2 and 5 is listed in Table~\ref{tab_power_case2579}. The time-averaged power is calculated between $t =$ 30 s to 40 s when the device dynamics become steady.  Other simulations produce similar trends, which for brevity are not shown.  Instead, the time-averaged powers are shown in Fig.~\ref{fig_regwave_power_bar_plot}. 

From the results presented in Fig.~\ref{fig_Case2_5_results} and Table~\ref{tab_power_case2579}, it is observed that for the small wave height Case 2, the BEM-LFK solver results are close to those of BEM-NLFK and CFD solvers. In contrast, for the large wave height Case 5, the dynamics and the power absorbed by the WEC device are largely over-predicted. Another important observation from Table~\ref{tab_power_case2579} and Fig.~\ref{fig_regwave_power_bar_plot} is that the MPC-LFK and MPC-NLFK solvers produce almost the same time-averaged powers, when used either with the BEM or the CFD solver. It can also be observed that the BEM-NLFK and CFD solver results are in good agreement. 

The results of this section provide two meaningful insights: (\textbf{1}) the main cause of discrepancy between the BEM-LFK and the CFD (or the BEM-NLFK) solver is the manner in which wave excitation forces are computed; and (\textbf{2}) there is a little advantage to increasing the complexity of the hydrodynamical model within MPC. The latter also implies that the simpler and computationally faster LFK model is sufficiently acurate for the model predictive control of WECs.


One can also note that by using $\lambda_2 = 0.2$ s,  the negative part of the power cycle is largely eliminated for all WSI solvers. This can be verified from the instantaneous power curves of Figs.~\ref{fig_power_CF100_1st_H1_Tp2} and \ref{fig_power_CF300_1st_H5_Tp2}. Similar observation can be made for the irregular wave cases that are presented in the next section.

\subsubsection{Comparing the predictions with irregular waves}
\label{subsec_irregwave_solver_comp}

Next, the controlled heave dynamics of the WEC device operating in irregular sea conditions are compared. Cases 6 and 7 in Table~\ref{tab_wave_cases} are of irregular waves of small significant wave height $\cH_s = 0.15$ m and peak time period $\cT_p = 1.7475$ s, with control force limits of $\pm 25$ N, and $\pm 100$ N, respectively. Cases 8 and 9 concern irregular waves of moderate significant wave height $\cH_s = 0.3$ m and (the same) peak time period $\cT_p = 1.7475$ s, with control force limits of $\pm 25$ N and $\pm 100$ N, respectively. 

\begin{figure}[]
   \centering
   \subfigure[Heave displacement for Case 7]{
   	\includegraphics[scale= 0.33]{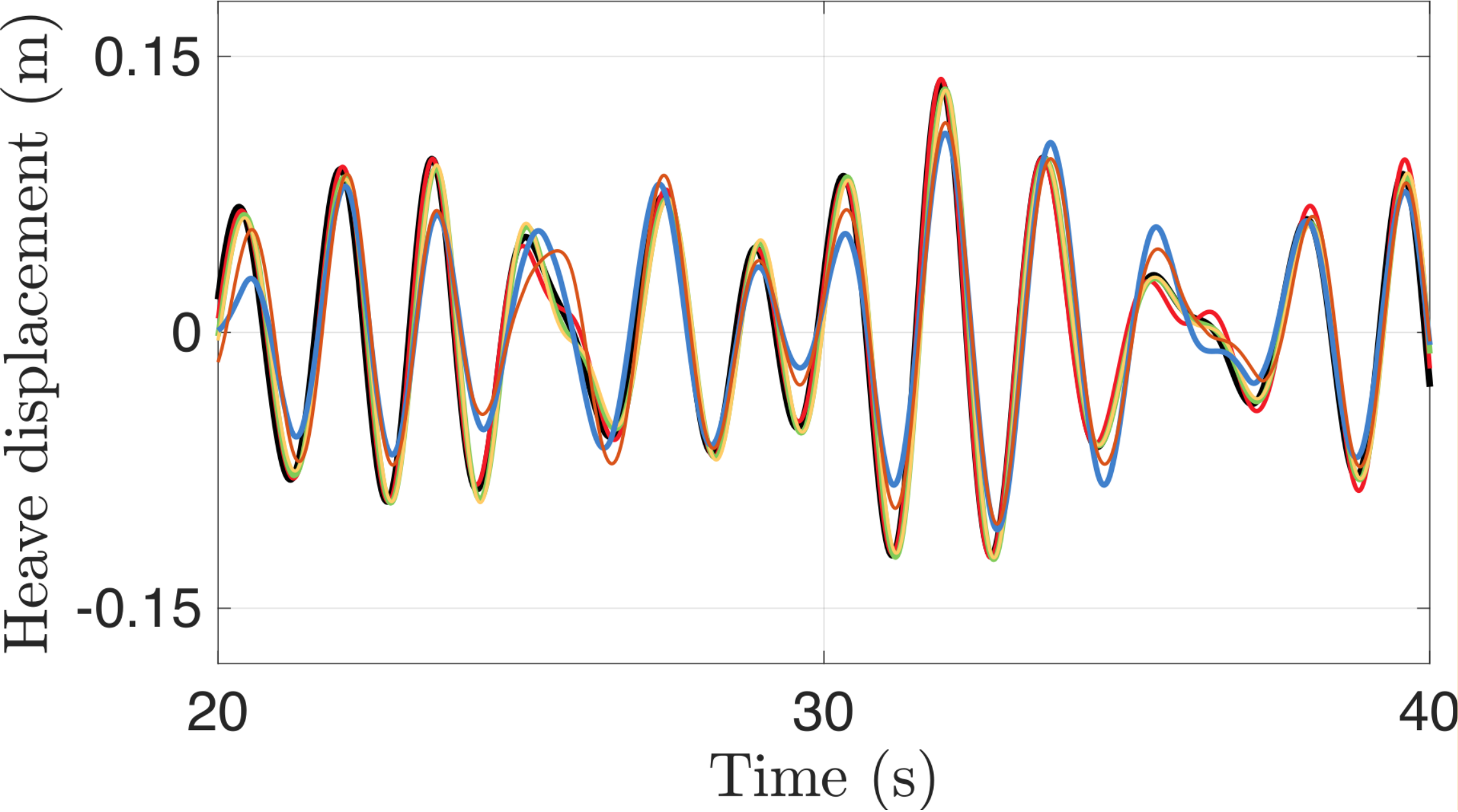}
	\label{fig_heave_CF100_irreg1}
   }
      \subfigure[Heave displacement for Case 9]{
   	\includegraphics[scale= 0.33]{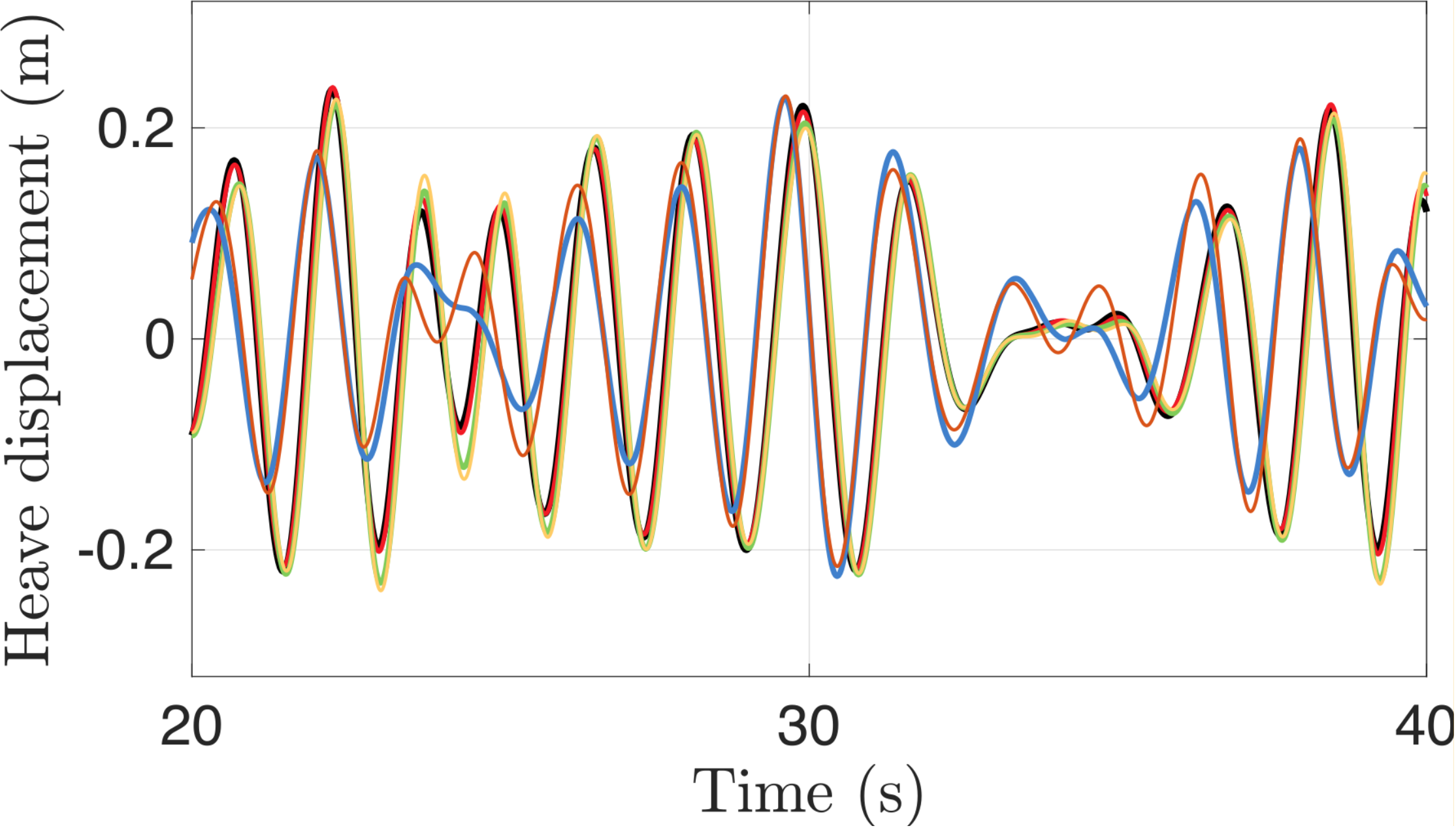}
	\label{fig_heave_CF100_irreg2}
   }
   \subfigure[Control force for Case 7]{
   	\includegraphics[scale = 0.33]{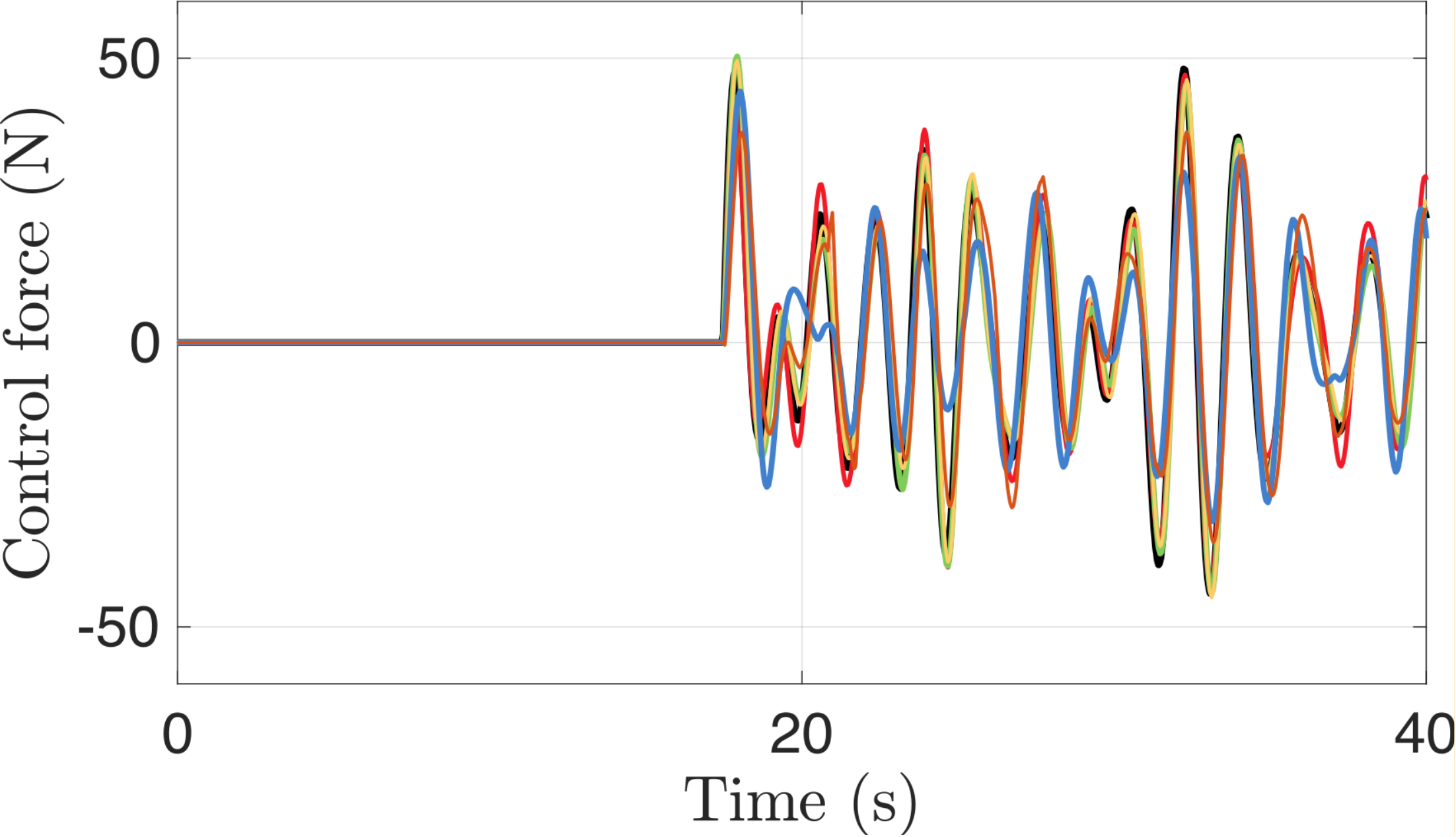}
	\label{fig_CF_CF100_irreg1}
   } 
      \subfigure[Control force for Case 9]{
   	\includegraphics[scale = 0.33]{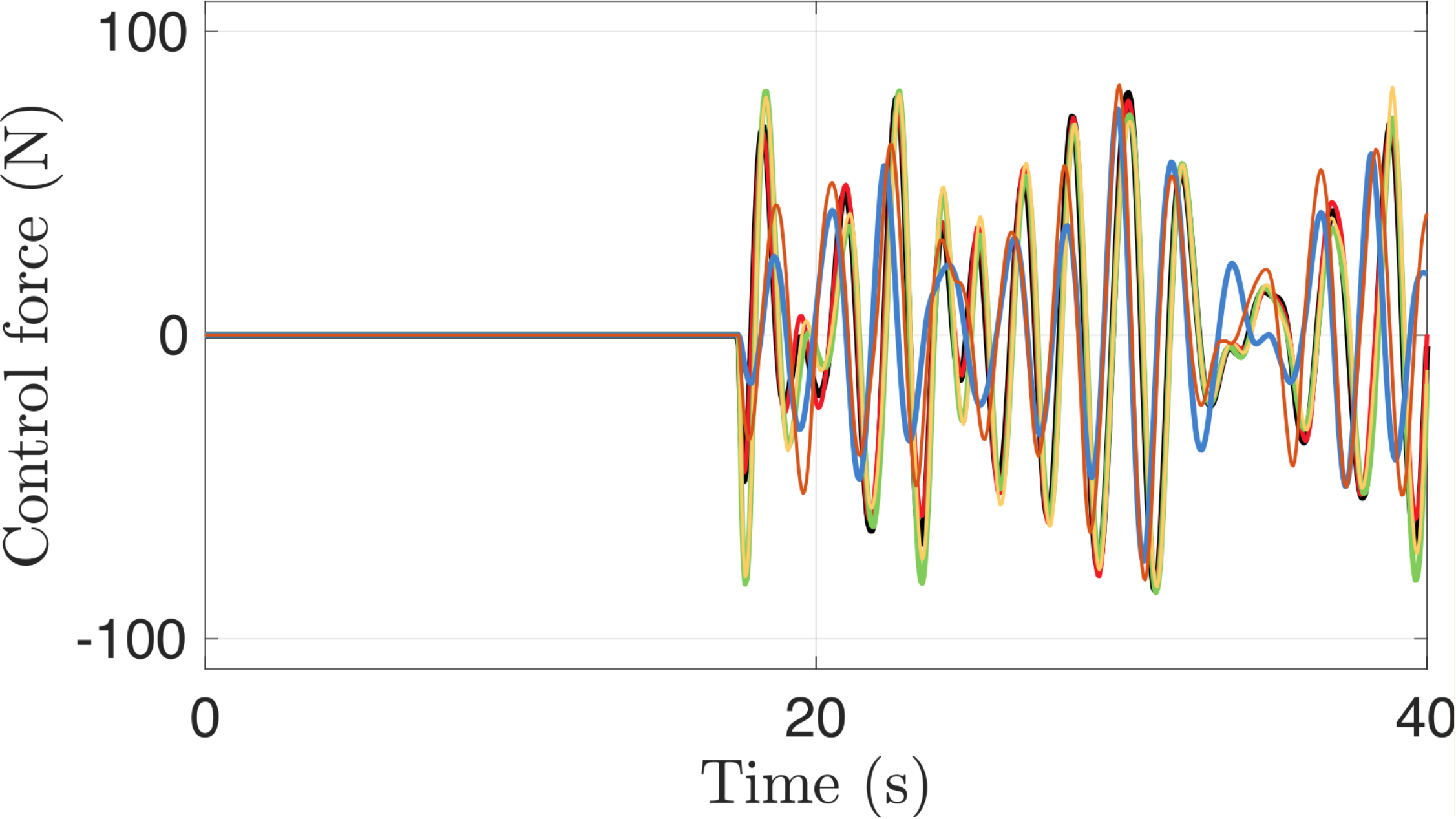}
	\label{fig_CF_CF100_irreg2}
   }
   \subfigure[Power for Case 7]{
   	\includegraphics[scale= 0.33]{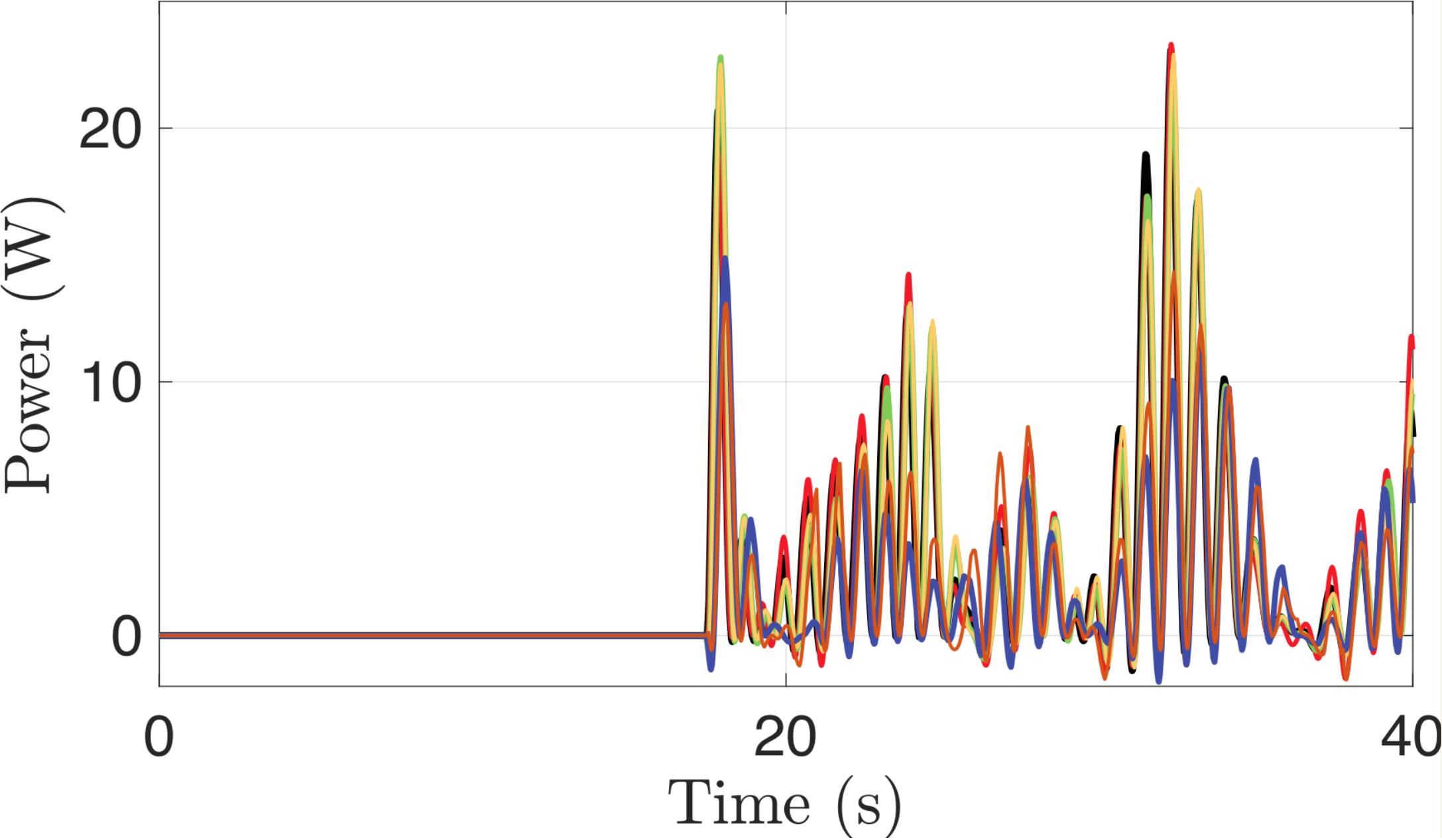}
	\label{fig_power_CF100_irreg1}
   }
   \subfigure[Power for Case 9]{
   	\includegraphics[scale= 0.33]{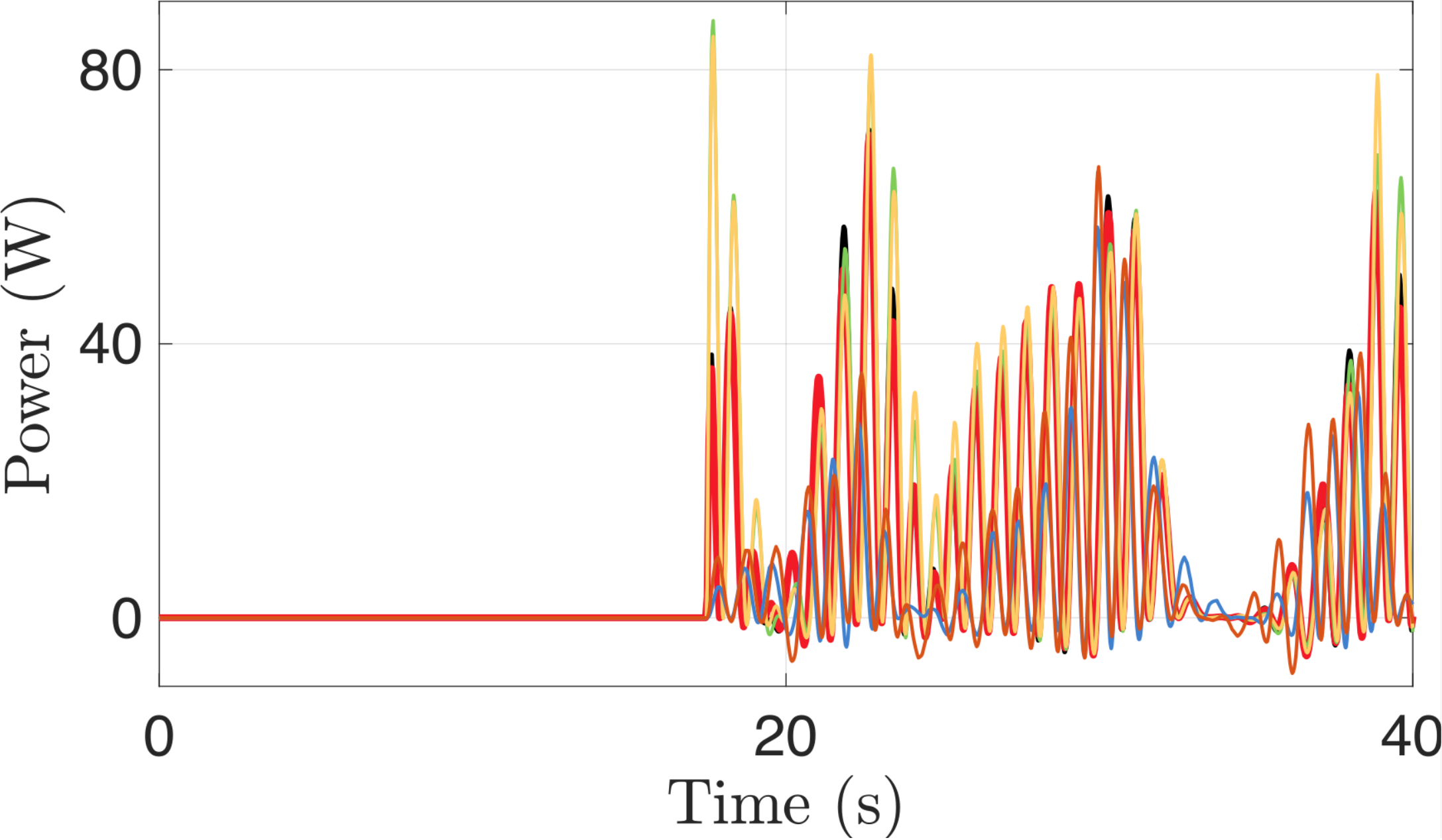}
	\label{fig_power_CF100_irreg2}
   }
   \caption{Comparison of the controlled heave dynamics of the 3D vertical cylinder WEC device with irregular waves. Case 7 and Case 9 of Table~\ref{tab_wave_cases} are considered here. The WSI and MPC solver combinations are:  BEM-LFK and MPC-LFK  (\textcolor{Black}{\textbf{-----}}, black), BEM-LFK and MPC-NLFK  (\textcolor{Red}{\textbf{-----}}, red), BEM-NLFK and MPC-LFK  (\textcolor{Green}{\textbf{-----}}, green), BEM-NLFK and MPC-NLFK  (\textcolor{Goldenrod}{\textbf{-----}}, mustard), CFD and MPC-LFK  (\textcolor{Blue}{\textbf{-----}}, blue), and CFD and MPC-NLFK  (\textcolor{Orange}{\textbf{-----}}, orange).}
   \label{fig_Case7_9_results}
\end{figure}

\begin{figure}[]
   \centering
   \subfigure[Regular waves]{
   	\includegraphics[scale= 0.34]{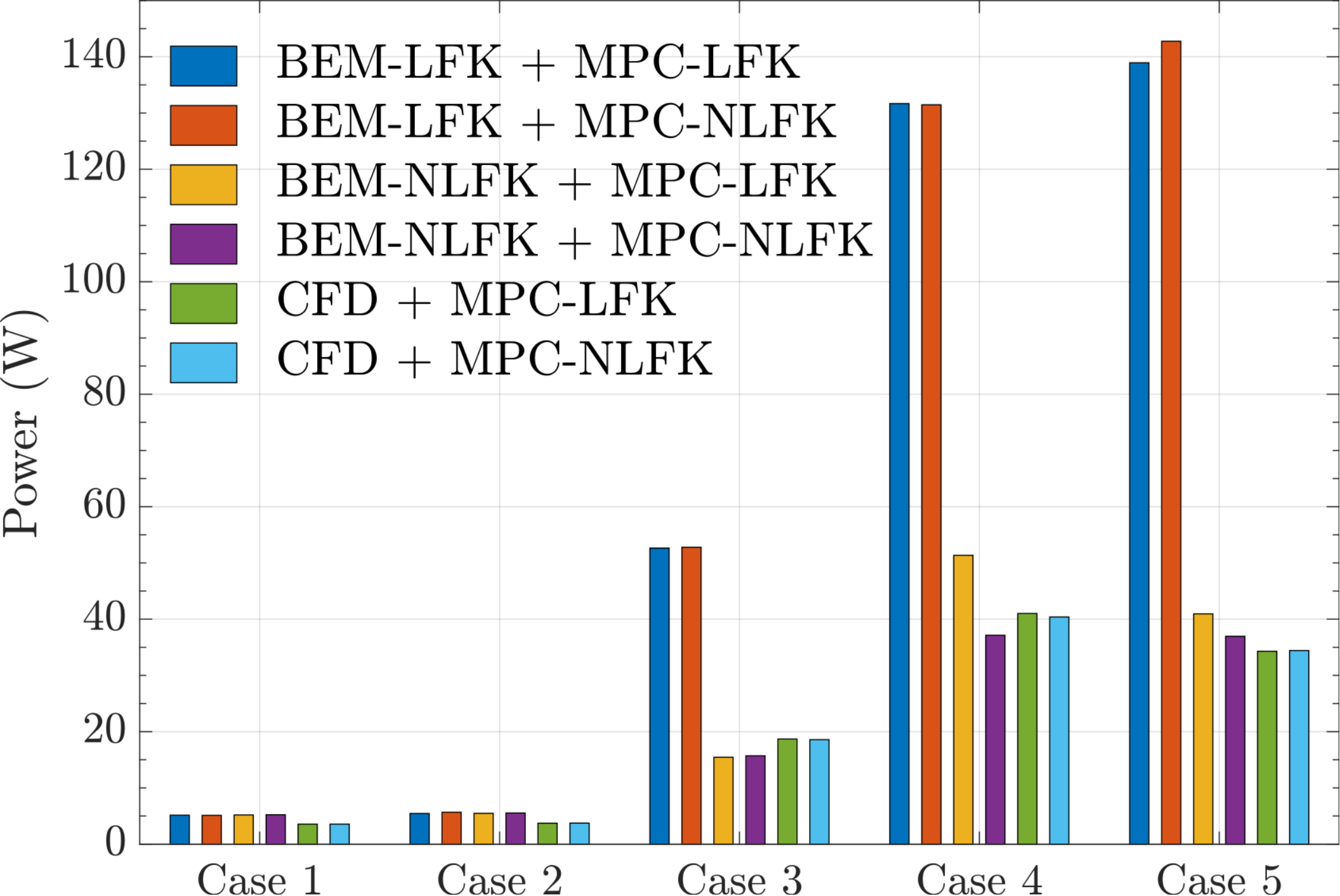}
	\label{fig_regwave_power_bar_plot}
   }
   \subfigure[Irregular waves]{
   	\includegraphics[scale = 0.37]{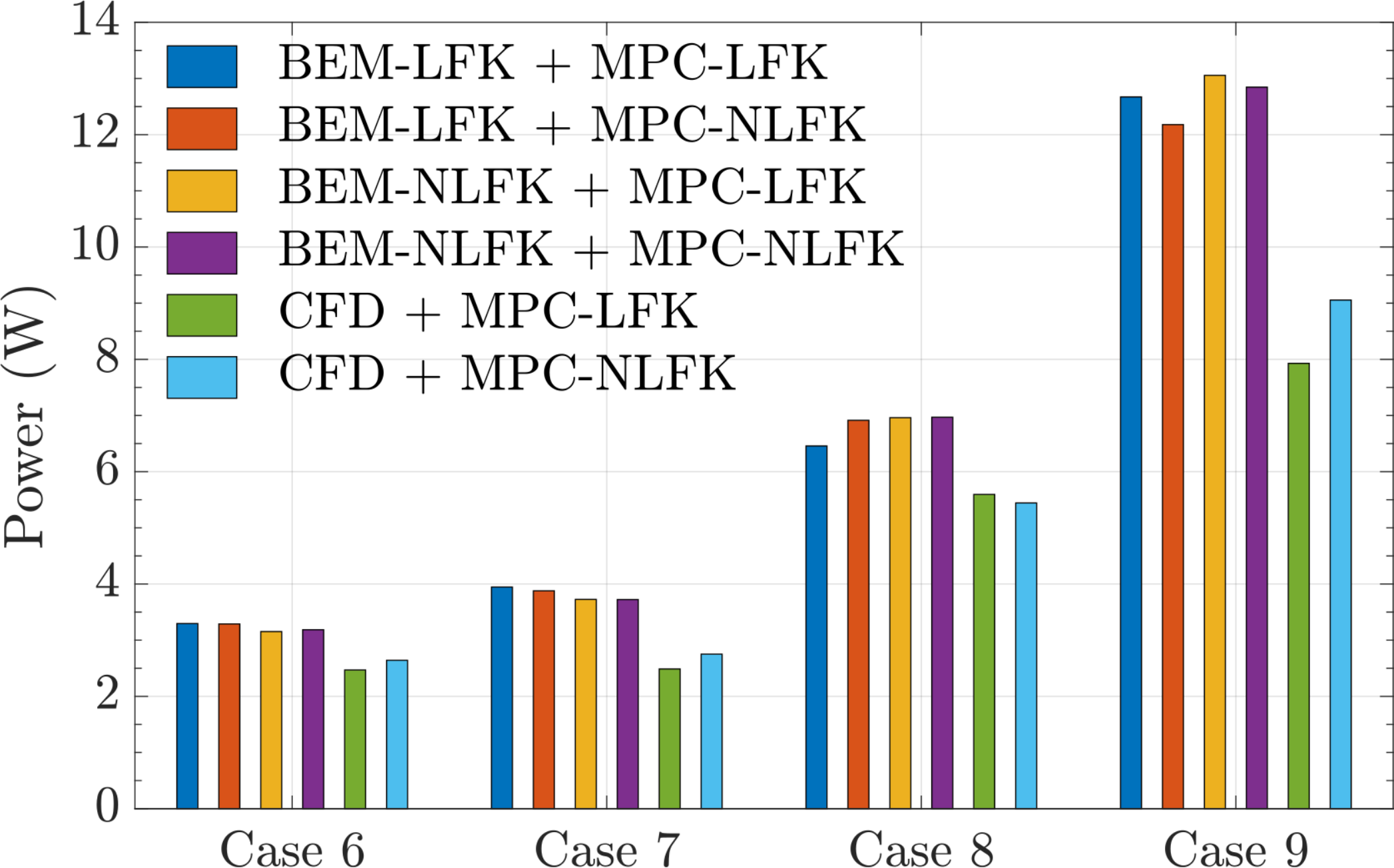}
	\label{fig_irregwave_power_bar_plot}
   }
   \caption{Comparison of time-averaged powers for cases given in Table~\ref{tab_wave_cases}}
   \label{fig_bar_plots_solver_comp}
\end{figure}

Fig.~\ref{fig_Case7_9_results} presents the WEC dynamics for Cases 7 and 9. Results for Cases 6 and  8 are not presented for brevity, as they show similar trends.  Figs.~\ref{fig_heave_CF100_irreg1} and~\ref{fig_heave_CF100_irreg2} compare the heave dynamics,  \ref{fig_CF_CF100_irreg1} and \ref{fig_CF_CF100_irreg2} compare the optimal control force, and \ref{fig_power_CF100_irreg1} and \ref{fig_power_CF100_irreg2}  compare the instantaneous power absorbed by the device using different WSI and MPC solvers for Case 7 and  9, respectively.  The time-averaged power of the device is listed in Table~\ref{tab_power_case2579} and is  calculated between $t$ = 30 s to 40 s when the device dynamics become steady. Simulations of the other cases produce similar trends and are not shown for brevity. Instead, the time-averaged powers are plotted in Fig.~\ref{fig_irregwave_power_bar_plot}.

As shown in Fig.~\ref{fig_Case7_9_results} and Table~\ref{tab_power_case2579}, all WSI and MPC solvers perform almost the same, though the CFD solver predicts slightly lower power for Case 9 than the BEM-LFK and BEM-NLFK solvers.  This is not surprising since the wave heights considered in this section are relatively low. At larger (significant) wave heights, we expect the differences between BEM-LFK and CFD (or BEM-NLFK) solvers to increase; this is confirmed in the next section. Waves with large significant wave heights are not considered here, since the CFD solver requires very small time steps to maintain the numerical stability. As a result, the 3D simulation will take very long to run, which is something we cannot afford at the moment.


The results of Secs.~\ref{subsec_regwave_solver_comp} and \ref{subsec_irregwave_solver_comp} suggest that the BEM-LFK solver may give too optimistic results, especially when the hydrodynamic nonlinearities increase. Conversely, the CFD solver can resolve hydrodynamical non-linearities with high-fidelity, albeit at an increased computational cost, and provides more realistic results.  Between these two extremes is the BEM-NLFK solver, which yields somewhat optimistic power values, but not quite as large as the BEM-LFK solver.  In addition, either MPC-LFK or MPC-NLFK is equally effective for a specific WSI solver since they give very close results. Since the MPC-LFK technique is computationally faster than MPC-NLFK, it is better suited for practical control of WEC devices. 


\subsubsection{Comparing the predictions with varying wave periods}
\label{subsec_varying_H_T}

This section compares the predictions of the BEM-LFK and BEM-NLFK solvers for varying wave periods. Regular and irregular sea conditions are considered. For the two WSI solvers, MPC-LFK is used. Due to the high computational cost associated with simulating waves of longer durations and wavelengths, CFD simulations are not performed here. 

\begin{figure}
   \centering
   \subfigure[Regular waves]{
   	\includegraphics[scale= 0.34]{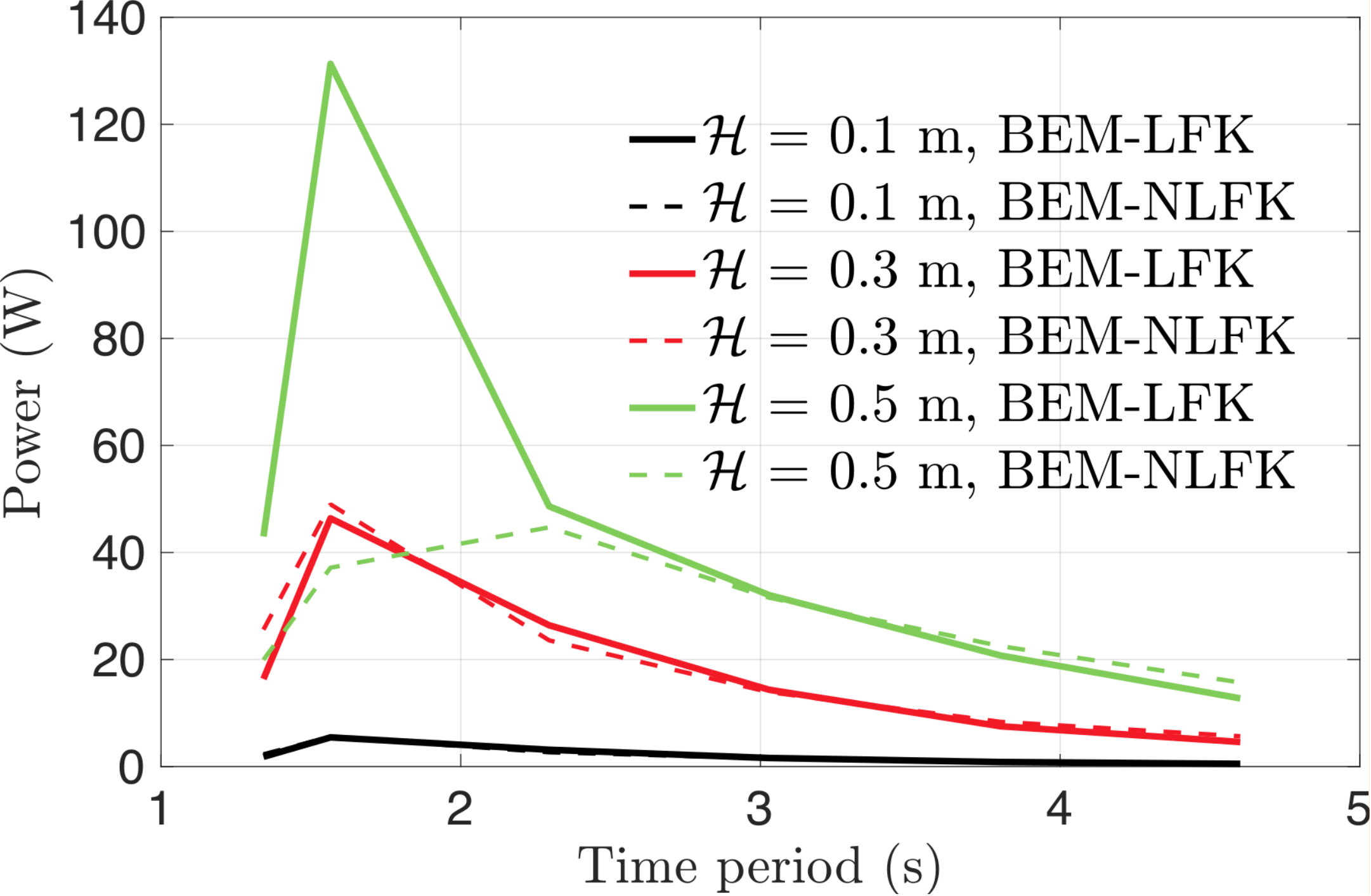}
	\label{fig_power_regwave_H_Tp_variation}
   }
   \subfigure[Irregular waves]{
   	\includegraphics[scale = 0.34]{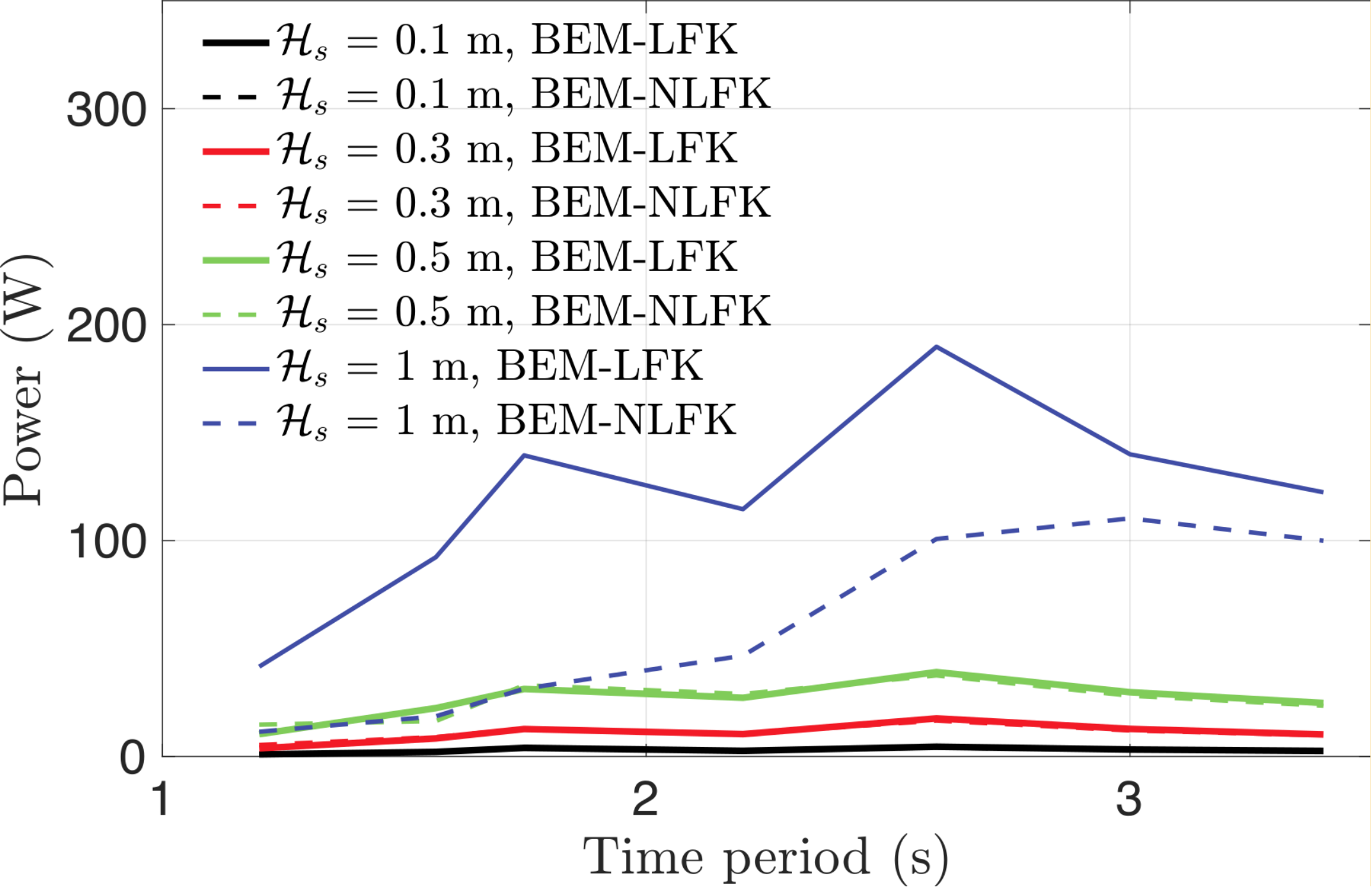}
	\label{fig_power_irregwave_H_Tp_variation}
   }
   \caption{Comparison of time-averaged power absorbed by the WEC device operating in~\subref{fig_power_regwave_H_Tp_variation} regular and~\subref{fig_power_irregwave_H_Tp_variation} irregular sea conditions with varying wave periods and heights. The BEM-LFK  (BEM-NLFK) solver results are shown with solid (dashed) lines.}
\end{figure}

Results compare the time-averaged power absorbed by the WEC device for regular waves in Fig.~\ref{fig_power_regwave_H_Tp_variation} and for irregular waves in Fig.~\ref{fig_power_irregwave_H_Tp_variation}. The regular waves have wave heights of $\cH$ = 0.1 m, 0.3 m, and 0.5 m, with time periods varying from 1.2 s to 4.6 s. The irregular waves considered here have significant wave heights of $\cH_s$ = 0.1 m, 0.3 m, 0.5 m, and 1 m, with peak time periods varying from 1.2 s to 3.4 s.   

The results show that the BEM-LFK solver over-predicts the time-averaged power absorbed by the device for large waves; for regular waves, $\cH$ = 0.5 m and for irregular waves, $\cH_s$ = 1 m. Further, for both regular and irregular waves, the difference between the two solvers' predictions is greater at smaller time periods than at larger time periods. This is because the natural period of oscillation of the device is 1.54 s, which falls in the small time period region, where the device oscillates with large amplitude due to the waves and actuator induced resonance. The BEM-LFK solver inherently violates the small motion assumption used in its formulation near or at resonance, and therefore provides inaccurate power estimates. A separate CFD simulation was used to determine the natural period of oscillation of the device; those simulation results aren't discussed here for brevity. 


\subsection{CFD simulations with AR-enabled wave predictions} 
\label{subsec_real_time_control}

\begin{figure}[]
   \centering
   \subfigure[AR model prediction for regular waves]{
   	\includegraphics[scale= 0.33]{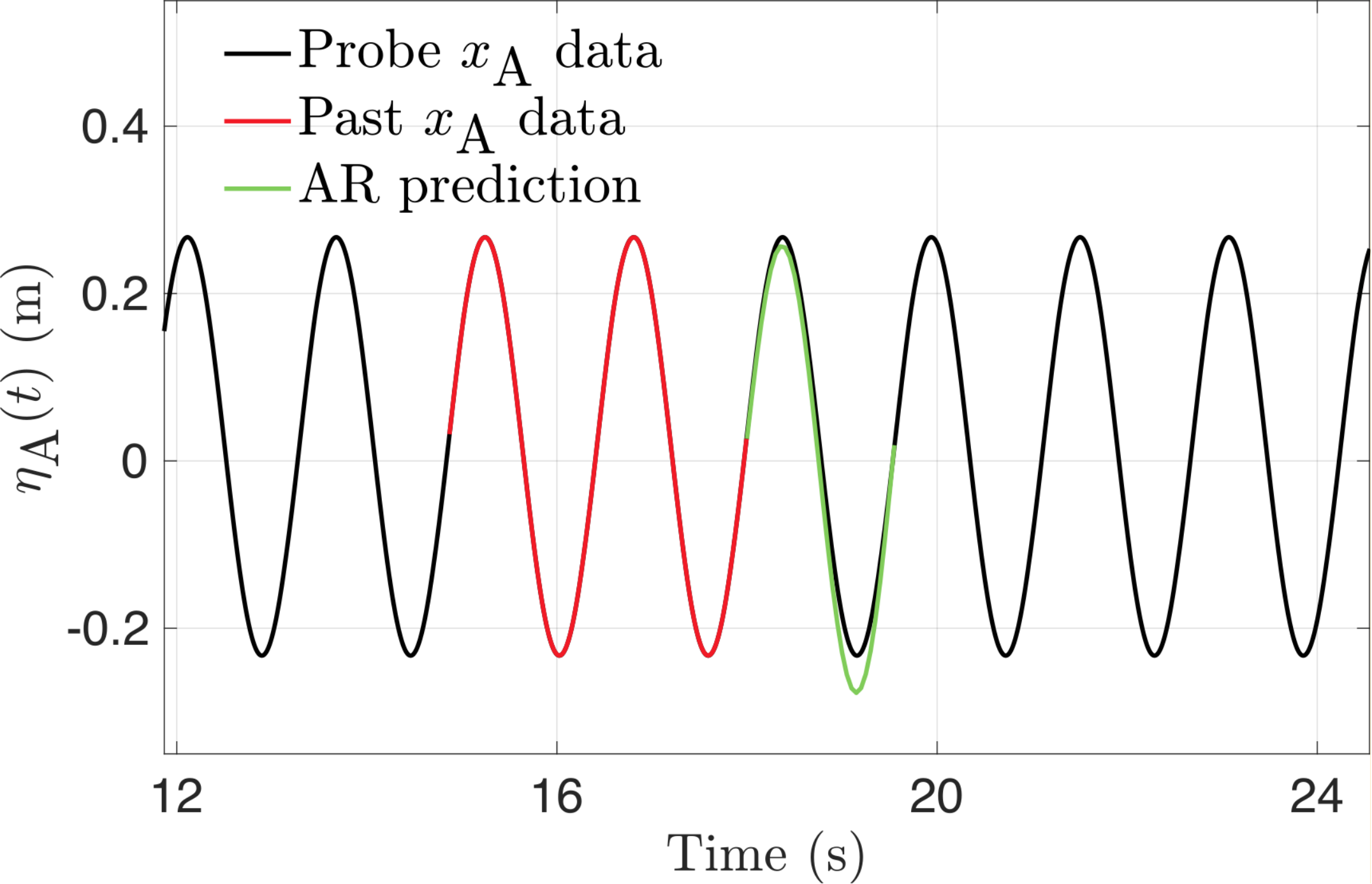}
	\label{fig_AR_regwave_pred}
   }
    \subfigure[AR model prediction for irregular waves]{
   	\includegraphics[scale= 0.33]{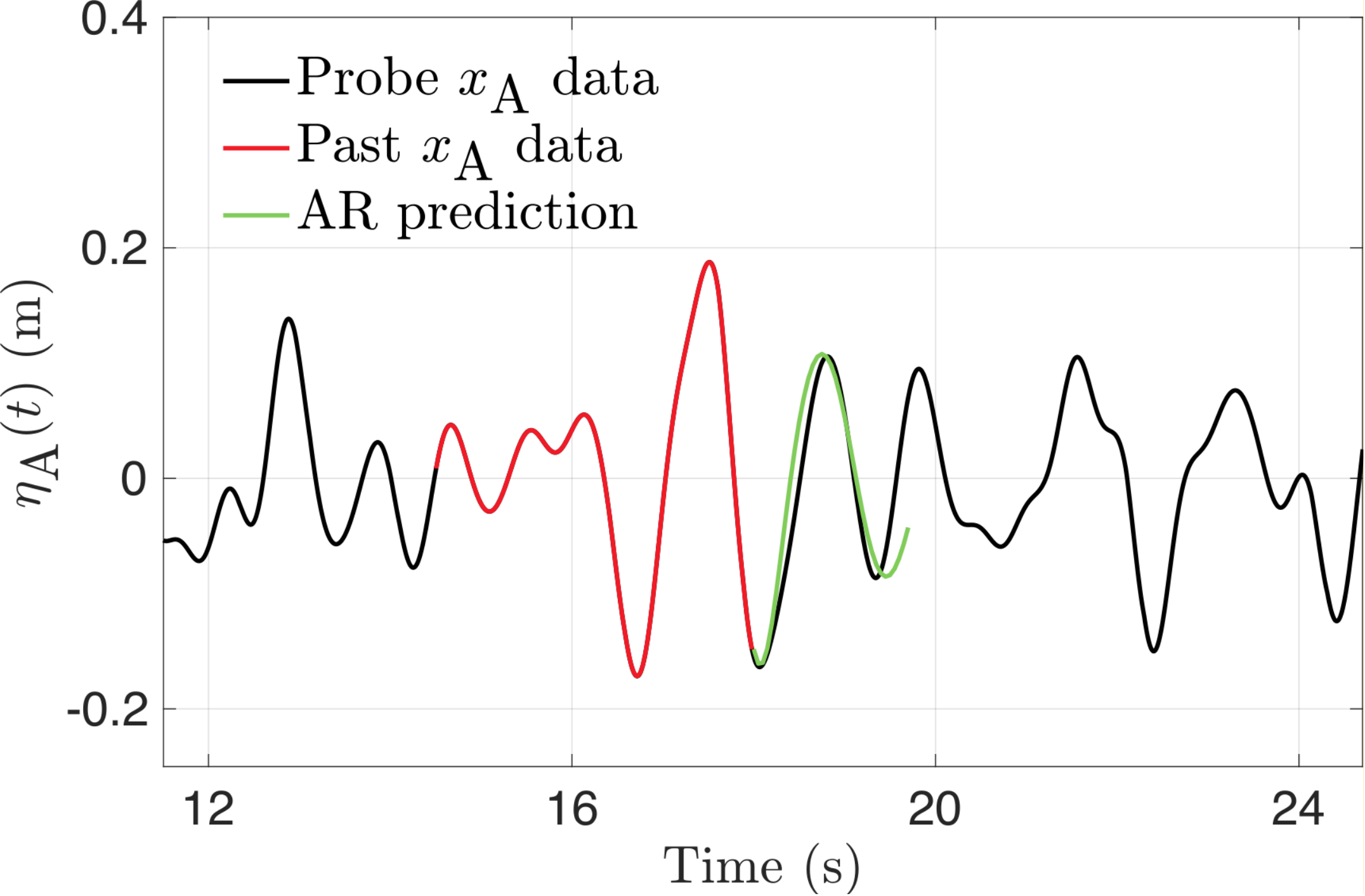}
	\label{fig_AR_irregwave_pred}
   }
   \caption{AR model predictions (\textcolor{Green}{\textbf{-----}}, green) of~\subref{fig_AR_regwave_pred} regular and~\subref{fig_AR_irregwave_pred} irregular waves for one wave period into the future using the past two wave period elevation data  (\textcolor{Red}{\textbf{-----}}, red).}
   \label{fig_AR_model_predictions}
\end{figure}

In this section, we examine the effect of AR predictions on MPC performance. In this test, we use the MPC-LFK and CFD solvers with regular waves of height $\cH$ = 0.5 m and time period $\cT$ = 1.5652 s, and with irregular waves of significant wave height $\cH_s$ = 0.3 and peak time period $\cT_p$ = 1.7475 s. We set  $AR\_start\_time$ equal to MPC start time: $ t = 10 \, \cT$ (or $10 \, \cT_p$). Therefore, the controller and the AR predictions will begin once the device exhibits steady-state oscillations under the influence of incoming waves.  MPC and NWT interaction is schematically represented in Fig.~\ref{fig_CFD_integrated_MPC}. In particular, wave elevation data at an up-wave probe point $A$ ($\eta_A$) for the past two wave periods is collected and sent to the AR model to allow for wave elevation prediction over one wave period into the future (at the same location $A$).  For predicting regular and irregular waves, we use AR models of order 3 and 5, respectively.  Figs.~\ref{fig_AR_regwave_pred} and~\ref{fig_AR_irregwave_pred} illustrate that the chosen AR models are sufficiently accurate for predicting regular and irregular waves, respectively.  Based on the past and predicted wave data, the wave excitation force $\text{F}_\text{exc}$ acting on the device is calculated using Eq.~\ref{eq_exc_force3}. 

\begin{figure}[]
   \centering
   \subfigure[Heave displacement (regular waves)]{
   	\includegraphics[scale= 0.33]{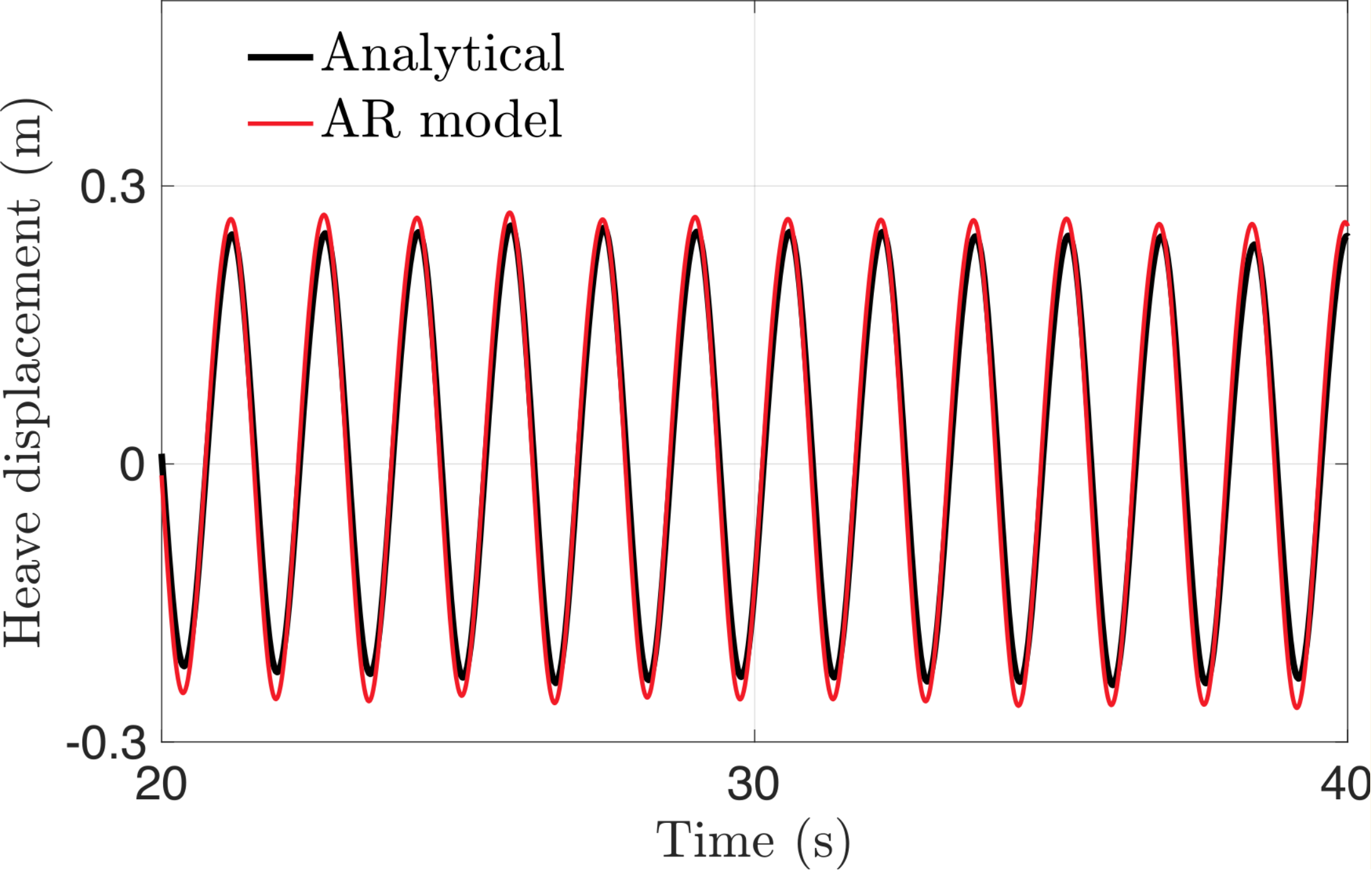}
	\label{fig_heave_CF100_regwave_AR_model}
   }
    \subfigure[Heave displacement (irregular waves)]{
   	\includegraphics[scale= 0.33]{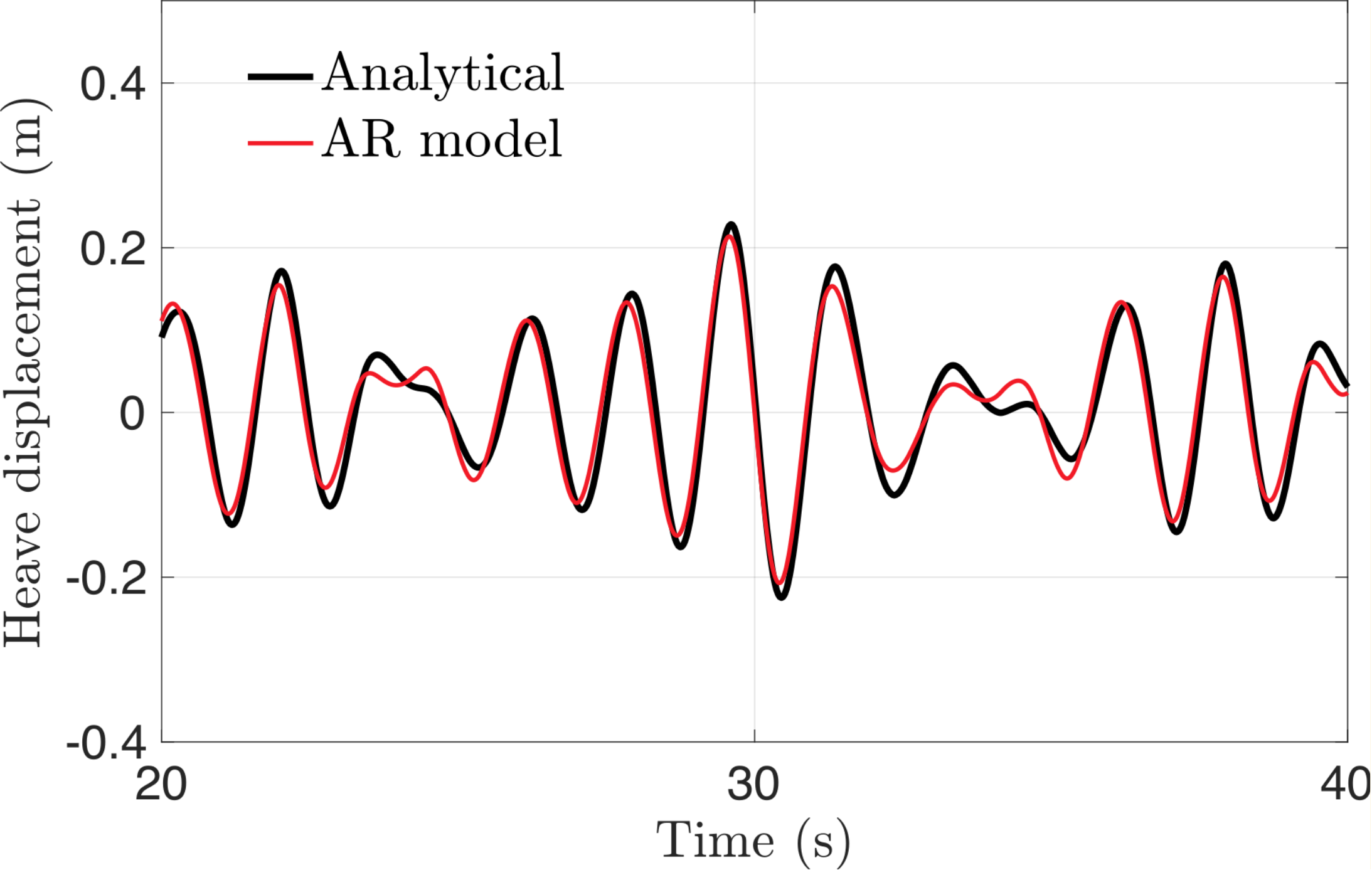}
	\label{fig_heave_CF100_irregwave_AR_model}
   }
   \subfigure[Control force (regular waves)]{
   	\includegraphics[scale = 0.33]{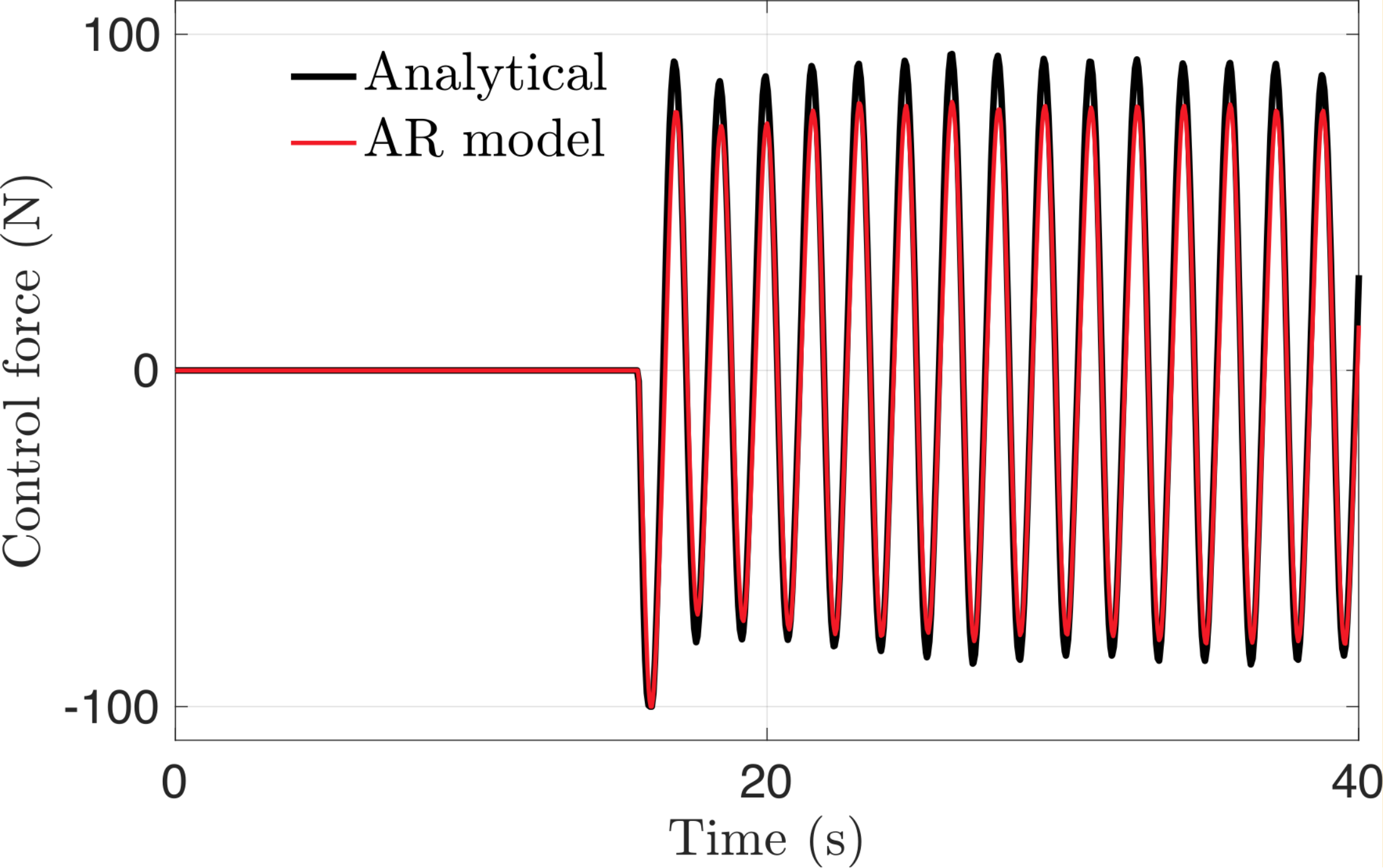}
	\label{fig_CF_CF100_regwave_AR_model}
   }
    \subfigure[Control force (irregular waves)]{
   	\includegraphics[scale = 0.33]{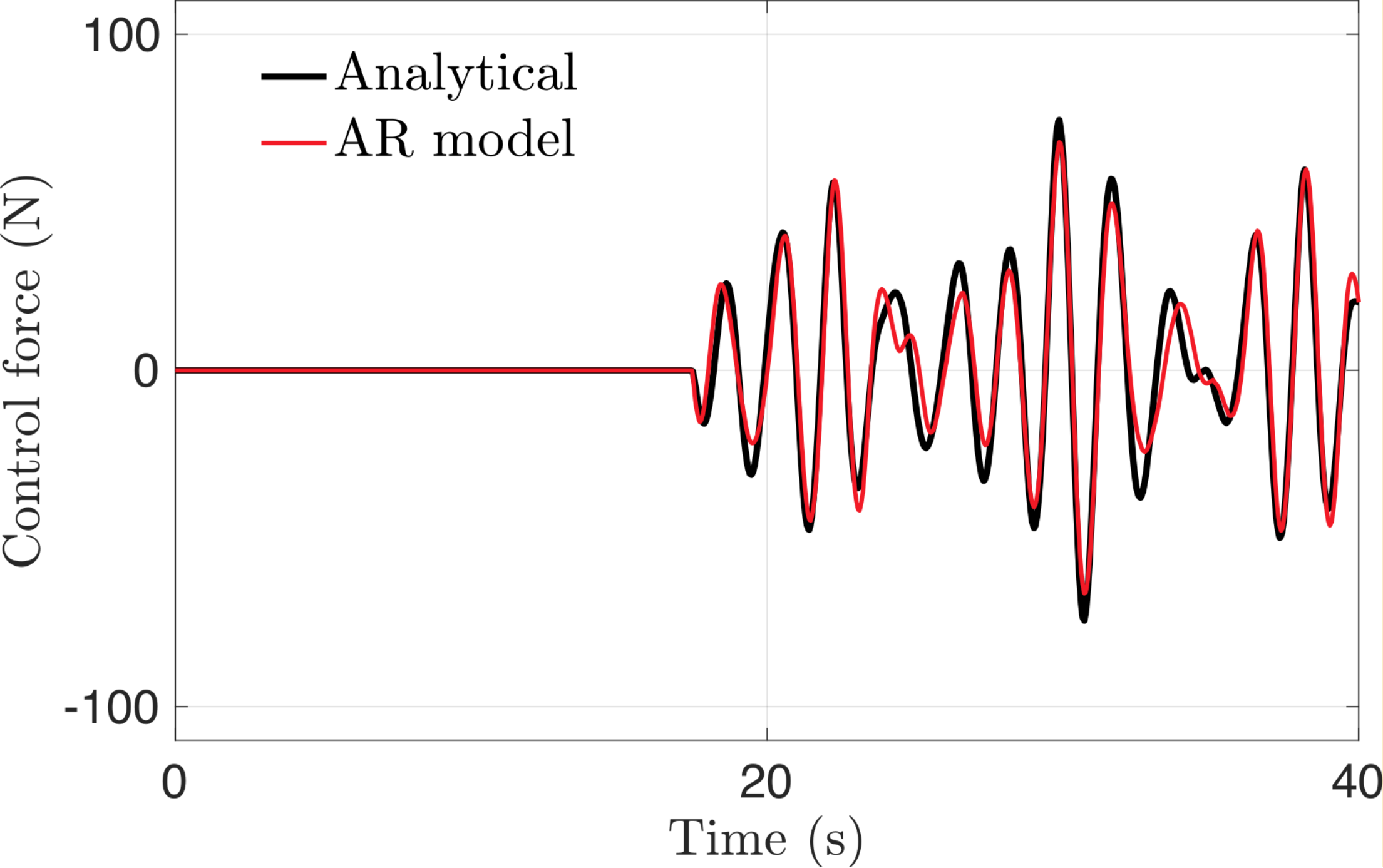}
	\label{fig_CF_CF100_irregwave_AR_model}
   }
   \subfigure[Power (regular waves)]{
   	\includegraphics[scale= 0.33]{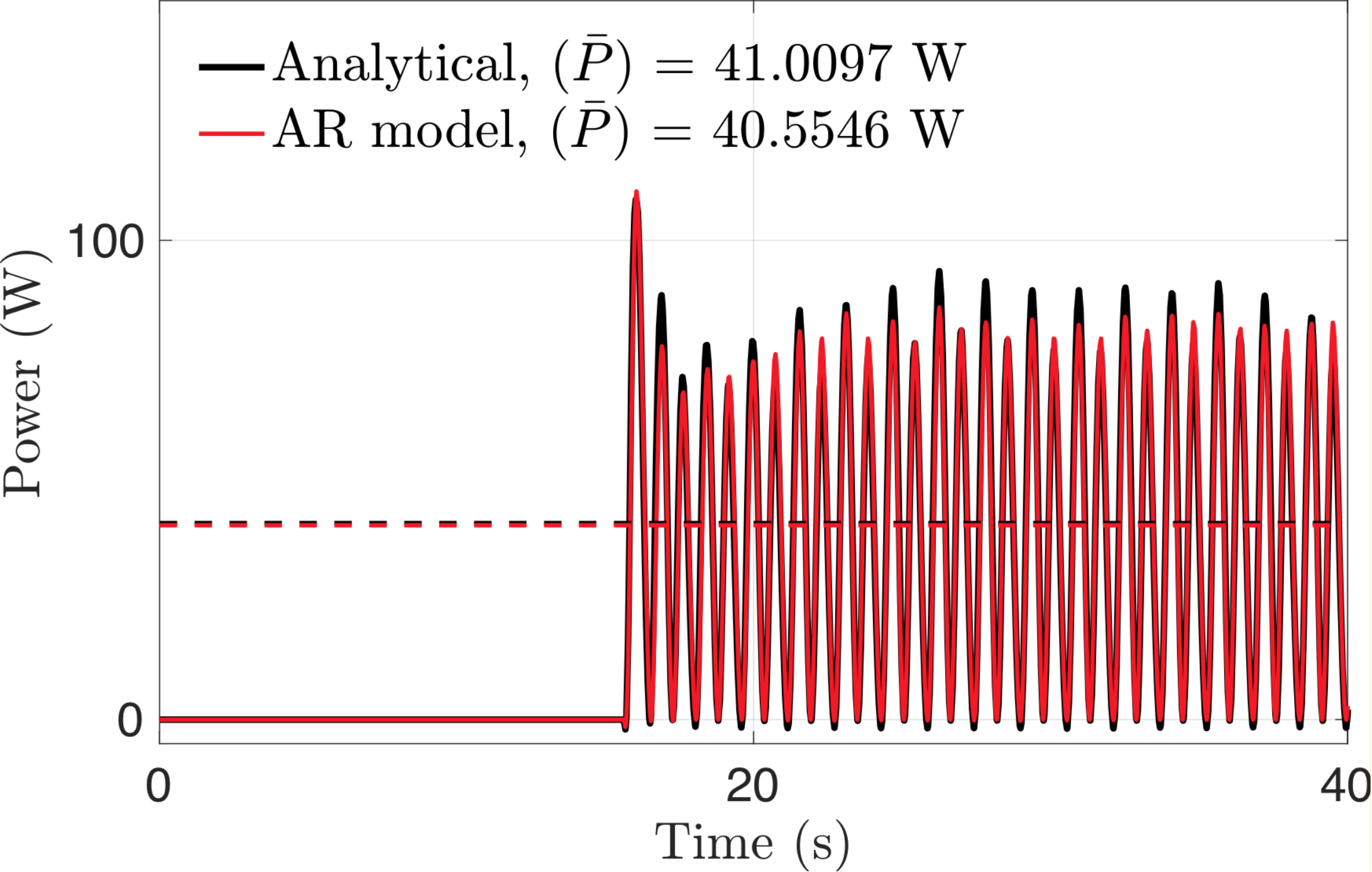}
	\label{fig_power_CF100_regwave_AR_model}
   }
    \subfigure[Power (irregular waves)]{
   	\includegraphics[scale= 0.33]{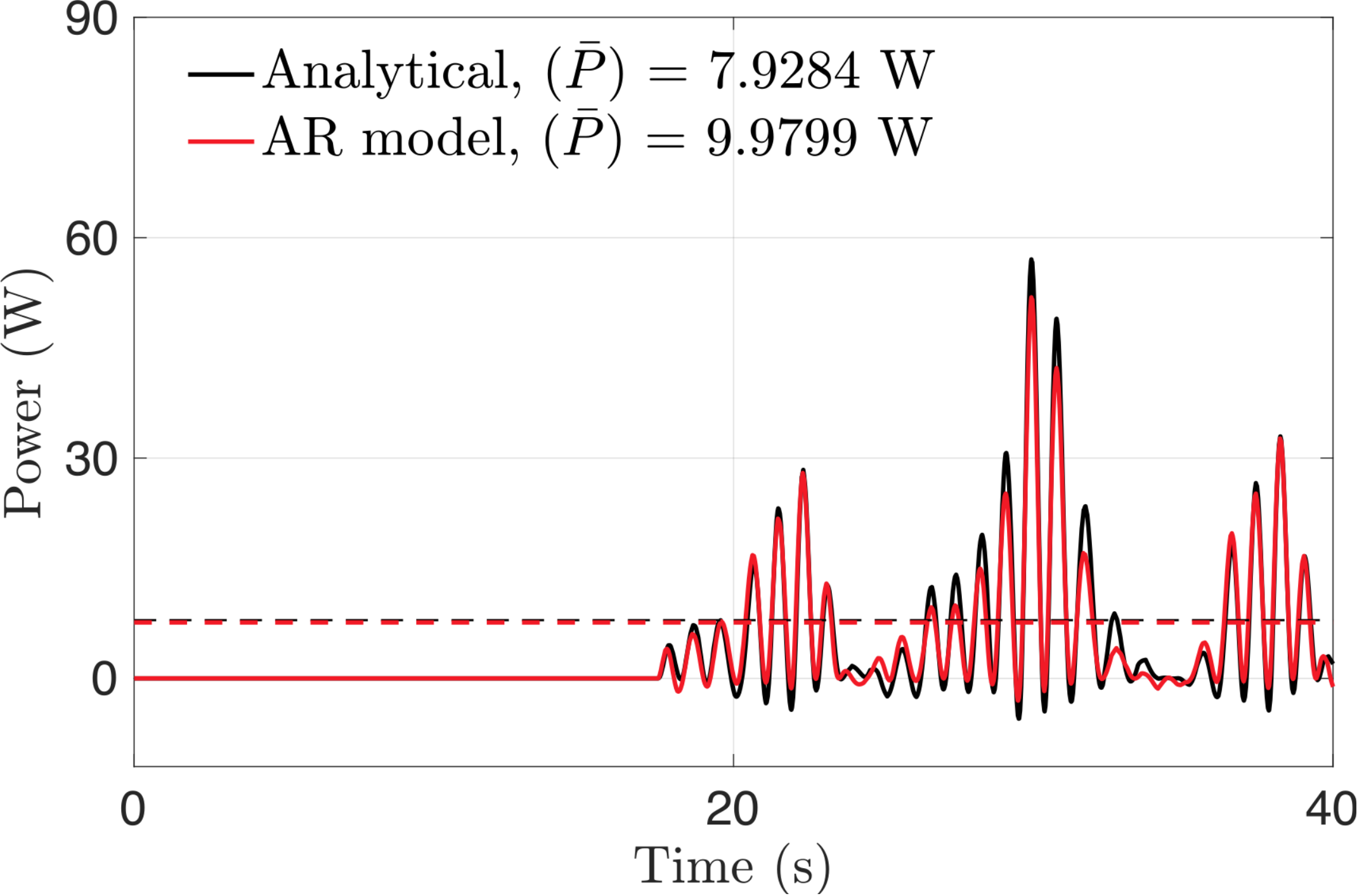}
	\label{fig_power_CF100_irregwave_AR_model}
   }
   \caption{Comparison of the controlled heave dynamics of the 3D vertical cylinder with and without AR predictions. The WEC dynamics are simulated using the CFD and MPC-LFK solver.  For regular water waves of height $\cH$ = 0.5 m and time period $\cT$ = 1.5652 s results are compared for \subref{fig_heave_CF100_regwave_AR_model} heave displacement, \subref{fig_CF_CF100_regwave_AR_model} control force,  and~\subref{fig_power_CF100_regwave_AR_model} instantaneous power. For irregular water waves of significant wave height $\cH_s$ = 0.3 m and peak time period $\cT_p$ = 1.7475 s results are compared for \subref{fig_heave_CF100_irregwave_AR_model} heave displacement, \subref{fig_CF_CF100_irregwave_AR_model} control force, and \subref{fig_power_CF100_irregwave_AR_model} instantaneous power. In all cases the control force limits are set to $\pm$ 100 N.}
   \label{fig_AR_model_results}
\end{figure}

As a test of the accuracy of the AR-integrated MPC solver, the results are compared with those obtained using analytical forcing, which was also used in Sec.~\ref{subsec_LFK_NLFK_CFD_MPC_comparison}. As for regular waves, Figs.~\ref{fig_heave_CF100_regwave_AR_model}, \ref{fig_CF_CF100_regwave_AR_model}, and \ref{fig_power_CF100_regwave_AR_model} compare the heave displacement, control force, and the instantaneous power absorbed by the device, respectively.  Figs.~\ref{fig_heave_CF100_irregwave_AR_model}, \ref{fig_CF_CF100_irregwave_AR_model}, and~\ref{fig_power_CF100_irregwave_AR_model} compare these quantities for irregular waves.  The results show that the device dynamics are very close with or without the AR predictions. The time-averaged power absorbed by the WEC device subject to regular waves is 40.5546 W when the AR model is enabled. The value of 41.0097 W obtained by analytical forcing agrees well with this result. In the case of irregular waves, these values are 9.9799 W and 7.9284 W, which also match fairly well. Further improvements can be obtained for the irregular wave case by using a better method of time-series forecasting or by fine-tuning the AR model. 

We conclude from the results of this section that our technique of collecting wave elevation data from an up-wave location in the NWT and predicting future waves based on it (through an AR model) works well with the CFD/MPC-LFK solver combination. 

\subsection{Power transfer from waves to the PTO system: Verifying the relationships with CFD simulations} 
\label{subsec_power_transfer_verification}

We re-analyze the AR-enabled CFD simulations of the previous section to verify the power transfer relations in Sec.~\ref{sec_power_transfer_pathway}. In the case of regular waves of height  $\cH$ = 0.5 m and time period $\cT$ = 1.5652 s, the power transferred by the waves to the device (or the work done by the hydrodynamic forces) is $\widebar{P}_\text{waves $\rightarrow$ cyl}$ = 38 W and that absorbed by the PTO unit is $\widebar{P}_\text{PTO}$ = 39 W. A time average is taken from $t$ = 30 s to 31.5652 s, i.e., for one wave period. Based on these power values, we conclude that the power transfer Eq.~\eqref{eq_KE_eq3} is verified in the case of regular waves.  In the case of irregular waves, we calculate the left and right sides of Eq.~\eqref{eq_KE_eq2} separately. $t = $ 30 s to 40 s is chosen as the time interval for time-averaging the terms of the equation. Accordingly, the two sides of the equation evaluate to 72.06 W and 71.47 W, respectively, which also match reasonably well.  
  
Based on the results of this section, we conclude that our CFD simulations satisfy the power transfer relationships of Sec.~\ref{sec_power_transfer_pathway}.


\subsection{MPC adaptivity} 
\label{subsec_mpc_adaptivity}


To test the adaptive capability of MPC for WEC devices, we simulate the dynamics of the 3D vertical cylinder subject to changing sea states. Specifically, three consecutive sea states are considered within a single CFD simulation: sea state 1 consisting of first-order regular waves  of height $\cH$ = 0.1 m and time period $\cT$ =  1.5652 s between $t_1 = 0$ s to $t_2 = 40$ s, sea state 2 consisting of first-order regular waves  of height $\cH$ = 0.2 m and time period $\cT$ = 2 s between $t_2 = 40$ s to $ t_3 = 60$ s, and sea state 3 consisting of first-order regular waves  of height $\cH$ = 0.15 m and time period $\cT$ = 1.7475 s between $t_3 = 60$ s to  $t_4 = 120$ s. The wave elevation is smoothly varied from one sea state to the other using the following expression:
\begin{equation}
	\eta_{i, i+1}(t) = \eta_i(t) + (\eta_{i+1}(t) - \eta_i(t)) \cdot (1 + \tanh(t - (t_{i+1} - t_\text{half-interval})) / 2,
\label{eq_eta_adapt_case}
\end{equation}
in which $\eta_i(t) = (\cH_i/2)\cos(\kappa_i x - \omega_i t)$ and $t_\text{half-interval} = 5$ s is the transition time between sea state $i$ to $i+1$. AR predictions are also enabled for the CFD simulation. For MPC, each sea state uses a pre-configured AR model that is optimized offline. While this is inconvenient, it is necessary to allow accurate predictions of wave excitation forces.

\begin{figure}
   \centering
   \subfigure[Heave displacement]{
   	\includegraphics[scale= 0.34]{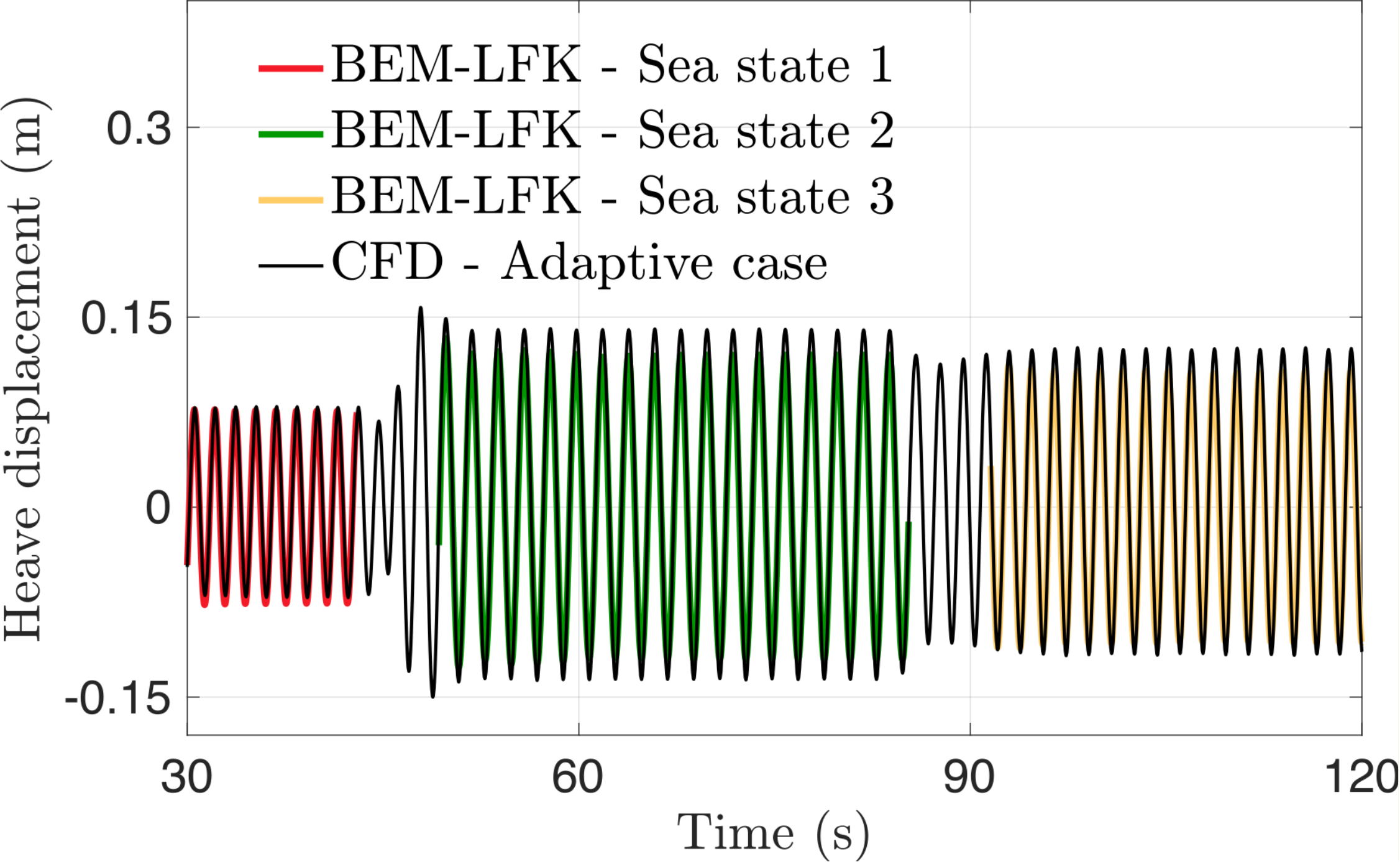}
	\label{fig_heave_regwave_adaptive_case}
   }
   \subfigure[Heave velocity]{
   	\includegraphics[scale = 0.34]{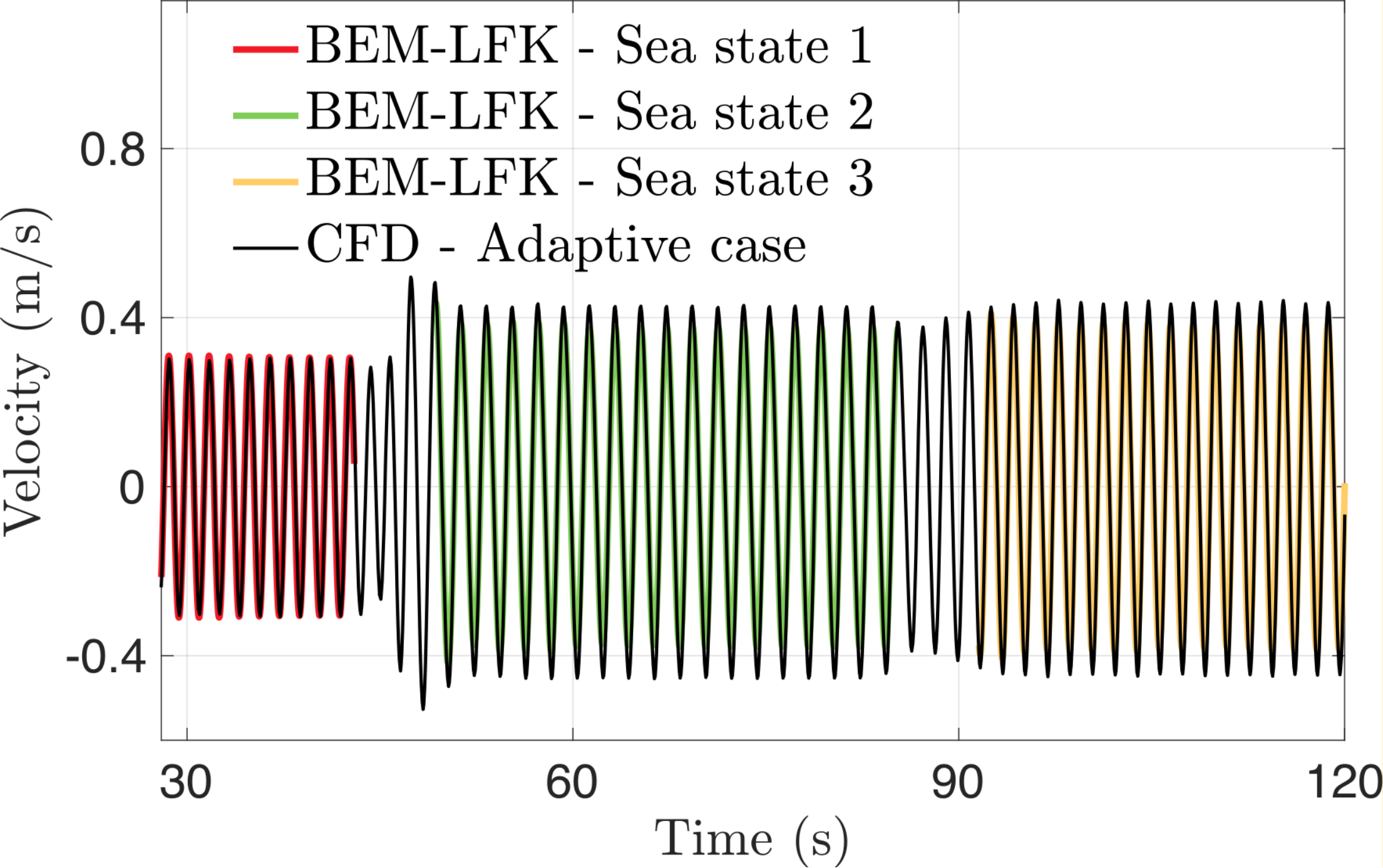}
	\label{fig_velocity_regwave_adaptive_case}
   }
   \caption{Comparison of the~\subref{fig_heave_regwave_adaptive_case} heave displacement and \subref{fig_velocity_regwave_adaptive_case} velocity of the device subject to changing sea states using CFD and BEM-LFK solvers. The BEM-LFK solver solves the three sea states separately, whereas the CFD solver considers them consecutively.}
   \label{fig_adaptive_case}
\end{figure}

Fig.~\ref{fig_adaptive_case} shows the temporal evolution of the heave displacement and velocity. We compare the CFD results with three separate BEM-LFK simulations for different sea states. Because all three sea states have small amplitude waves, the BEM-LFK solver is expected to be accurate. Indeed, it is observed that the adaptive CFD simulation agrees well with the BEM-LFK solver results, which indicates that the MPC algorithm is able to adapt according to the current sea state and produces an optimal solution in each case. 

%% file: Conclusions.tex

In this study, we simulated the controlled dynamics of a heaving 3D vertical cylinder WEC device using  BEM and multiphase IB solvers. A MPC strategy was used to maximize the energy absorption capacity of the WEC device under regular and irregular sea conditions.

We validated our BEM-LFK and MPC-LFK implementations by simulating a benchmarking case from Cretel et al.~\cite{Cretel2011} in Sec.~\ref{sec_validation_and_motivation}. The scaled-down version of the same device was then simulated using the multiphase IB solver, and its wave excitation forces were significantly greater than those predicted by the BEM solvers. A more surprising result was that the WEC device drew a large amount of power from the grid instead of producing energy, as predicted by the BEM solvers. Moreover, it was observed that $J_3$ is a better choice for the model predictive control of WECs compared to $J_2$, as the latter can provide misleading power output.  To understand the main cause of the discrepancy, we examined six different combinations of the WSI and MPC solvers using $J_3$ as the cost function. It is found that when the sea state is calm and the wave height is small, the BEM solvers' predictions match well with the CFD solver's. However, in agitated sea conditions, the BEM solvers over-predict the device performance, which can be misleading to the device designer. On the other hand, the CFD solver provides realistic results both in calm and agitated sea conditions. It is evident that resolving the hydrodynamic non-linearities associated with the WSI is essential to obtaining realistic estimates of the device's power. It is further confirmed by the results of the BEM-NLFK solver, which are closer to those of the multiphase IB solver. Therefore, we recommend using the BEM-NLFK solver to study the controlled dynamics of WECs when computational resources are limited to employing a CFD solver. In addition, it is straightforward to switch to the BEM-NLFK solver by using the static grid technique described in Sec.~\ref{sec_NLFK_calculation}.  Additionally, we found that the choice between MPC-LFK or MPC-NLFK is irrelevant, as both algorithms give very similar results. Nevertheless, MPC-LFK solver is computationally-efficient and is proposed as a practical model-based control for WECs. 

We also compared MPC-LFK performance with and without AR predictions in Sec.~\ref{subsec_LFK_NLFK_CFD_MPC_comparison}. We found that the AR prediction strategy worked well in both regular and irregular waves. The AR model can be tuned further or a different time-series forecasting algorithm can be used for further improvements. The pathway of energy transfer from waves to the PTO unit for the heaving WEC device was also derived and confirmed. By simulating three different sea states consecutively within a single CFD simulation, we tested the adaptive capabilities of MPC of WECs. The MPC is shown to adapt to different sea states and find the optimal solution for each situation, thus living up to its reputation as the ``Tesla" of control approaches.
